\documentclass[twocolumn,prb,aps,amsmath,amssymb,floatfix,showpacs,longbibliography,superscriptaddress]{revtex4-1}

\usepackage{color}
\usepackage{graphicx}
\usepackage{dcolumn}
\usepackage{bm}
\usepackage{subfigure}
\usepackage{array}
\newcolumntype{L}{>{\centering\arraybackslash}m{0.33\textwidth}}
\usepackage[colorlinks=true,linkcolor=blue,citecolor=blue,urlcolor=blue]{hyperref}
\usepackage[normalem]{ulem}

\graphicspath{{./figs/}}
\begin{document}

\newcommand{\bo}{\boldsymbol}
\newcommand{\boq}{\mathbf{q}}
\newcommand{\bok}{\mathbf{k}}
\newcommand{\bor}{\mathbf{r}}
\newcommand{\boG}{\mathbf{G}}
\newcommand{\boR}{\mathbf{R}}
\newcommand\2{$_2$}
\newcommand\4{$_4$}

\newcommand\theosmarvel{Theory and Simulation of Materials (THEOS), and National Centre for Computational Design and Discovery of Novel Materials (MARVEL), \'Ecole Polytechnique F\'ed\'erale de Lausanne, CH-1015 Lausanne, Switzerland}
\newcommand\geneva{Department of Quantum Matter Physics, University of Geneva, CH-1211 Geneva, Switzerland}

\title{Mobility of 2D materials from first principles in an accurate and automated framework}

\author{Thibault Sohier}
\affiliation{\theosmarvel}
\author{Davide Campi}
\affiliation{\theosmarvel}
\author{Nicola Marzari}
\affiliation{\theosmarvel}
\author{Marco Gibertini}
\affiliation{\theosmarvel}
\affiliation{\geneva}

\date{\today}


\begin{abstract}
We present a first-principles approach to compute the transport properties 
of 2D materials in an accurate and automated framework. We use density-functional 
perturbation theory in the appropriate bidimensional setup with open-boundary 
conditions in the third direction. The materials are charged by field effect via planar counter-charges. 
In this approach, we obtain electron-phonon matrix elements in which dimensionality and doping effects are inherently accounted for, without the need for post-processing corrections. 
This treatment highlights some unexpected consequences, such as an increase of 
electron-phonon coupling with doping in transition-metal dichalcogenides.
We use symmetries extensively and identify pockets of relevant electronic states 
to minimize the number of electron-phonon interactions to 
compute; the integrodifferential Boltzmann transport equation is then linearized and solved beyond the relaxation-time approximation. We apply the entire protocol to a set of much studied materials with diverse electronic and vibrational band structures: 
electron-doped MoS\2, WS\2, WSe\2, phosphorene, arsenene, and hole-doped phosphorene. 
Among these, hole-doped phosphorene is found to have the highest mobility, with a room temperature value around $600$ cm$^2\cdot$V$^{-1}\cdot$s$^{-1}$. 
Last, we identify the factors that affect most phonon-limited mobilities, such as the number and the anisotropy of electron and hole pockets, to provide a broader understanding of the driving forces behind high mobilities in two-dimensional materials.
\end{abstract}

\maketitle

\section{Introduction}

The scientific and engineering community is devoting a major effort towards the identification and 
fabrication of novel 2D materials\cite{Efetov2010,Radisavljevic2011,Lebegue2013,Hanlon2015,Yasaei2015,Gibaja2016,Ares2016,Bandurin2016,
Ashton2017,Cheon2017,Mounet2018} and their application in electronic devices\cite{Wang2012,Chhowalla2016}. 
To accelerate the discovery of the best candidates for electronic transport, systematic and 
accurate methods to compute phonon-limited resistivity and mobility from first principles would be 
very beneficial. 
Nevertheless, these quantities are not straightforward to compute and various degrees of 
approximations are often needed. 

A very successful approach relies on the solution of the Boltzmann transport equation\cite{Ziman,Allen1978} (BTE), 
in which the electron-phonon coupling (EPC) enters the expression for the  scattering rates\cite{Grimvall,Giustino2017} and 
can be computed from first principles within density-functional perturbation theory 
(DFPT)\cite{Savrasov1994,Liu1996,Mauri1996,Bauer1998,Baroni2001} or with finite-differences schemes\cite{Dacorogna1985,Gunst2016a}. 
Although both methods proved to be very successful, DFPT is more efficient in dealing with long-wavelength effects since it does not need large supercells to accommodate the perturbation, and will thus be the method of choice in the following.

Either way, the computation of EPC in 2D materials from first principles presents some challenges. Being ruled by long-range Coulomb interactions, the long-wavelength electron-phonon dynamics are highly dependent on dimensionality. The most obvious example is the Fr\"ohlich interaction in polar materials, which diverges in the 
long-wavelength limit in 3D while remains finite in 2D \cite{Sohier2016}. This fundamentally different behaviour is difficult to capture in standard electronic-structure codes based on plane-wave basis sets because of the spurious interaction with artificial periodic images, which leads to erroneous results when ${\bm q}\to0$. Thus, special care must be adopted when computing the response of a 2D system to long-wavelength perturbations\cite{Kozinsky2006,Sohier2015,Sohier2016,Sohier2017nl}.

An appropriate treatment of doping is an additional challenge. While field-effect charging is 
omnipresent in experimental setups, it is usually not included in first-principles 
simulations. Often, calculations are performed for the neutral case, and the 
Fermi level is shifted within a frozen-band approximation 
when integrating the EPC matrix elements to 
obtain a certain scattering-related quantity. This is approximate because the EPC 
itself depends on doping, most notably via screening, but also via other 
mechanisms that will be discussed later. 
A more realistic description is now possible thanks to a recent development of DFPT for gated 2D materials\cite{Sohier2017}, where the doping charge is neutralized by adding planar distributions of counter-charges, mimicking experimental conditions where 2D materials are doped electrostatically through gates.

Another source of challenge is that, in general, a very large number of EPC matrix elements must be 
computed to obtain a fine enough sampling of the electron-phonon dynamics over the Brillouin zone, 
which is computationally very expensive. To boost efficiency, one may use  Wannier interpolations 
\cite{Lee2005,Yates2007,Giustino2007,Calandra2010,Marzari2012,Ponce2016} to 
obtain inexpensively the EPC over a dense mesh. However, interpolations work seamlessly only
with short-range interactions. Long-range effects, such as Fr\"ohlich or piezo-electric coupling, need 
to be modeled and treated separately; this has been solved for three-dimensional 
systems\cite{Sjakste2015,Verdi2015}, while in 2D, 
Fr\"ohlich coupling has been modeled\cite{Sohier2016} and efficient ways to interpolate 
the phonon dispersions have been put forward\cite{Sohier2017nl}, but a thorough description of EPC 
interpolation is still missing.  
Nonetheless, the absence of interpolation techniques might not be, in the long run, a major constraint. Indeed, the reduced dimensionality entails one less dimension to sample, with a drastic reduction of electronic states and phonons to consider. Thus, it 
can be argued that for 2D materials, one should focus on addressing dimensionality 
and charging first, rather than interpolation.

Here, we thus choose a direct approach and compute all EPC matrix elements from DFPT in 
the appropriate boundary and charging conditions\cite{Sohier2017}.
The use of symmetries and energy selection rules, in addition to 
the momenta being restricted to two dimensions, makes the process computationally feasible. 

For the solution to the Boltzmann transport equation for electrons and holes, 
different approaches involving various degrees of approximation have been put forward. 
In the case of metals, Allen derived an approximate solution\cite{Allen1978} by suitably modifying Eliashberg theory\cite{Grimvall} to transport. This approach has been successfully applied in the early efforts to compute mobilities from first principles\cite{Savrasov1996,Bauer1998} and it is now customary in many available transport codes such as EPW\cite{Ponce2016}. 
In the general case, including semiconductors, the major challenge is that the integrodifferential BTE does not have a closed-form solution.  
Most approaches use then some form of relaxation-time 
approximation\cite{Restrepo2009,Shishir2009,Borysenko2010,Jin2014,Restrepo2014,Park2014,
Gunst2016a,Jin2016,Rudenko2016,Trushkov2017,Gaddemane2018,Ponce2018} to obtain a closed-form result.  

Still, the errors associated with the various approximate solutions to the BTE in the literature 
are often difficult to quantify. 
An iterative scheme inspired by the original Rode's method~\cite{Rode1970a,Rode1970b} to solve the inelastic part of the 
BTE has been recently proposed\cite{Kaasbjerg2012a,Kaasbjerg2013,Li2015,Ma2018}, while a very efficient preconditioned conjugate-gradient approach has been reported in Ref.~\onlinecite{Fiorentini2016} following a recipe introduced in the context of the phonon Boltzmann equation for thermal transport\cite{Fugallo2013,Fugallo2014,Cepellotti2015}. 
Ref.~\onlinecite{Li2015} also offers an interesting comparison of the different 
methods to solve the BTE and shows a broad agreement between these for the case of MoS\2; still, this might not be the case for all materials. A full numerical solution of the BTE beyond the relaxation-time approximation was introduced in Ref.~\onlinecite{Sohier2014a} for graphene. An advantage of this latter method is that it does not require a broadening of the $\delta$ functions enforcing energy selection rules, contrary to the other methods mentioned above. It was, however, tailored specifically for graphene. Here, we propose a more general approach with iterative scheme combined with the use of the triangles method\cite{Ashraff1987,Pedersen2008} for accurate integrations of the $\delta$ functions.

Relatively few 2D materials have had mobilities theoretically investigated up to now. Graphene has been studied 
extensively\cite{Ando2006,Hwang2007,Manes2007,Samsonidze2007,Stauber2007,Fratini2008,Hwang2008a,Shishir2009,Hwang2009,Borysenko2010,Mariani2010,DasSarma2011,DasSarma2011,Kaasbjerg2012,Hwang,Restrepo2014,Park2014,Sohier2014a,Gunst2016a,Gunst2017}, showing excellent agreement\cite{Park2014} with experiments\cite{Efetov2010} and  a detailed understanding of the main processes limiting mobility\cite{Borysenko2010}, including the effects of 
dimensionality and charging by field effect\cite{Gunst2017,Sohier2017}.
MoS\2, another prototypical 2D material,  
has also been studied in several works
\cite{Gunst2016a,Kaasbjerg2012a,Kaasbjerg2013,Li2015,Jin2014,Restrepo2014}, as well as
phosphorene \cite{Gaddemane2018,Qiao2014,Jin2016,Trushkov2017,Rudenko2016,Liao2015}, 
arsenene \cite{Wang2015,Pizzi2016nc,Wang2017},
silicene \cite{Gunst2016a}, and other TMDs \cite{Jin2014}. 
On the other hand, the effects of periodic images and dimensionality are explicitly treated only in some of the 
first-principles efforts\cite{Gunst2016a,Kaasbjerg2012a,Kaasbjerg2013}. Charging is most often treated as a rigid shift of the Fermi energy in the computation of the transport properties, 
but not included in the computation of the matrix elements themselves. 
Consequently, the influence of doping on the EPC is generally neglected. 
In some instances, electronic screening is accounted for analytically\cite{Kaasbjerg2012a,Kaasbjerg2013}, 
but this necessarily entails the use of models and approximations (not obvious in 2D). 
Also, as will be shown here, screening is not the only effect of doping on EPC matrix elements. Last, computational accuracy is often limited by the very expensive nature of the Brillouin zone (BZ) integrals. For all these reasons, a full treatment of doping and periodicity is necessary, together with an efficient and automatic implementation of all BZ sums.

All these points are addressed in this paper, that is structured as follows. In the first section, we describe the formal framework of the BTE and identify the 
quantities needed to solve it. In the second section, we present the workflow of first-principles calculations used to 
compute the physical quantities associated with phonon-limited transport.
Then, we discuss the results of this workflow applied to a set of known prototypical materials, 
i.e., three electron-doped TMDs in their 2H form (MoS\2, WS\2, WSe\2), as well as electron-doped (gray) arsenene, and both hole- and electron-doped (black) phosphorene.

\section{Boltzmann Transport Equation}
\label{sec:theory}

In this first section, we review the framework of the Boltzmann transport equation for electrons in an effort to settle the context and introduce the various quantities used in the rest of the paper. Similar derivations can be found in the literature, e.g. in Ref.~\onlinecite{Ziman,Ashcroft1976,Nag1980}.   
We consider any 2D material to lie in the $x-y$ plane with an applied electric field in the same plane, in the $-\bo{u}_{\bo{E}}$ direction. The electric field favors states with wavevectors $\bok$ in the opposite direction, $\bo{u}_{\bo{E}}$, thus taking the electronic distribution $f$ out of its equilibrium Fermi-Dirac ground-state $f^0$. Phonon scattering acts to bring the system back towards its unperturbed equilibrium state; then, a steady-state regime is reached and a net electric current $\bo{j}$ emerges. The conductivity, or the inverse of the resistivity, is then defined as\cite{Ziman,Ashcroft1976,Nag1980}:
\begin{align} \label{eq:conductivity}
\sigma=\frac{1}{\rho}= \sum_{\alpha} \frac{2e}{|\bo{E}|} \int_{k \in \alpha} \frac{d\bok}{(2\pi)^2} f(\bok) \bo{v}(\bok) \cdot \bo{u}_{\bo{E}}
\end{align}
where the factor 2 accounts for spin degeneracy, $e>0$ is the Coulomb charge, $\bo{E}$ the electric field, $\alpha$ is an index representing the different valleys in the Brillouin zone, $f(\bok)$ is the steady-state occupation function, and $\bo{v}(\bok)$ the band velocity of the electronic state. Here, rather than integrating over all the Brillouin zone, we limit ourselves to the relevant electron or hole pockets that are occupied by temperature or doping. In semiconductors, these pockets form valleys which we will further define later, and wavevectors $\bok$ are taken from these valleys. Mobilities can then be obtained from the Drude model as $\mu=\sigma/n$, where $n$ is the electron density (replace with hole density $p$ in the case of hole doping). 
The central quantity to obtain is thus the occupation distribution $f(\bok)$. Assuming this to be spatially uniform and time-independent (steady-state), the Boltzmann transport equation states that the change of the occupation distribution driven by the electric field must be compensated by  scattering\cite{Ziman,Ashcroft1976,Nag1980}:
\begin{align} \label{eq:BTE0}
-\frac{e\bo{E}}{\hbar} \cdot \frac{\partial f}{\partial \bok}= \left( \frac{\partial f}{\partial t} \right)_{\rm{scatt}}(\bok)
\end{align} 
The collision integral (right-hand side) can be found using Fermi's golden rule:
\begin{align} \label{eq:coll}
\begin{split}
\left( \frac{\partial f}{\partial t} \right)_{\rm{scatt}}(\bok) = \sum_{\bok'} & P_{\bok'\bok} f(\bok') \left(1-f(\bok)\right)  \\
& - P_{\bok\bok'} f(\bok) \left(1-f(\bok')\right),
\end{split}
\end{align} 
where $P_{\bok\bok'}$ is the scattering probability from state $\bok$ to state $\bok'$; $P_{\bok\bok'}$ should include in general all relevant scattering processes. Here, we sum over all electron-phonon scattering probabilities associated with the phonons modes of the system, i.e. $P_{\bok\bok'}=\sum_{\nu} P_{\bok\bok', \nu}$, where $P_{\bok\bok', \nu}$ is the sum of phonon emission and absorption terms\cite{Grimvall} for phonon mode $\nu$ of momentum $\boq=\bok'-\bok$:
\begin{align}\label{eq:Pkkp}
\begin{split}
P_{\bok \bok+\boq, \nu}= \frac{2\pi}{\hbar} \frac{1}{N} & |g_{\bok \bok+\boq, \nu}|^2 
\{
n_{\boq, \nu} \delta(\varepsilon_{\bok+\boq}-\varepsilon_{\bok}-\hbar \omega_{\boq, \nu}) \\
+&(n_{\boq, \nu}+1) \delta(\varepsilon_{\bok+\boq}-\varepsilon_{\bok}+\hbar \omega_{\boq, \nu})
\}.
\end{split}
\end{align} 
In the expression above, $g_{\bok \bok+\boq, \nu}$ is the electron-phonon coupling (EPC) matrix element and $n_{\boq, \nu}$ is the phonon occupation, which we assume to be the equilibrium Bose-Einstein distribution. The $\delta$ functions stem from the energy selection rules involved in scattering, forcing the need to evaluate the EPC on very fine momentum grids, making the calculations challenging. 
Phonon scattering as described in Eq. \ref{eq:Pkkp} involves three quasi-particles: an initial electronic state, a final electronic sate, and a phonon.
The EPC matrix element connecting the initial and final states can be computed within DFPT\cite{Baroni2001,Giustino2017} and reads:
\begin{align} \label{eq:EPC}
g_{\bok, \bok+\boq,\nu}= \sum_{a,i} \bo{e}^{a,i}_{\boq,\nu} \sqrt{\frac{\hbar}{2M_a \omega_{\boq,\nu}}} \langle \bok+\boq | \frac{\partial V_{\rm{KS}}(\bor) }{\partial \bo{u}_{a,i}(\boq)} |\bok\rangle 
\end{align}
where $a$ is an atomic index, $i$ a cartesian index, $\bo{e}^{a,i}_{\boq,\nu}$ is the phonon eigenvector, $M_a$ is the mass of atom $a$, $|\bok\rangle ,|\bok+\boq\rangle $
are the initial and final electronic states, and $ \frac{\partial V_{\rm{KS}}(\bor) }{\partial \bo{u}_{a,i}(\boq)}$ is the derivative of the Kohn-Sham potential with respect to a periodic displacement of atom $a$ in direction $i$.

We adopt a perturbation approach at first order in electric field and write :
\begin{align} \label{eq:f0f1}
f(\bok)=f^0(\bok)+f^1(\bok)
\end{align}
where $f^0(\bok)$ is the Fermi-Dirac function and $f^1(\bok)$ is the linear perturbation proportional to the electric field, for which it is convenient to make the following general ansatz
\begin{align} \label{eq:ansatz}
f^1(\bok)= e|\bo{E}|\bo{u}_{\bo{E}} \cdot\bo{F}(\bok)\frac{\partial f^0(\bok)}{\partial \varepsilon} 
\end{align} 
where $\bo{F}(\bok)$ is a vectorial quantity with units of length that can be understood as a mean free displacement\cite{Li2015}. In the following the only approximation we make is to assume that this mean free displacement is along the band velocity, that is we write:
\begin{align}\label{eq:approx}
\bo{F}(\bok) = \bo{v}(\bok) \tau(\bok)
\end{align}
where $\tau(\bok)$ is unknown and has the dimension of time. We will call it scattering time, although it is {\it not} the  scattering time as often understood in the context of the relaxation-time approximation.
This is similar to what has been proposed by Rode~\cite{Rode1970a,Rode1970b}, and reduces to his trial solution when $\bo{v}(\bok)\propto \bok$.

By replacing the occupation distribution of Eq.~\ref{eq:f0f1} with  the ansatz Eq. \ref{eq:ansatz} and \ref{eq:approx} in Eq.~\ref{eq:BTE}, and keeping only first-order terms in the electric field, we get to what we will refer to as the linearized BTE:
\begin{align} \label{eq:BTE}
\begin{split}
(1-f^0(\bok))  \bo{v}(\bok) \cdot \bo{u}_{\bo{E}} = & \sum_{\bok'} P_{\bok\bok'} (1-f^0(\bok')) \times  \\
& \left\{ \bo{v}(\bok) \cdot \bo{u}_{\bo{E}} \tau(\bok)  -  \bo{v}(\bok') \cdot \bo{u}_{\bo{E}} \tau(\bok')  
\right\} 
\end{split}
\end{align}
where we have used the detailed balance condition
$P_{\bok'\bok}f^0(\bok') (1-f^0(\bok))=P_{\bok\bok'}f^0(\bok) (1-f^0(\bok'))$ and the fact that $\frac{\partial f^0(\varepsilon_{\bok})}{\partial \varepsilon}=-\frac{f^0(\bok)\left(1-f^0(\bok)\right)}{kT}$. In Eq.~\ref{eq:BTE} we keep the term $(1-f^0(\bok))  \bo{v}(\bok)$ on the left-hand side to avoid having it on the right-hand side as a denominator, which would bring some numerical instability in the process of solving the equation numerically. Putting back the distribution Eq. \ref{eq:f0f1} in the expression for the conductivity Eq. \ref{eq:conductivity} we get:
\begin{align} \label{eq:conductivity2}
\sigma=\frac{1}{\rho}=  \sum_p 2e^2 \int_{k \in \alpha}  \frac{d\bok}{(2\pi)^2}   
\left( \bo{v}(\bok) \cdot \bo{u}_{\bo{E}}\right)^2 \tau(\bok) \frac{\partial f^0}{\partial \varepsilon}
\end{align}
A very minimal set of approximations have been made to get to Eq. \ref{eq:BTE} and \ref{eq:conductivity2} (linear order in electric field, steady state, equilibrium phonon distribution, mean free displacement along band velocity). 
We will compute the conductivity at this level of approximation. 
This goes beyond the relaxation time approximation, which would correspond to neglecting the second term in brackets on the right-hand side of Eq.~\ref{eq:BTE}, and differs from the approaches in Ref.~\onlinecite{Li2015} and \onlinecite{Fiorentini2016} only in the assumption of Eq.~\ref{eq:approx}.
In our approach we solve the linearized BTE (Eq. \ref{eq:BTE}) iteratively starting from a guess inspired by the relaxation-time approximation (see App. \ref{app:closed_form}), as suggested initially by Rode\cite{Rode1970a,Rode1970b}. In addition to some differences in the formulation of the BTE and the sampling of electronic and phonon momenta, our improvement with respect to the existing literature is the use of the triangles method \cite{Ashraff1987,Pedersen2008} to integrate the delta functions of Eq. \ref{eq:Pkkp}.

\section{Workflow / Calculations}

\begin{figure*}[h]
\includegraphics[width=0.97\textwidth]{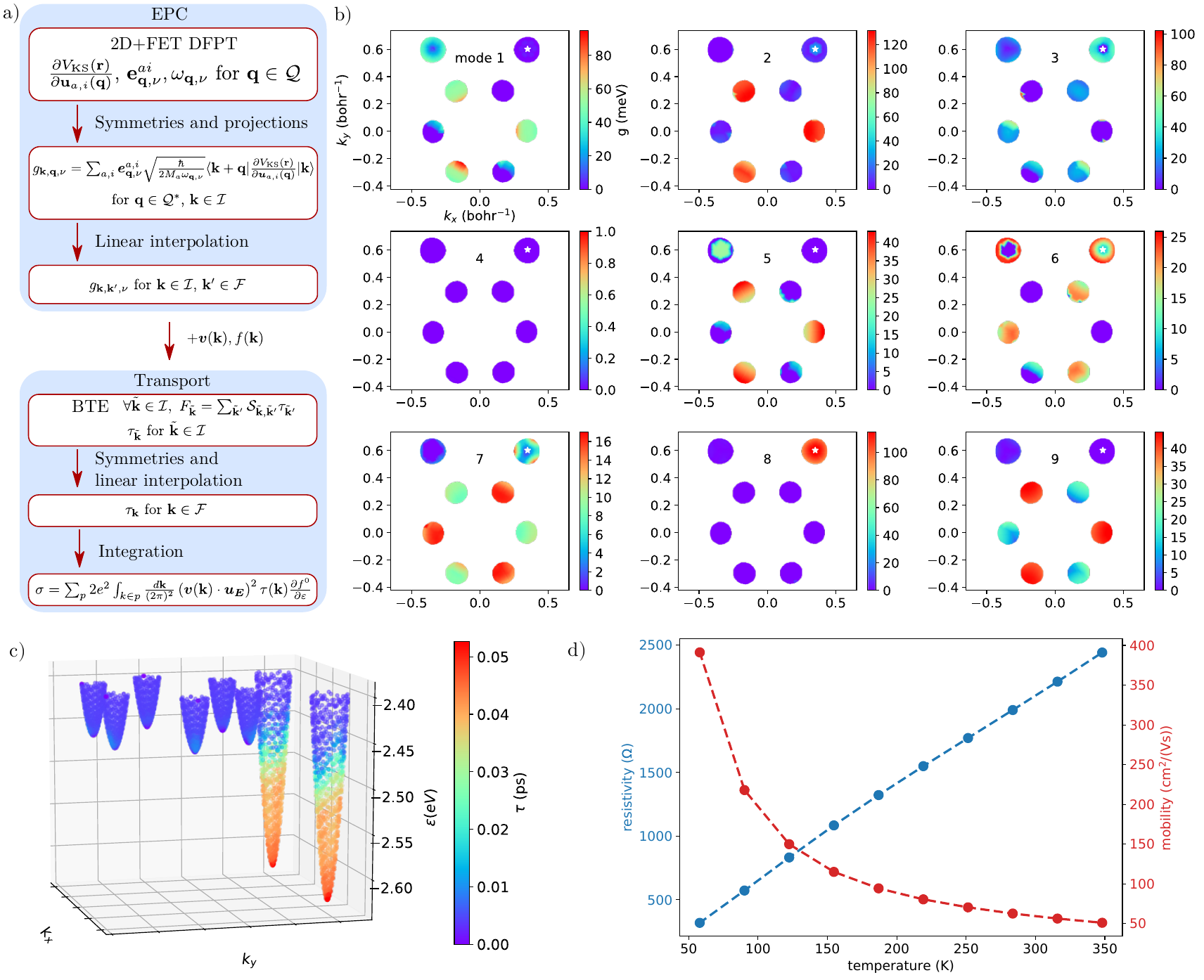}
\caption{a) Schematic description of the first-principles workflow for electronic transport
b) Interpolated electron-phonon couplings $g_{\bok, \bok'}$ for electron-doped WS$_2$. The initial state $\bok$ considered here is indicated by a white star; the other points are the possible final states in the finely sampled pockets, where the color of the point indicates the strength of the electron-phonon coupling matrix element. The index of the phonon mode indicated at the top of each subplot refers to a purely energetic ordering of the phonon modes associated with each transition. This implies that crossings in phonon dispersions may lead to discontinuities in the plots. This explains the set of seemingly out-of-place EPC matrix elements in the K' valley for modes 5 and 6, and in lesser measure in the Q valleys for modes 2 and 3. c) Scattering times interpolated on the fine grid of electronic states $\mathcal{F}$, shown using a color scale for each electronic state in the valleys of WS$_2$. d) Temperature-dependent resistivity and mobility of electron-doped WS\2 ($n=5 \ 10^{13}$ cm$^{-2}$).}
\label{fig:WF}
\end{figure*}

In this section we describe the computational workflow developed to calculate the transport properties of 2D materials. 
Electron-doped WS$_2$ ($n = 5\times10^{13}$ cm$^{-2}$) is used as a case study; this is a relatively complex system due to its multi-valley nature.
As illustrated in Fig. \ref{fig:WF}a), the process can be separated into the two following steps:
\begin{itemize}
\item EPC: We compute the linear response of the system with respect to a set of phonon momenta $\mathcal{Q}$ (defined in the following section, along with all other  sets of wave vectors, $\mathcal{I}$ and $\mathcal{F}$). The resulting EPC matrix elements $g^2_{\bok, \bok'}$ are then projected on a set $\mathcal{I}$ of relevant initial states $\bok \in \mathcal{I}$ and interpolated on a set $\mathcal{F}$ of final states $\bok' \in \mathcal{F}$. The initial states are where we want to evaluate the scattering time. 
Final states are all the states accessible from the initial states via phonon scattering.  
\item Transport: The matrix elements are then used to compute the scattering probabilities $P_{\bok, \bok'}$ and solve the BTE, which yields the scattering time for each initial state in  $\mathcal{I}$. The scattering time is then interpolated for all states in $\mathcal{F}$, as shown in Fig.~\ref{fig:WF}c). The integration of the scattering times gives the transport quantities, like the temperature-dependent mobilities and resistivities shown in Fig. \ref{fig:WF} d).
\end{itemize}
In the following we first detail and justify the sampling choices for the momentum sets $\mathcal{I}$, $\mathcal{F}$ and $\mathcal{Q}$. Then we describe each of the two steps outlined above.

\subsection{Sampling}

Phonon scattering involves three quasiparticles: an initial electronic state, a final electronic state and a phonon. We use a different momentum sampling over the Brillouin zone (BZ) for each of those quasiparticles, according to their use in the workflow and the cost of the associated calculations.

Only a subset of the electronic states are relevant for transport.
We see from the expression of conductivity (Eq. \ref{eq:conductivity2}) that one needs to find the dependency of the perturbed distribution only for electronic states for which $\frac{\partial f^0(\varepsilon_\bok)}{\partial \varepsilon}$ is significantly different from zero: this represents a set of electronic states with energies in a range of a few $kT$ around the Fermi level.
Furthermore, we see in the collision integral that the scattering time at $\bok$ depends on the scattering time of possible final states at $\bok'$. This means that the energy 
range where we need to evaluate $\tau(\bok)$ must be extended by the maximum phonon energy 
$ \hbar \omega_{\rm{max}}$ above and below. Such constraints define pockets or valleys in the BZ, and in WS$_2$ we keep all electronic states up to an energy  $0.24$ eV above the bottom of the conduction band.     
Panel c) of Fig.\ \ref{fig:WF} shows these valleys in the Brillouin zone for WS$_2$. 
We distinguish two kinds of valleys: two K-valleys around the high-symmetry points K 
and K', and six valleys around the Q points, situated approximately halfway between $\Gamma$ and K.

\begin{figure}[h]
\includegraphics[width=0.47\textwidth]{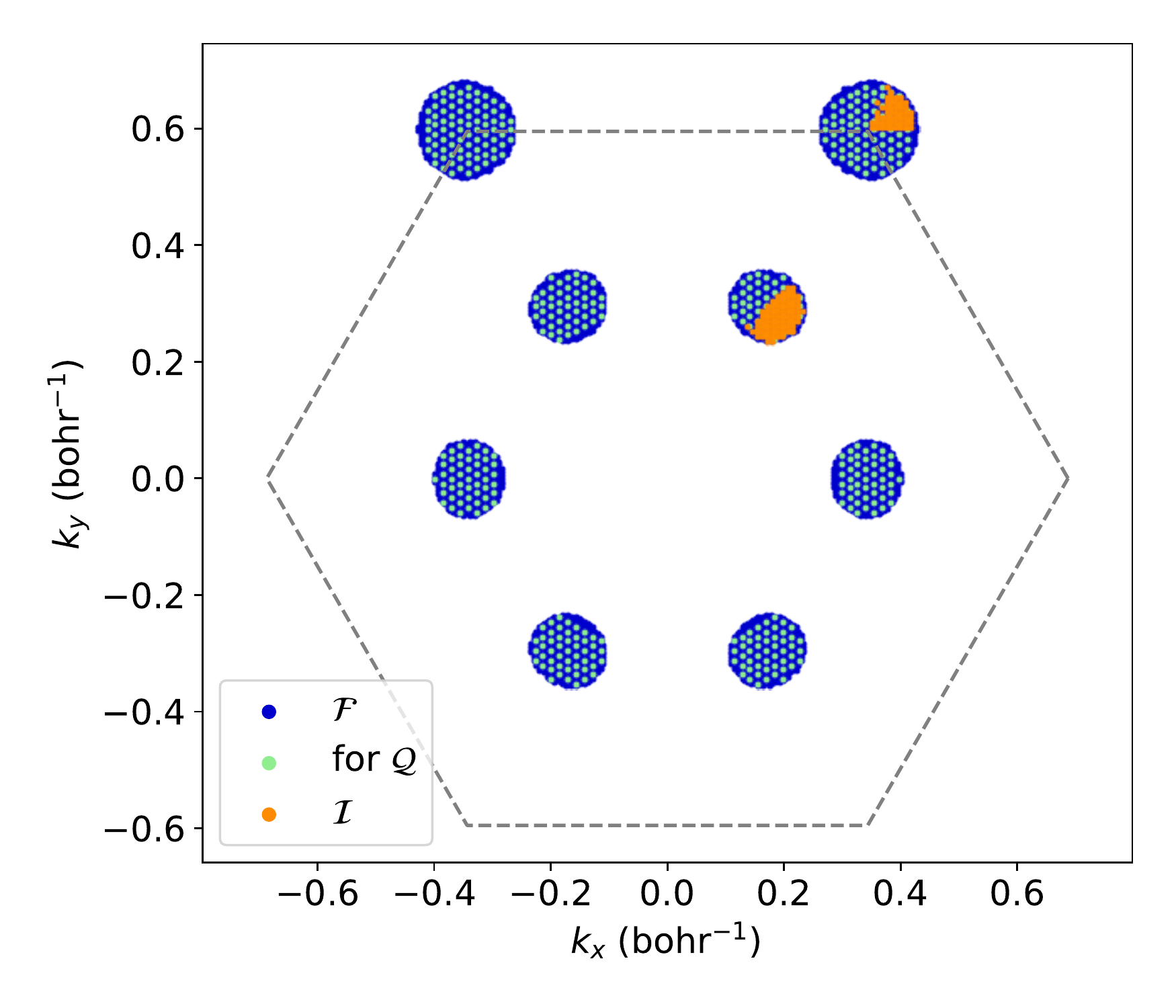}
\caption{Representation of the different sampling grids. $\mathcal{F}$ is the fine grid used for the final integration of the conductivity and the sum over final states in the BTE. $\mathcal{I}$ is a coarser grid of irreducible states on which we solve the BTE and compute the scattering times. The grid $\mathcal{Q}$ of phonon momenta on which the linear response is computed is obtained by finding the irreducible set of momenta linking the electronic states from an even coarser grid, represented here in lime green.}
\label{fig:kpoints}
\end{figure}

We define a fine grid $\mathcal{F}$ on the pockets, see Fig. \ref{fig:kpoints}, on which quantities related to the band structure, i.e. energies and velocities, are computed. A fine grid is needed to evaluate the velocities (as gradients of the energies) properly and to converge the integrals in Eq. \ref{eq:BTE} and Eq. \ref{eq:conductivity2}. To compute the eigenenergies on a fine grid, the ground-state electronic density is first computed on a relatively coarse grid ($32 \times 32 \times 1$); then, non self-consistent calculations are performed on a finer grid. Since the second step is relatively inexpensive, one can afford very fine sampling. We always define the grid such that the distance between two consecutive k-points is $0.01$ \AA$^{-1}$ (or as close as possible to this value using a Monkhorst-Pack mesh). For example, in WS\2, this corresponds to 226$\times$226 k-points. The scattering times are interpolated on this grid and integrated to get the conductivity, resistivity, or mobility, as shown in Fig.\ \ref{fig:WF}c-d). In this work, more focused on accurate linear response, band structures are computed within DFT. However, the same approach could use bands computed at a higher level of theory, such as hybrid functionals, many-body perturbation theory (e.g. GW associated with a Wannier interpolation technique \cite{Marzari2012}), or applying a correction to the fundamental band-gap obtained from an evaluation of the derivative discontinuity as in GLLBsc functionals\cite{gllb1,gllb2}, or even bands fitted to experiments \cite{Cohen1966,Wang1995}.

As written in Eq. \ref{eq:BTE}, the linearized BTE gives the value of $\tau(\bok)$ at an initial 
state $\bok$ as a function of $\tau(\bok')$  at all possible final states $\bok'$ linked 
to the initial state by a phonon-scattering event. 
As will be explained in Sec.\ \ref{sec:transport}, the sampling of initial and final states needs 
not be the same, and we simply need a map from the final states to their closest 
symmetry-equivalent initial state.
The set of final states ultimately defines the set of scattering events we account for, 
represented by the scattering probability $P_{\bok\bok'}$. These quantities have low symmetry 
since they combine the initial and final states with the phonon that links them. 
Thus, the final states 
of the BTE need to cover the full extent of the valleys. Furthermore, one needs to 
account for the sharp variations of $P_{\bok\bok'}$ due to the energy selection 
rules in Eq. \ref{eq:Pkkp}. We thus use the fine sampling of the pockets $\mathcal{F}$ for the final states in the 
BTE. This simply means that the EPC eventually needs to be interpolated on this grid.
The set of momenta $\mathcal{I}$ is where we want to compute the scattering times by 
solving the BTE, and it is convenient to note that the scattering times $\tau(\bok)$ have the symmetry 
of the electronic eigenenergies $\varepsilon_{\bok}$. This is not obvious because the computation of the resistivity involves the 
electric field and phonons which break the symmetries of the band structure. 
The integrand in Eq. \ref{eq:conductivity2}, for example, has the symmetry of the band 
structure plus the electric field. This quantity and the aforementioned $P_{\bok\bok'}$ have thus 
lower symmetry than the electronic states. 
On the other hand, the form of the ansatz for the perturbed distribution (Eqs. \ref{eq:ansatz} and \ref{eq:approx})
is such that the effect of the electric field is separated out and $\tau(\bok)$ essentially 
represents the dependency of the perturbed distribution on the electronic states only. 
Further analysis of the BTE (Eq. \ref{eq:BTE}) shows that $\tau(\bok)$ depends on a sum over 
all possible final states. While each term of the sum might not have the symmetries of the 
band structure ($P_{\bok\bok'}$, e.g., does not), the sum is invariant under the 
symmetry transformations associated with the BZ. So, since $\tau(\bok)$
has the symmetry of the band structure, the initial states $\mathcal{I}$ 
are sampled from an irreducible representation of the 
valleys. This sampling should be chosen fine enough to capture the variations of the 
scattering time; however, the size of $\mathcal{I}$ defines the size of the linear system 
to solve for the BTE and it is a factor in the total number of EPC matrix elements to 
compute. Those calculations have a non-negligible computational cost, and using the 
irreducible states in the fine grid of the pockets would be unnecessarily expensive. 
As shown in Fig. \ref{fig:kpoints}, we use a coarser grid: for all materials in this 
work we found that a grid with a k-points spacing approximatively $2.6$ times 
larger than the fine grid $\mathcal{F}$ leads to a converged solution to the 
Boltzmann transport equation at a reasonable computational cost. 
For example, in WS2, this leads to 126 irreducible initial states, while using 
the fine pockets grid would have led to more than 800 initial states.

The electronic wave functions in Eq. \ref{eq:EPC} can be easily recomputed from the 
ground-state charge density, while the rest is associated with a given phonon perturbation, and 
is the bottleneck in terms of computational cost (in particular the calculation of 
$\frac{\partial V_{\rm{KS}}(\bor) }{\partial \bo{u}_{a,i}(\boq)} $). 
There are methods\cite{Giustino2007,Calandra2010,Ponce2016,Sjakste2015} to interpolate the perturbation $\frac{\partial V_{\rm{KS}}(\bor) }{\partial \bo{u}_{a,i}(\boq)} $ to the Kohn-Sham potential, 
but the spirit of the current approach is to compute this quantity 
directly in DFPT. We will, however, take full advantage of symmetries and use 
a reasonable sampling of phonon momenta.
Each scattering event from an initial to a final states defines a phonon momentum $\boq=\bok_F-\bok_I$, where $\bok_F$ spans the pockets $\{\alpha\}$, and $\bok_I$ spans the irreducible states in the pockets. 
We use the same support as for $\mathcal{I}$ and $\mathcal{F}$, 
but define a coarser grid, as shown in Fig. \ref{fig:kpoints}. 
We then obtain the set of phonons momenta $\mathcal{Q}$, where we actually compute the linear response to the phonon perturbation, by the following process: 
first, we find all possible $\boq$ vectors connecting all final states and initial states in the 
coarse grid. For these, there are many duplicates, since a single momentum can link several 
pairs of initial and final states, and so we remove those duplicates.  
The $\boq$ points are then reduced by symmetry. Indeed, from one $\boq$-point, 
the linear response code (as implemented in Quantum ESPRESSO) allows, as a post-process, the computation of 
$\frac{\partial V_{\rm{KS}}(\bor) }{\partial \bo{u}_{a,i}(\boq^*)} $ 
for all $\boq^*$ points in the set $\mathcal{Q}^*$ defined as all $\boq^*=S(\boq)$ where $S$ is a 
symmetry operation of the crystal. The symmetry reduction is done with a tolerance of about 
the grid step, since the momenta in $\mathcal{Q}^*$ do not always fall exactly on the grid.
The number of phonons to compute thus depends on the initial grid chosen for the electronic 
states in a non-straightforward way. We go through the above process several times until 
we reach a number of phonons that is largely sufficient to capture the variations of EPC while staying reasonable in terms of computational cost. For example, in WS$_2$, this results in 200  q-points where to compute the linear 
response $\frac{\partial V_{\rm{KS}}(\bor) }{\partial \bo{u}_{a,i}(\boq)}$. For materials with 
less valleys (such as phosphorene studied in Section \ref{sec:Results}), less than 100 phonons are 
needed.

\begin{figure}[h]
\includegraphics[width=0.47\textwidth]{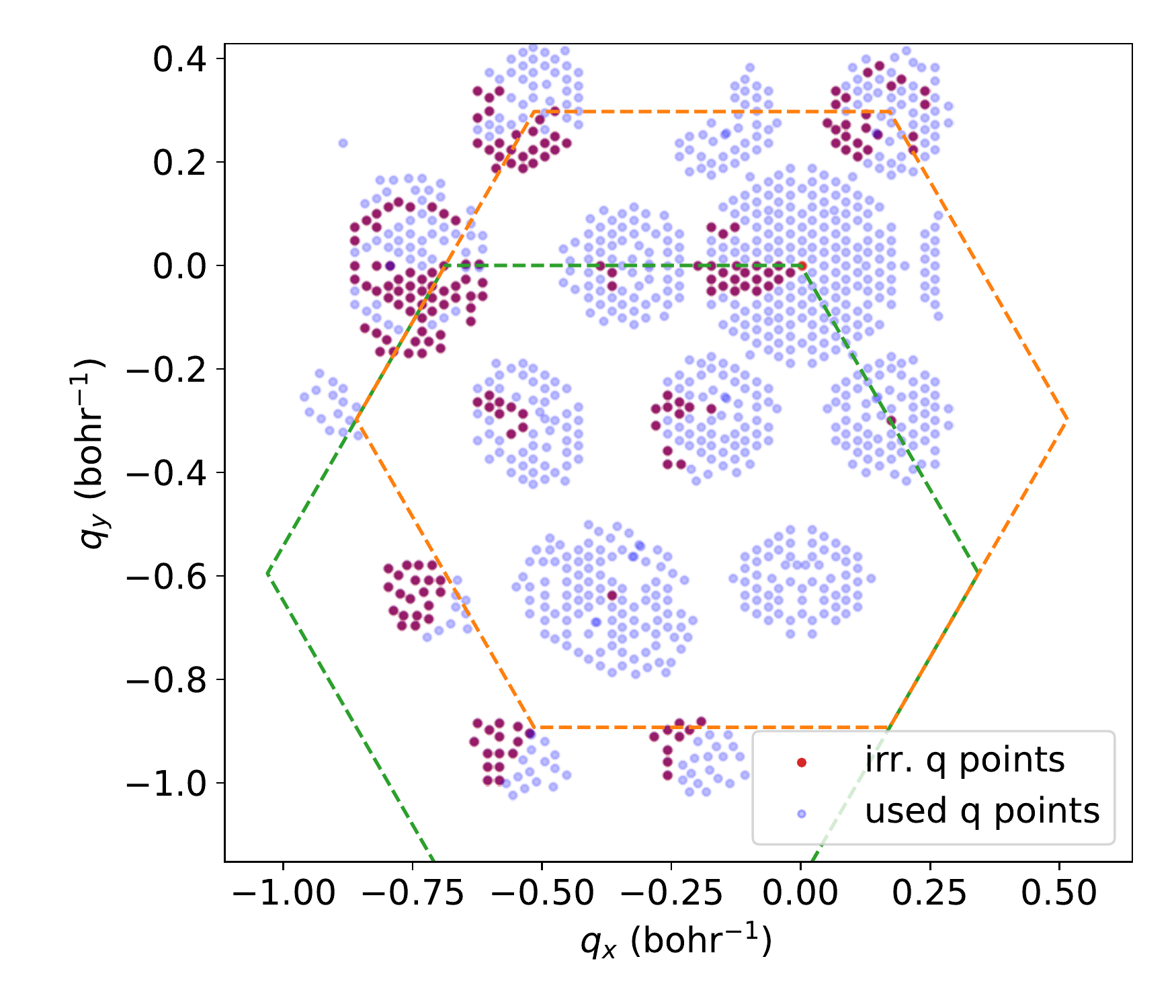}
\caption{Irreducible phonon momenta $\boq \in \mathcal{Q}$ (in red, those actually computed via DFPT), and all relevant phonons (in light blue) that can be obtained from the latter by symmetry transformations. To help  visualizing the corresponding pairs of initial and final electronic states, the dashed lines correspond to a BZ centered either the high-symmetry K point (green) or the bottom of the Q valley (orange) at the origin, the Q valley being approximately halfway between $\Gamma$ and Q.}
\label{fig:qpoints}
\end{figure}

In Fig \ref{fig:qpoints} we show both the set of irreducible phonon momenta $\mathcal{Q}$ and the relevant phonon momenta in $\mathcal{Q}^*$ that lead to final states in the BZ for at least one of the initial states.

\subsection{Phonons and EPC calculations} 

Phonons are computed using the recent implementation of DFPT for gated 2D heterostructures \cite{Sohier2017} 
in Quantum ESPRESSO\cite{Giannozzi2017,Giannozzi2009} (QE). This development includes two important modifications: i) a cutoff of the Coulomb interactions in the non-periodic direction\cite{Rozzi2006,IsmailBeigi2006} and ii) the inclusion of gates to simulate charging of the material in a field-effect setup (FET).
  The cutoff is necessary to properly account for dimensionality effects \cite{Sohier2016,Sohier2017nl}: as QE relies on 3D periodic-boundary conditions, there would always be artificial periodic images of the 2D system, and the Coulomb cutoff suppresses spurious interactions between them.
The FET setup allows the simulation of the charging of the material in an electrostatic environment that
aims to be more realistic than the standard approach of introducing a compensating background of charge uniformly distributed over the full simulation cell (vacuum included).
So, we compute the linear response to the phonon perturbations $ \frac{\partial V_{\rm{KS}}(\bor) }{\partial \bo{u}_{a,i}(\boq)}$ and the dynamical matrices for $\boq \in \mathcal{Q}$ using these
two unique features. 

These quantities are obtained for all  $\boq$ in $\mathcal{Q}^*$ by applying symmetry transformations. Then, we compute the EPC matrix elements in Eq. \ref{eq:EPC} for the $|\bok\rangle$ states that are in the irreducible initial states $\mathcal{I}$. 
This process is automatized via the AiiDA materials' informatics infrastructure\cite{Pizzi2016} to manage the calculations and store the data. Each symmetry transformation of  $\frac{\partial V_{\rm{KS}}(\bor) }{\partial \bo{u}_{a,i}(\boq)}$ and its application to the initial states represents a relatively fast run of the Phonon code of QE. Thanks to AiiDA, such operations 
can be performed in a highly parallelized fashion.  
For each initial state, a list of the $\boq$ vectors and the corresponding values of $g_{\bok, \bok+\boq,\nu}$ for each mode is stored; the  $g_{\bok, \bok+\boq,\nu}$ are then interpolated linearly to obtain $g_{\bok, \bok',\nu}$ for $\bok'$ on the finer grid of the pockets $\mathcal{F}$. 
Fig. \ref{fig:WF}b) shows the result of this procedure for electron-doped WS$_2$. 
The data storage and provenance provided by AiiDA ensure that all the information collected from the computationally demanding phonon calculations is safely kept and easily re-used for multiple applications (e.g. one can study phonon-mediated superconductivity starting from the same electron-phonon matrix elements).

\subsection{Transport} 
\label{sec:transport}

Suppose sampling $\mathcal{I}$ yields $N_i$ irreducible initial states, while sampling 
$\mathcal{F}$ yields $N_f$ final states.
Writing the BTE for the irreducible states yields $N_i$ equations relating $\tau(\tilde{k})$ for $\tilde{k}\in \mathcal{I}$ to  $\tau(\bok')$ for $N_f$ $\bok'\in \mathcal{F}$. 
\begin{align}\label{eq:iter}
\begin{split}
\forall \tilde{\bok} \in \mathcal{I}, & \ \\
\tau(\tilde{\bok})  \times & \sum_{\bok'} P_{\tilde{\bok}\bok'} \left( 1-f^0(\bok')\right) \bo{v}(\tilde{\bok}) \cdot \bo{u}_{\bo{E}}  = \\
& \left( 1-f^0(\tilde{\bok})\right) \bo{v}(\tilde{\bok}) \cdot \bo{u}_{\bo{E}}    \\
& + 
\sum_{\bok'} P_{\tilde{\bok}\bok'} \left( 1-f^0(\bok')\right)   ( \bo{v}(\bok') \cdot \bo{u}_{\bo{E}})  \tau(\bok').
\end{split}
\end{align}
This system can be solved iteratively, using the the closed-form solution in \ref{app:closed_form} as a starting point. Each iteration gives us the values of $\tau$ for $\tilde{\bok} \in \mathcal{I}$ grid. It is then rotated according to the symmetries of the band structure, interpolated linearly on the fine grid of the pockets $\bok' \in \mathcal{F}$, and put back into the next cycle, until self-consistency.

We address below a couple of technical issues that prove important if one wants to have an accurate and robust numerical solution. 
To arrive at the final form of the BTE, we address the fact that its solution $\tau(\bok)$ is ill-defined  when $\bo{v}(\tilde{\bok}) \cdot \bo{u}_{\bo{E}}$ approaches zero. 
Indeed, considering $\bo{v}(\tilde{\bok}) \cdot \bo{u}_{\bo{E}} = 0$, any 
$\tau(\tilde{\bok})$ satisfies Eq. \ref{eq:iter}. In practice, this brings numerical noise.
This situation can happen: 
i) if $\bo{v}(\tilde{\bok})\approx 0$, which is relatively rare in practice (e.g. only when $\tilde{\bok}$ is very close to the extremum of a valley); 
ii) if $\bo{v}(\tilde{\bok}) \perp \bo{u}_{\bo{E}} $, where $\bo{u}_{\bo{E}}$ is the direction of the electric field.
Situation i) can be treated approximately, without much consequences on the transport results 
(indeed, states with zero velocity do not contribute to  transport), and we compute the scattering time using the closed form of the BTE reported in the App. \ref{app:closed_form}.
In situation ii), we do need a consistent evaluation of $\tau(\bok)$. Indeed, even though $\tilde{\bok} \in \mathcal{I}$ does not contribute to the conductivity because its velocity is perpendicular to the field, this is not necessarily true for all $\bok \in \mathcal{F}$ that are equivalent to $\tilde{\bok}$ according to the symmetries of the BZ. 
We thus use the following technique: since $\tau(\tilde{\bok})$ has the symmetry of
the band structure, it does not depend on the direction of the electric field. 
In other words, any choice for the direction of the electric 
field gives the same $\tau$'s. Thus, we are free to use any direction for the electric
field to solve the BTE, and we can choose different directions for each of the 
$N_i$ equations corresponding to the $N_i$ $\tilde{\bok}$ points. 
For each $\tilde{\bok}$, we choose the direction 
$\bo{u}_{\tilde{\bok}}=\frac{\bar{\bok}}{|\bar{\bok}|}$, 
where $\bar{\bok}$ is $\tilde{\bok}$ taken from the extremum of its valley, since 
$\bo{v}(\tilde{\bok}) \perp \bo{u}_{\tilde{\bok}} $ virtually never happens. 

We then have a well-behaved set of equations.  
The other numerical issue is the treatment of the $\delta$ functions in Eq. \ref{eq:Pkkp}. The standard numerical procedure is to replace them by Gaussian functions and test the convergence of the solution with respect to the corresponding broadening. However, convergence is very slow, and even with the relatively fine grids used for here, the corresponding errors on the mobilities can reach a few percents. In addition, the automation of these convergence tests can be challenging.
An alternative is to use more sophisticated integration techniques like the so-called triangles method \cite{Ashraff1987,Pedersen2008} (two-dimensional equivalent of the tetrahedron method\cite{Blochl1994}),
which amounts to performing the sum analytically as an integral by interpolating the functions involved linearly within three points that form a triangle. 
Although slower, it gives more accurate results and this is what we implemented here. To the best of our knowledge, it is the first application of this method to the BTE. 
As a final remark, note that an iterative procedure is not needed in principle. The above system of equations can be solved algebraically. As discussed in App. \ref{app:matrix_sol}, however, the use of Gaussians and the associated choice of broadening parameter becomes necessary.

Once we have the angular and energy dependent $\tau(\bok)$, which does not depend on the electric field in itself, we can compute the conduction integral for an arbitrary direction of the electric field, thus probing any diagonal element of the conductivity tensor.

\section{Results}
\label{sec:Results}

We have applied our approach to a set of six different cases: five electron-doped materials (WS$_2$, MoS$_2$ WSe$_2$, arsenene, and phosphorene), as well as hole-doped phosphorene. These are common 2D systems often praised for their potential for transport applications.
Also, this set leads to an interesting diversity of band structures, in terms of valleys,  their symmetries and their energetic accessibility.
Let us stress that we are working here in the framework of DFT, 
rather than higher levels of theory, and
without spin-orbit coupling (SOC). Both approximations can of course 
affect the bands of the materials studied here and have an impact on mobility\cite{Ponce2018,Ma2018}. 
In particular, inclusion of many-body corrections within, e.g., the GW approximation, would not only correct quasiparticle energies of the pristine, undoped materials, but would also improve the description of band renormalization effects associated with free carriers, especially at high doping.
For SOC, we expect the variations to be relatively small for electron-doped arsenene and phosphorene\cite{Qiao2014}. 
The case of TMDs is a more delicate one: 
the relative positions of the K and Q valleys
seems to be sensitive to many aspects of the calculations, like the inclusion of SOC\cite{Brumme2015}, choice of pseudo potentials, lattice parameters, level of theory\cite{Shi2013a} (DFT versus GW) or doping\cite{Brumme2015}. 
Thus, while we will be able to compare between the three TMDs and discuss qualitative trends, 
we do not claim to be quantitative with respect to experiments.
That being said, the methodological approach presented in this work is very flexible. One can 
combine the electron-phonon coupling matrix elements found in DFPT with the band structure 
found by any mean; e.g., one can use a band structure computed at the highest level of 
theory, and/or correct it to better fit experiments, before inserting it in the workflow. 
Regarding SOC, it can be easily included in the band structure calculations and EPC calculations; further development is planned to include it in the solution to the BTE.

Another important approximation we make, common to most DFT/DFPT approaches to compute phonons and EPC from first principles\cite{Giustino2017}, is the adiabatic approximation. This means that electrons are assumed to be able to relax to their ground state during phonon perturbations, either as a result of their motion being faster than phonon oscillations (Fermi velocity higher than the phonon phase velocity) or because the electronic scattering times due to other momentum-changing sources of disorder are short enough to establish equilibrium\cite{Maksimov1996,Saitta2008}. This amounts to taking the zero-frequency limit -- allowing intraband transitions -- in the phonon self-energy\cite{Giustino2017,Grimvall}, which, among other things, results in discarding non-adiabatic screening effects\cite{Mahan}. The former approximation can sometimes break down, especially in metals, as is the case for the non-adiabatic Kohn anomalies in graphene\cite{Lazzeri2006,Pisana2007}. Although these effects can be predicted from first principles\cite{Lazzeri2006,Saitta2008,Calandra2010}, even accounting for correlations beyond DFT\cite{Lazzeri2008}, their inclusion in the solution of the BTE has never been performed in first-principles approaches and will not be considered here. Dynamical screening of electron-phonon interactions seems even more challenging, and it's typically overlooked in first-principles investigations\cite{Giustino2017} or at most introduced via model dielectric functions\cite{Verdi2017}.

In the calculations, we chose an electrostatic doping of $5\times10^{13}$ cm$^{-2}$. While undoubtedly high, this doping is experimentally achievable, at least with the use of ionic liquid gates \cite{Efetov2010,Bisri2017}. In addition, this regime is interesting from a fundamental point of view, because it allows to suppress extrinsic contributions, like charge impurities via screening, thus getting closer experimentally to phonon-limited transport. 
This high doping regime is poorly studied from first principles, and accessible thanks to our implementation of DFPT for gated materials without any analytical corrections.
Nonetheless, it also poses some additional challenges, such as the emergence of band-renormalization effects, electron-electron and electron-plasmon scattering, which has been posited to be relevant at high doping\cite{Hu1996,Caruso2016}; these are beyond the scope of the present manuscript. 
As far as applications are concerned, the relevant doping regime varies. High doping is relevant for several possible application, such as high frequency electronics. For logic-gate transistors operating at very low densities, the zero doping limit is more relevant. The method presented here could be applied also in this case, but we chose to propose a new perspective with respect to the theoretical literature already treating this limit. As a consequence, in the following, comparisons with theoretical works in the zero doping limit must be taken carefully. More generally, many devices might operate at moderate doping densities ($\sim10^{12}$ cm$^{-2}$). In this regime, the application of our approach would still be possible, although it would require the use of finer k-point sampling for the phonon calculations, which we chose to avoid here given the number of systems studied. Nevertheless, for moderate doping, the high doping results presented here might be more relevant than undoped results.

As previously mentioned, first-principles calculations are performed with the Quantum ESPRESSO 
distribution \cite{Giannozzi2017,Giannozzi2009} including a 2D Coulomb cutoff and the 
possibility to charge the material with gates \cite{Sohier2017}, 
and using the SSSP Accuracy (version 0.7) 
library\cite{Prandini2018a,Hamann2013,Vanderbilt1990,DalCorso2014}.
Structures are taken from the database described in Ref.\ \onlinecite{Mounet2018}. 
To build this database, structural relaxations were performed in the neutral material, using the 
SSSP library as well and a k-point sampling corresponding to a spacing of 0.2 \AA$^{-1}$ in each direction.
We use a symmetric double-gate setup to charge the materials with a density of
$5\times10^{13}$ cm$^{-2}$ electrons or holes for all systems except for electron-doped phosphorene, where we choose a lower density of $5/3 \times10^{13}$ cm$^{-2}$ to avoid raising the Fermi level too high in the conduction band, where additional valleys appear. Each gate carries half 
the opposite charge of the material, such that the electric field has equal norm but 
opposite direction on each side of the material. 
We could also have used a single-gate setup, with an electric field 
only on one side, but no large difference is expected since the impact
of the electric field setup should come mainly from its effect on 
spin-orbit coupling, which is not included here. 
Barrier potentials are added to avoid leakage of electrons towards the gates. 
These also lead to a hardening of the ZA phonons in the long-wavelength 
limit, with a non-zero frequency at $\Gamma$ around $10 - 25$ cm$^{-1}$ 
for the materials considered. This emulates a relatively soft out-of-plane mechanical interaction with gate dielectrics or substrate compared to interlayer interactions in the layered parent 3D material, since breathing ZA modes are usually in the 20-50 cm$^{-1}$ range. 
Ground state and linear-response calculations 
on the charged materials are performed with a $32 \times 32 \times 1$ k-point grid and $0.02$ Ry Methfessel-Paxton smearing to sample 
the Fermi surface. Note that we are 
working at relatively high doping, such that the Fermi surface is large enough to be sampled correctly. Non self-consistent calculations are performed to obtain the band structure on 
the fine grid $\mathcal{F}$.
Band structures and phonon dispersions are reported in the App. \ref{app:bands_phonons}.

\begin{figure*}[h]
\includegraphics[width=0.85\textwidth]{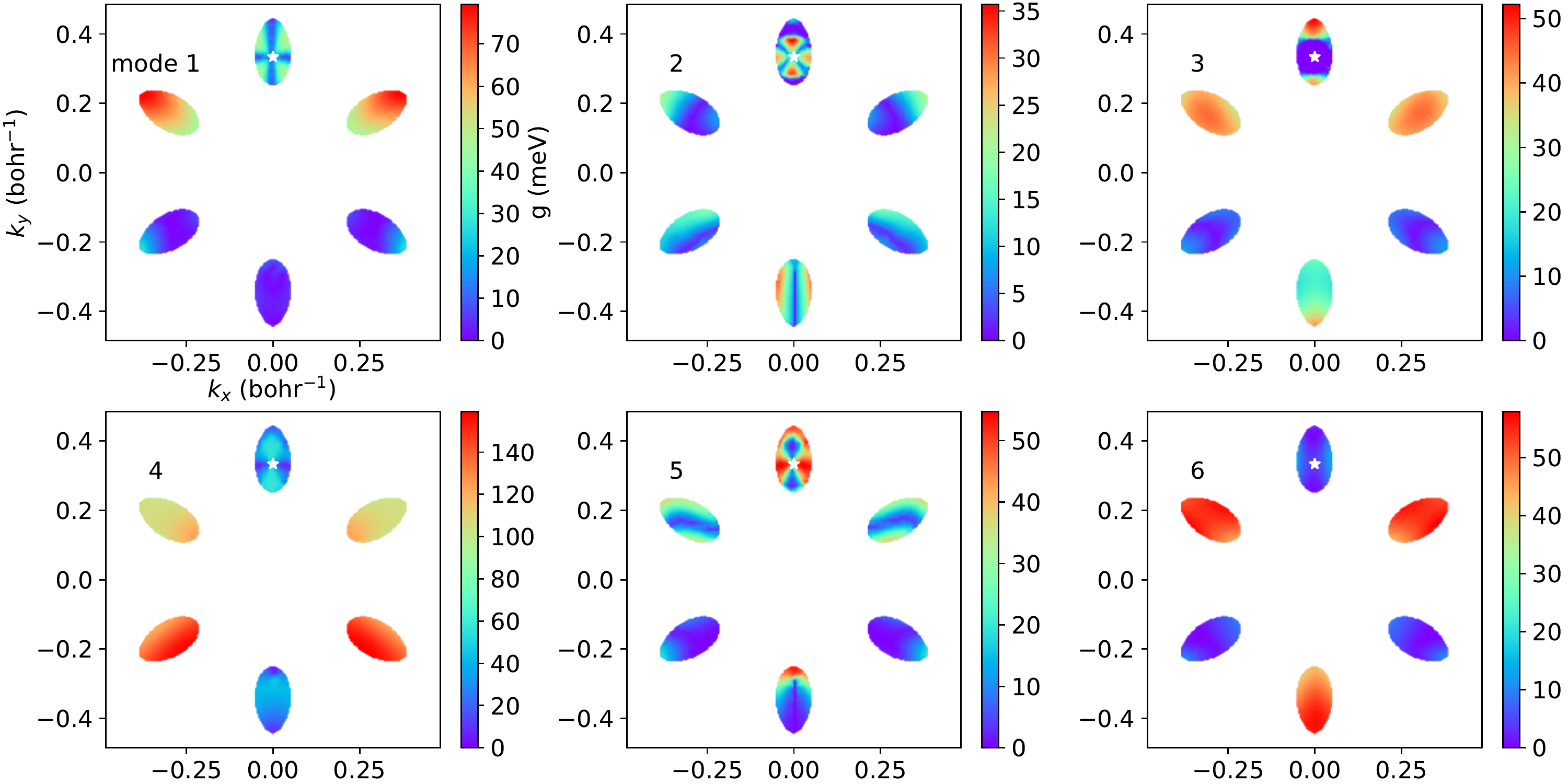}
\caption{Interpolated $g_{\bok \bok'}$ for electron-doped arsenene. The initial state considered is indicated by a white star. The rest of the points are the possible final states in the finely sampled pockets and the color of the point indicates the strength of the electron-phonon coupling matrix element. The index of the phonon mode indicated at the top of each subplot refers to a purely energetic ordering of the phonon modes associated with each transition.}
\label{fig:EPC_As}
\end{figure*}

\begin{figure*}[h]
\includegraphics[width=0.85\textwidth]{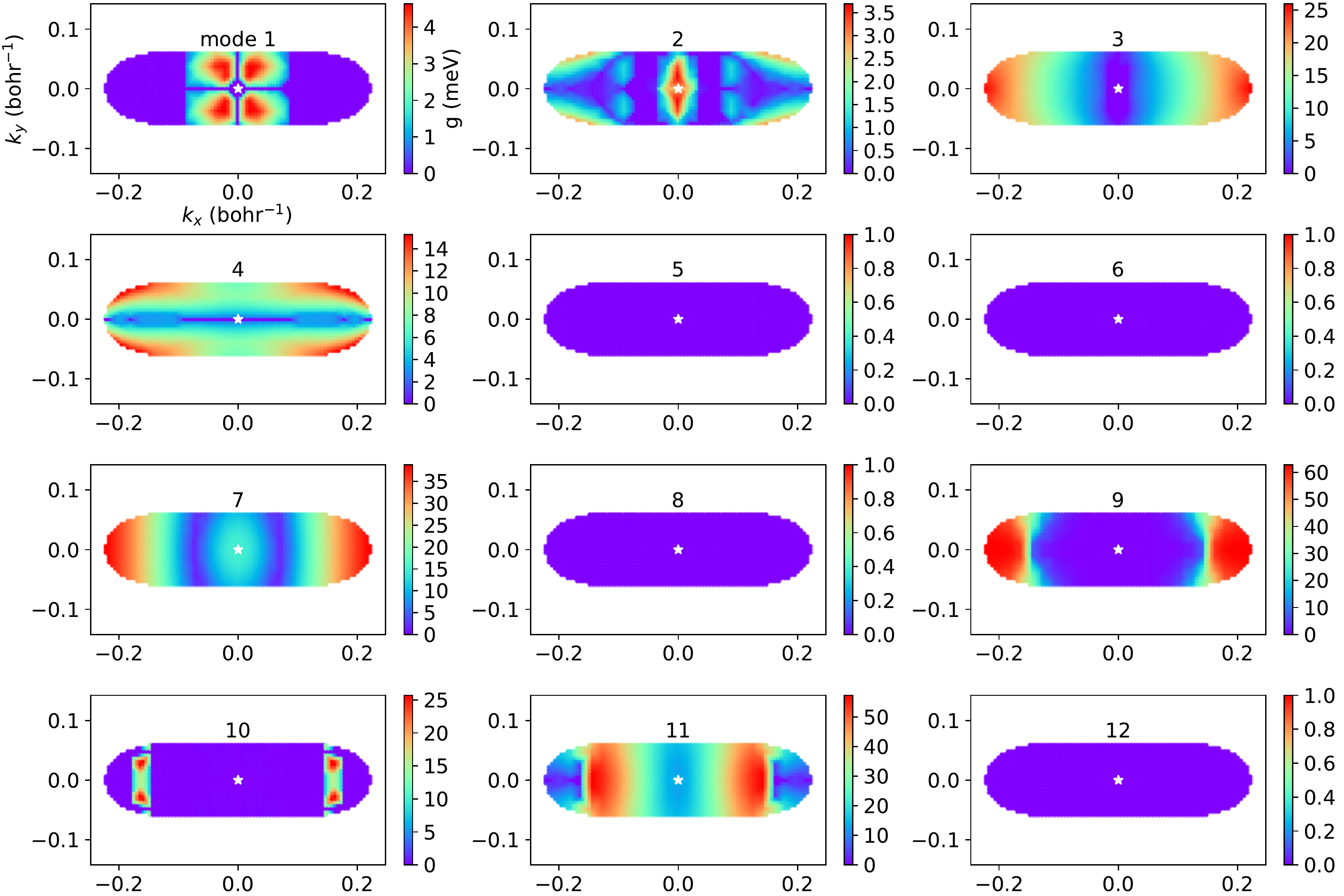}
\caption{Interpolated $g_{\bok \bok'}$ for hole-doped phosphorene. The initial state considered is indicated by a white star. The rest of the points are the possible final states in the finely sampled pockets and the color of the point indicates the strength of the electron-phonon coupling matrix element. The index of the phonon mode indicated at the top of each subplot refers to a purely energetic ordering of the phonon modes associated with each transition.}
\label{fig:EPC_P4h}
\end{figure*}

The interpolated EPC matrix elements for electron-doped WS\2 are shown in Fig.\ \ref{fig:WF}; for electron-doped arsenene in Fig.\ \ref{fig:EPC_As}; and for hole-doped phosphorene in Fig.\ \ref{fig:EPC_P4h}. 
Equivalent plots for electron-doped MoS$_2$ and WSe$_2$
can be found in the App. \ref{app:EPC}; they are very similar to WS$_2$.
The angular dependencies of the EPC are non-trivial: the EPC can undergo some rather sharp variations, most often for intravalley transitions via acoustic modes. For intervalley transitions, we observe overall smoother variations if we ignore the discontinuities coming from phonon crossings. However, intervalley scattering is activated or not depending on the valleys and the mode in rather non-trivial ways.
These plots also serve to give a visual confirmation that we use sufficient sampling to capture the variations of the EPC.
One important aspect to keep in mind when interpreting these plots and the transport properties of the system is that energy and momentum conservation conditions drastically reduce the final states effectively relevant for a given initial state. 

\begin{figure}[h]
\includegraphics[width=0.45\textwidth]{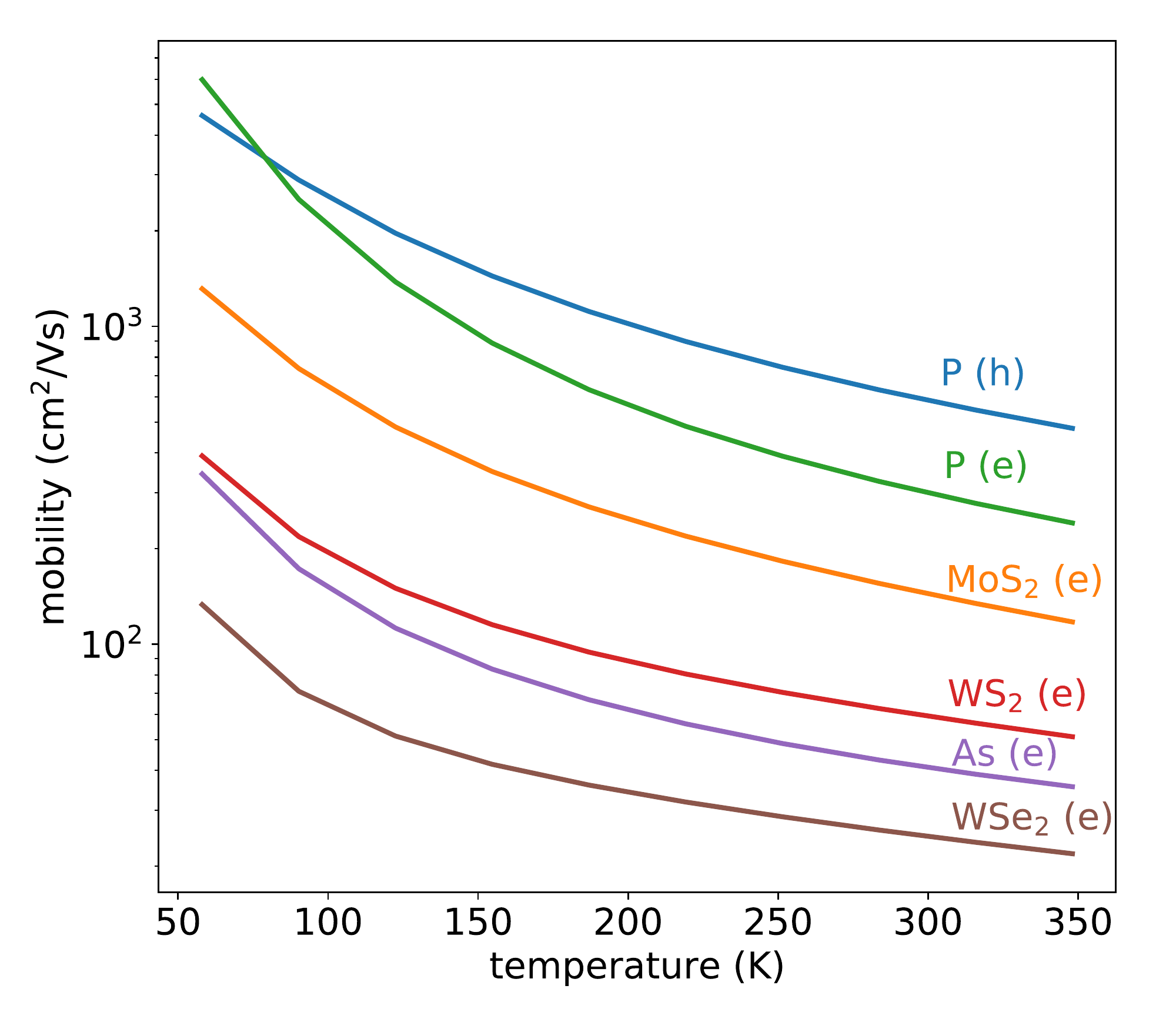}
\caption{Mobilities as a function of temperature for electron-doped of MoS$_2$, WS$_2$, WSe$_2$, arsenene and phosphorene, as well as for the hole side of phosphorene. The carrier density is $5\times10^{13}$ cm$^{-2}$ for all systems except electron-doped phosphorene where it is $5/3\times10^{13}$ cm$^{-2}$.}
\label{fig:mobs}
\end{figure}

The linearized BTE is solved in all these systems, giving the mobilities shown in Fig.\ \ref{fig:mobs}. 
These cover three orders of magnitude, and the hole side of phosphorene shows mobilities 
ten times larger than the rest, almost reaching $10^3$ cm$^2$/Vs at 
room temperature. For comparison, electron-doped graphene has mobilities on the order of 
$10^4$ to $10^5$ cm$^2$/Vs depending on the density. Table \ref{tab:mobilities} summarizes our findings, focusing on room-temperature results, and 
compares them with some values available in the literature. 
To characterize the temperature dependence of mobilities, we report the parameter $\gamma$ in Table \ref{tab:Tdep}, defined as $\mu\propto T^{-\gamma}$.
Below, we identify the general trends and discuss three important factors in the prediction of 
transport properties in 2D materials. Finally, we argue that those three factors might account 
for the discrepancies observed in Table \ref{tab:mobilities}.

\begin{table*}[h]
\caption{Room-temperature mobilities of the 6 systems considered (at the DFT/DFPT level without SOC). Mobilities are given in cm$^2$/Vs, doping densities are indicated in parentheses, in units of 10$^{13}$ cm$^{-2}$. Question marks for densities of experimental works indicate that the information was not provided. For theoretical works, "0" means that the mobility was derived in the zero doping limit (this includes calculations based on the density-independent formulaes). Note that SOC is not included in any of those works. Results on phosphorene are reported in both the armchair (ac) and zig-zag (zz) directions.
}
\label{tab:mobilities}
\begin{tabular}{ c c L L }
\hline
Material (e or h doped) & Present result $\mu$($n$) & Other first-principles results  $\mu$($n$) & Experimental  $\mu$($n$) \\ 
\hline
\hline
MoS$_2 (e)$ & 144 (5) & 265 (5)\cite{Gunst2016a}, 150 (0) \cite{Li2015} , 130 (0) \cite{Li2013},  320 (0.01) \cite{Kaasbjerg2013}, 410(0.01)-340(1) \cite{Kaasbjerg2012a}, 340 (0) \cite{Zhang2014}  
& 217 (0.46)\cite{Radisavljevic2011}, 150 (0.7) \cite{Yu2016}, 81 (0.7) \cite{Yu2014}, 63 (1.35) \cite{Radisavljevic2013},  \\
\hline
WS$_2$ (e)  & 60  (5) & 320 (0) \cite{Jin2014}, 1103 (0) \cite{Zhang2014}  
&  44 (10) \cite{Jo2014}, 44 (?) \cite{Ovchinnikov2014}, 45 (0.6) \cite{Aji2017}, 25-83 (0.7) \cite{Cui2015}, 214 (?)\cite{Iqbal2015} \\ 
\hline
WSe$_2$ (e) & 25  (5) & 30 (0) \cite{Jin2014}, 705 (0) \cite{Zhang2014} 
& 7 (0) \cite{Huang2014}, 30 (1-1.5?) \cite{Allein2014}  \\
\hline
As  (e)    & 41 (5) & 21 (-) \cite{Wang2017}, 1700 (0) \cite{Pizzi2016nc} & - \\
\hline
P (h-ac)      & 586 (5) & 640-700 (0) \cite{Qiao2014}, 460 (0) \cite{Jin2016}, 292 (5) \cite{Rudenko2016}, 140 (1) \cite{Liao2015}, 19 (0) \cite{Gaddemane2018} \\
\hline
P (h-zz) & 44  (5) & 10000-26000 (0) \cite{Qiao2014}, 90 (0) \cite{Jin2016}, 157 (5) \cite{Rudenko2016}, 15 (1) \cite{Liao2015}, 3 (0) \cite{Gaddemane2018} \\
\hline
P  (e-ac)     & 302 (5/3) & 1100-1140 (0) \cite{Qiao2014}, 210 (0) \cite{Jin2016}, 738 (5) \cite{Rudenko2016}, 430 (5) \cite{Trushkov2017}, 140 (1) \cite{Liao2015}, 20 (0) \cite{Gaddemane2018}  \\
\hline
P  (e-zz)    & 35 (5/3) & 80 (0) \cite{Qiao2014}, 40 (0) \cite{Jin2016}, 114 (5) \cite{Rudenko2016}, 80 (5) \cite{Trushkov2017}, 25 (1) \cite{Liao2015}, 10 (0) \cite{Gaddemane2018}  \\
\hline
\hline
\end{tabular}
\end{table*}

\begin{table}[h]
\caption{Temperature dependence of the mobility, represented by the factor $\gamma$ such that $\mu \propto T^{-\gamma}$.} 
\begin{tabular}{ c c c }
\hline
Material (e or h doped) & $\gamma$ ($T \leq 150$ K) & $\gamma$ ($T > 150$ K)  \\ 
\hline
\hline
MoS2(e) & 1.33 & 1.34 \\
WS2(e) & 1.28 & 0.99 \\
WSe2(e) & 1.28 & 0.79 \\
As(e) & 1.49  & 1.04 \\
P (h-ac) & 1.14 & 1.36 \\
P (e-ac) & 1.96 & 1.60 \\
\hline
\hline
\end{tabular}
\label{tab:Tdep}
\end{table}

\subsection{Intervalley scattering}

\begin{figure*}[h]
\includegraphics[width=0.97\textwidth]{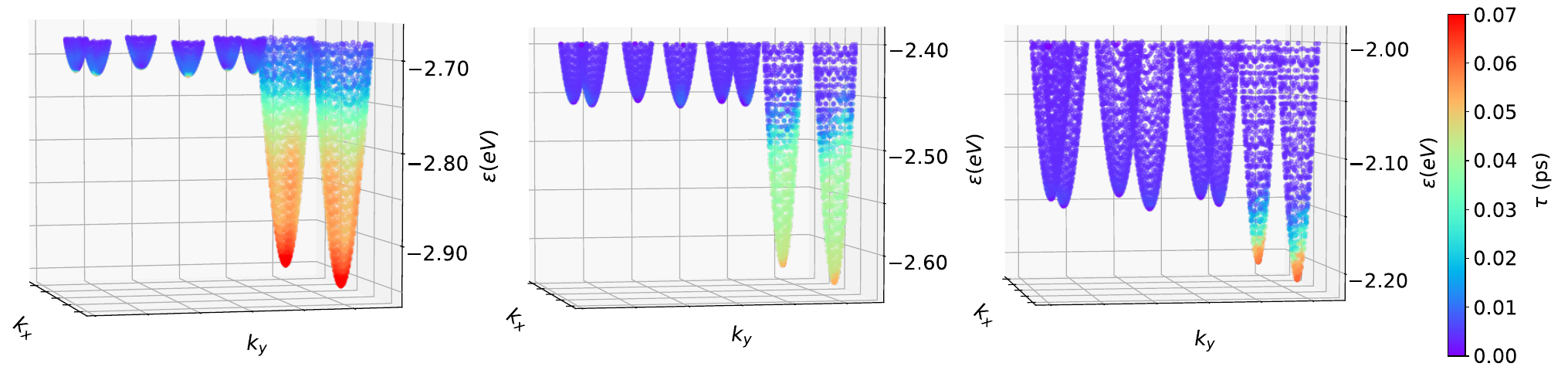}
\caption{Scattering times in MoS$_2$, WS$_2$, and WSe$_2$ (from left to right). The color scale is the same for all subplots. The scattering rate ($\hbar/\tau$) clearly increases for states with energy high enough for the Q valley to be accessible. The Fermi levels are : $-2.83$, $-2.46$, and $-2.13$ eV, respectively.}
\label{fig:taus}
\end{figure*}

The ranking of the above materials in terms of mobility reflects the importance of intervalley 
scattering. This is clearly demonstrated by considering the three TMDs: 
these have essentially the same type of electronic and phonon band structures, as well as similar EPC matrix elements. Yet, we obtain an order of magnitude variation in the mobilities. This stems from the position of the Q valley: indeed, as the Q valley approaches the 
Fermi level, it offers an additional scattering channel for the electrons. 
As shown in Fig.\ \ref{fig:taus}, 
the scattering time gets shorter (more scattering) at energies close to the bottom of the Q valleys.
The largest contributions to the resistivity of TMDs ($\approx 80\%$\footnote{This number is estimated by solving the BTE for each mode successively, setting the coupling to the other modes to zero. This is simply an educated estimation. Indeed, the process is not strictly valid quantitatively since the solution to the BTE including all phonon modes is not the sum of the contributions from each mode.}) comes from the scattering with the LA and TA modes: LA at $\Gamma$ (intravalley scattering) and M (K$\leftrightarrow$Q scattering) and TA at K (K$\leftrightarrow$K' scattering). This is not obvious from the plots of the electron-phonon matrix elements, because the dispersions of the three acoustic modes cross each other between $\Gamma$ and K, and between $\Gamma$ and Q. However, looking at the phonon displacements, it is quite clear that in-plane acoustic modes are associated with the regions of strong electron-phonon coupling in the first three sub-plots of Fig.~\ref{fig:WF}b, and in Figs. \ref{fig:EPC_MoS2}, \ref{fig:EPC_WSe2} and \ref{fig:EPC_WS2} of App. \ref{app:EPC}. 

The multi-valley nature of a material does not necessarily deteriorate the mobility in itself. The presence of multiple valleys increases 
both the accessible phase space for scattered states and the density of states, and the corresponding effects on the mobility cancel each other.
Rather, it is the existence of a strong EPC between the 
valleys that increases scattering and lowers the mobility. 
For the small subset studied here, all multi-valley 
materials showcase strong intervalley EPC. 
It may be argued, in general, that intervalley EPC is often strong 
compared to intravalley EPC. This could first be explained by the fact that intervalley EPC is not bound to vanish at long wavelengths, contrary to the coupling to acoustic phonons. Second, it involves 
larger phonon momenta and the EPC tends to be less screened, since the dielectric function goes to one in the short-wavelength limit. 

\subsection{Symmetries of the valleys}

\begin{figure}[h]
\includegraphics[width=0.47\textwidth]{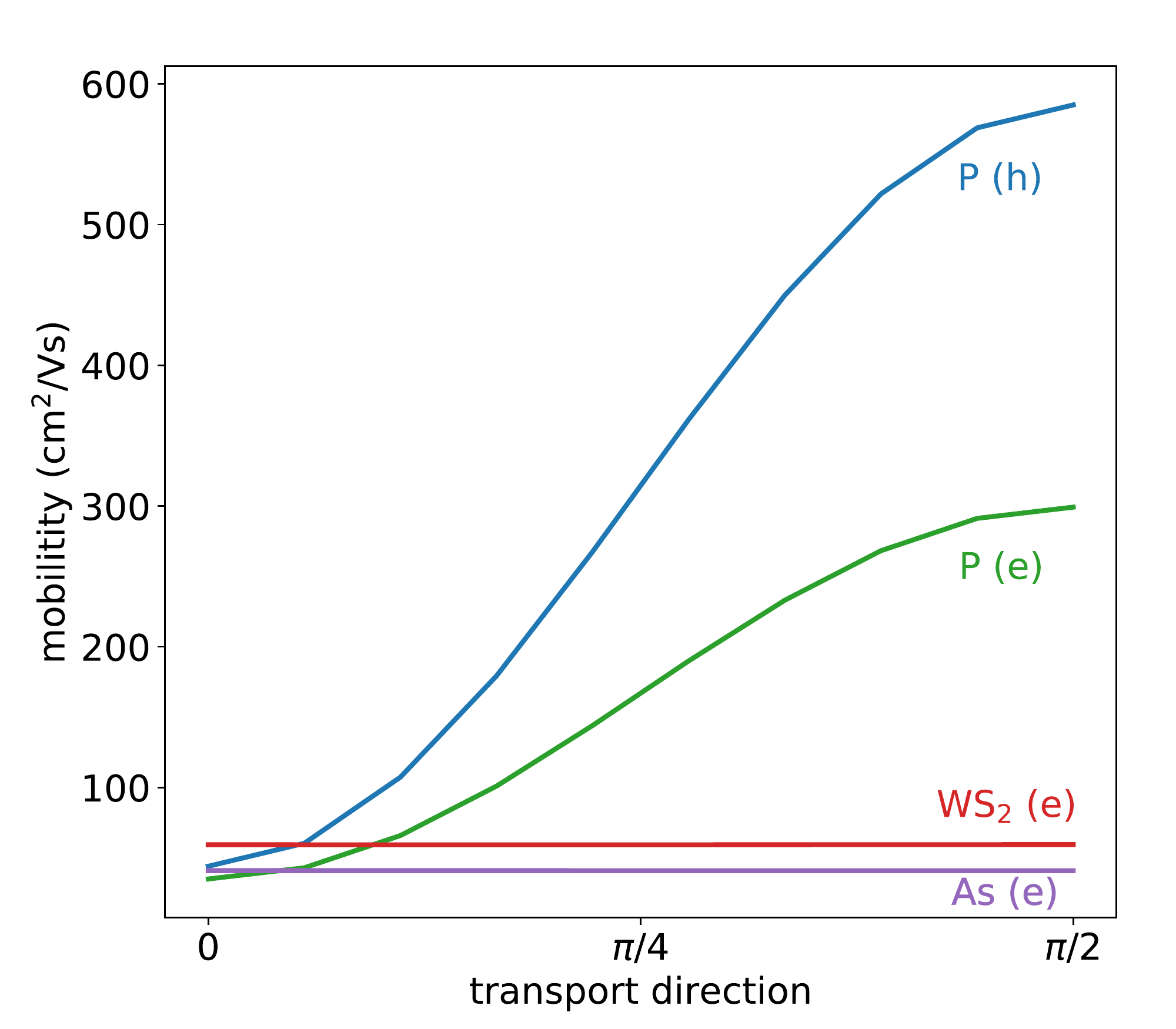}
\caption{Angle-dependent mobilities at room temperature for hole-doped and electron-doped phosphorene, and electron-doped arsenene and WS$_2$. The angle corresponding to the transport direction refers to the direction of the in-plane electric field driving the current with respect to the $\bo{x}$ direction indicated in the plots of EPC. For phosphorene, $0$ ($\pi/2$) corresponds to the zig-zag (armchair) direction.
While phosphorene shows highly anisotropic transport, only very small variations ($\approx 10^{-2}$ cm$^{2}/$(Vs), below numerical accuracy) can be observed in WS$_2$ and arsenene.}
\label{fig:ang_mobs}
\end{figure}

Effective masses are among the most influential features to consider when studying mobilities,  
and anisotropic effective masses can be very beneficial. Indeed, small effective masses have the 
benefit of bringing large carrier velocities while large effective masses have the benefit of 
bringing high carrier densities. Thus, one might combine those benefits having a small effective 
mass in the transport direction and a large one in the perpendicular direction. This contributes to phosphorene's good transport performance. 
In fact, phosphorene is the only material in the present study showing significant transport anisotropy. The transport 
properties of TMDs are isotropic because the bottom of the K and Q valleys are roughly
isotropic. In the case of arsenene, each single valley is quite anisotropic, but when summing up 
the six equivalent valleys the angular dependency of the 
transport averages out and vanishes\cite{Pizzi2016nc}. Fig. \ref{fig:ang_mobs} shows the mobility of P (h), As (e) and WS$_2$ (e) as a 
function of the direction of the electric field. 
The mobility of phosphorene is highly directional, while all others are isotropic.

\subsection{Doping effects on electron-phonon interactions}

The effects of doping on electron-phonon scattering are many. 
The first one is to move the Fermi surface within the electronic landscape. This leads to 
variations of the density of states and determines whether certain valleys are accessible 
via phonon scattering or not. 
As we saw in TMDs, the activation of intervalley scattering can have drastic consequences on 
transport. Those effects come from the energy selection rules of Eq. \ref{eq:Pkkp}. Doping also has consequences on the strength of EPC matrix elements themselves (Eq. \ref{eq:EPC}). 
A well-known and important effect is the additional screening coming from the electrons added in the conduction 
band or holes in the valence band. In our computational framework, this is inherently accounted for since we compute the linear-response of the charged system. 
Any EPC related to a periodic variation of the effective scalar potential in which the electrons move will be screened. This includes:
i) a variation of the charge state via a variation of the area of the unit cell, as induced by longitudinal acoustic phonons; 
ii) any EPC related to dipole or Born effective charges, like Fr\"ohlich or piezo-electric EPC, in which phonons interact with electrons via the generation of electric fields;
iii) other less straightforward mechanisms, like the gate-induced coupling to 
flexural phonons in graphene \cite{Gunst2017,Sohier2017}. 
In general, considering only the above types of EPC is largely insufficient in doped materials, as electronic screening strongly reduces their contribution, and bare interactions or weakly screened intervalley couplings dominate.

\begin{figure}[h]
\includegraphics[width=0.47\textwidth]{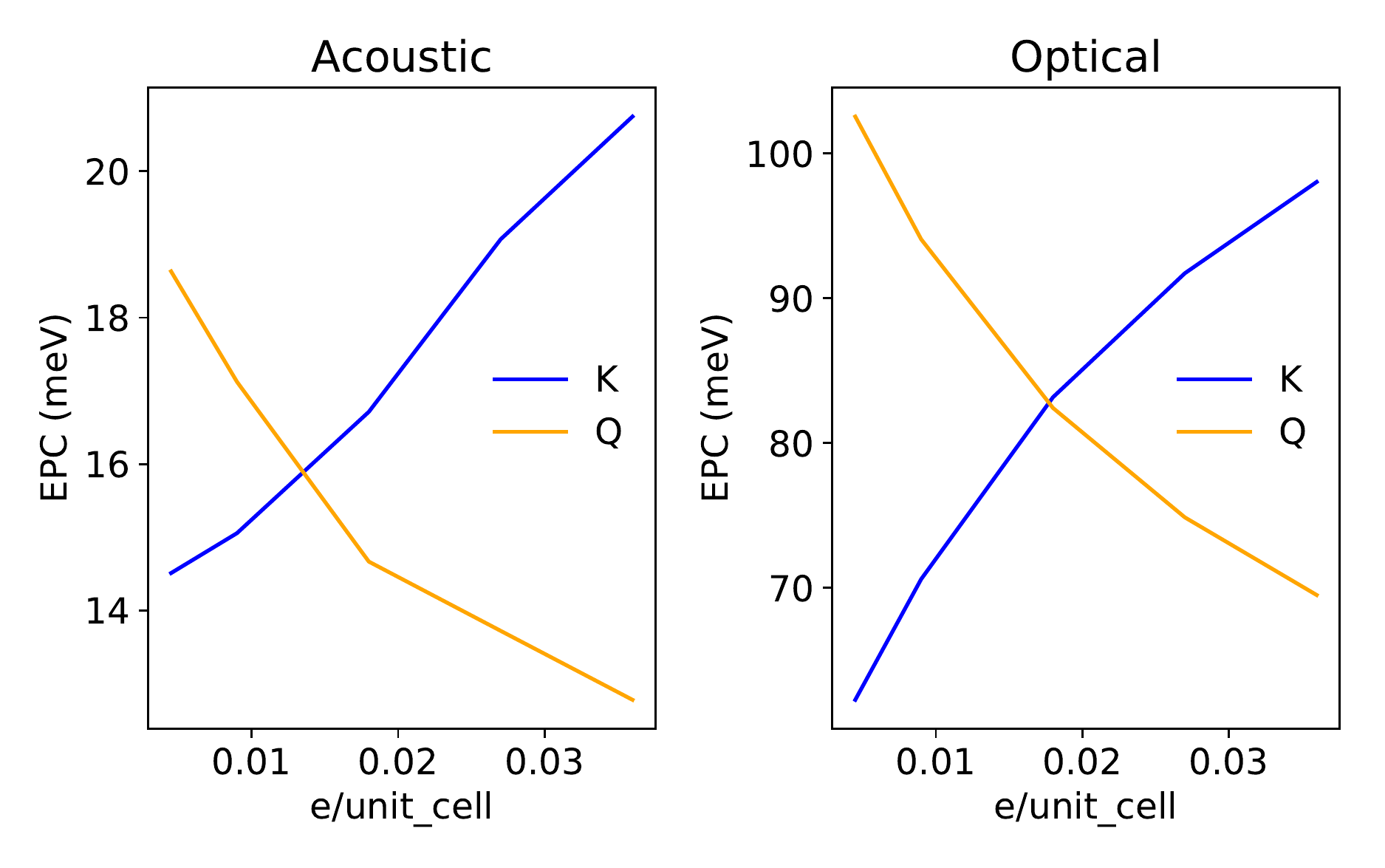}
\caption{Doping dependency of intravalley scattering in both electron valleys of WS$_2$. The EPC is 
measured by the quantity $\langle g_{\alpha}^2 \rangle$, described in the text.}
\label{fig:g_dop_WS2}
\end{figure}

Doping also affects EPC beyond screening: as the occupations of the valleys vary, certain orbitals/bands in the material get populated or depleted, which can directly change the amplitude of the coupling. This happens in TMDs, as shown in Fig.~\ref{fig:g_dop_WS2} in which we computed intravalley scattering for several doping levels. In particular, we compute the average of the long-wavelength coupling $\langle g^2 \rangle$ along a fixed iso-energetic line at $\varepsilon=E$, by 
taking a few initial states on the iso-energetic section of each the K and Q valleys, and six phonon momenta with a fixed small norm.
Linear-response calculations are then performed for each doping to capture non-trivial dependencies of the EPC matrix elements. We average on the phonon momenta and on the initial states to get $\langle g_{\alpha}^2 \rangle$ for each valley $\alpha = K,Q$. We sum the contributions from acoustic and optical phonon modes separately. 
Fig.~\ref{fig:g_dop_WS2} shows that as doping increases in WS$_2$, the intravalley coupling in K increases while it decreases in Q. The couplings to both acoustic and optical phonons show variations of $50\sim 60 \%$. Similar trends where observed in the literature\cite{Chakraborty2012}, but this kind of effect is poorly understood and difficult to predict, highlighting the importance of explicitly including doping in the linear-response calculations.

We note that the way doping is induced can be important. An example is the gate-induced coupling to flexural phonons in graphene
\cite{Gunst2017,Sohier2017}, related to a broken mirror symmetry with respect to the graphene plane. However, we do not expect such effects to be significant for the subset of materials studied here.

Last, we mention that doping obviously leads to a finite density of free carriers. This can give rise to additional loss mechanisms such as electron-electron and electron-plasmon scattering, which can in principle affect the transport properties of a material\cite{Hu1996,Caruso2016} but are not included here.

\subsection{Comparison with literature}

We first discuss the comparison with the other theoretical works reported in Table \ref{tab:mobilities}.
Due to the diversity of the techniques employed, both at the BTE and EPC level, 
discrepancies between different first-principles results can have many different explanations. 
However, comparison with the literature points to the three factors 
discussed above being quite relevant.
Indeed, intervalley scattering, doping and anisotropy, which we identified to be 
essential in determining the transport properties of 2D materials, also turn out to be treated quite differently in different works.

First and foremost, let us note that in general the mobility $\mu = \sigma / n$ depends strongly on carrier density $n$. In the context of doped semiconductors, the conductivity $\sigma$ tends to increase (or possibly decrease in more peculiar cases, like TMDs, where intervalley scattering can be activated above a certain doping) rather slowly as a function of $n$, so the mobility tends to decrease. 
Thus, direct comparison of mobilities calculated or measured at different carrier densities has limited meaning. One must keep in mind that in the small doping limit, the mobility might be increased by a factor $3 \sim 5$ with respect to the present high-doping results.
In the references mentioned in Table \ref{tab:mobilities}, 
first-principles calculations are done in neutral materials. 
The calculated mobilities marked by a "0" in Table \ref{tab:mobilities} apply to the undoped limit, as clearly stated in the references using Boltzmann transport\cite{Gaddemane2018,Li2015}, and implied in references based on Monte-Carlo transport\cite{Jin2014,Jin2016,Li2013}, and Takagi's formula\cite{Takagi,Qiao2014,Pizzi2016nc,Wang2017,Zhang2014}. For those works, the comparison with the high-doping mobilities presented here should be taken with caution. We report them nonetheless to highlight the importance of doping and the ambiguity surrounding it, given that those values are sometimes compared with experimental measurements at finite doping.
For the theoretical works where a finite carrier density is indicated, doping is sometimes included only a posteriori as a shift of the Fermi level in the computation of transport properties. In some instances, analytical models of screening are used, but those are not well established in 2D materials. In any case, we have seen that the effect of doping goes beyond shifting the Fermi level and electronic screening. Thus, performing EPC calculations with doping explicitly included is important to evaluate the transport properties of doped materials. This appears to be particularly important in phosphorene, where we find a larger mobility compared to most of the results in literature\cite{Jin2016,Rudenko2016,Liao2015,Gaddemane2018}. In particular we find that, due to the doping induced screening, the LA mode is less effective in scattering electrons than what previously reported and only contributes for approximately 50\% of the total scattering processes. Screening  effects might be large enough to compensate the $1/n$ dependency of the mobility and the small doping mobility could be smaller or of the same order of magnitude as the high doping mobility.

Many models include intravalley coupling only (e.g.\ ``deformation potential" models), thus neglecting intervalley transitions. Furthermore, given the complexity of EPC in multivalley materials, most analytical models of EPC with fitted parameters from first-principles are likely to be incomplete. 
It is no coincidence that works employing such approaches show the greatest differences with the 
current results (even assuming a factor 10 increase of the mobility with respect to our result in the low doping limit), often largely overestimating mobility, like the results on TMDs reported in Ref.~\onlinecite{Zhang2014} or on arsenene in Ref.~\onlinecite{Pizzi2016nc}.    
When intervalley scattering is  accounted for, it can still be a source of discrepancy depending on the relative positions of the valleys. This is similar to what happens in GaAs, where the value of mobility at high-temperature crucially depends on the relative energy between the $\Gamma$ and L valley\cite{Rode1971,Zhou2016,Ma2018}.
In 2D, this is relevant for TMDs, in which  the position of the Q valley with respect to the bottom of the K valley is difficult to determine, as mentioned before, from first-principles. For example, the energy separation between the two extrema in MoS$_2$ ranges between 70\cite{Li2013} and 260 meV\cite{Li2015,Kaasbjerg2012a} with the latter results closer to our calculations. This difference brings a large intervalley contribution for MoS$_2$ in Ref.~\onlinecite{Li2013} that we don't observe, and that we find instead in the W-based TMDs. The relative energy separation of the K and Q valley affects also our ranking in terms of electron mobility among the TMDs, with MoS$_2$ performing better than WS$_2$, contrary to the findings in Ref.~\onlinecite{Jin2014}. 
In Ref.~\onlinecite{Jin2014} the energy separation between K and Q is 80 meV in MoS$_2$ and 67 meV in WS$_2$ resulting in a similar intervalley scattering between the two materials and a better mobility for WS$_2$ on the basis of its lighter effective mass. In our case instead the K-Q energy separation shows a larger variation, from 257 meV in MoS$_2$ to only 178 meV in WS$_2$ resulting in an increased intervally scattering that undermines the advantage of a lighter effective mass.   

The anisotropy of the valleys affects the validity of approximated solutions to the Boltzmann transport equation. 
In particular, for anisotropic phosphorene, the energy relaxation time approximation (see App. \ref{app:closed_form}) gives results with up to $30\%$ errors with respect to the full solution. For the other materials studied here, with isotropic transport properties, this error reduces to a few percents.
Although these errors can be acceptable in most cases, a full solution to the BTE beyond the relaxation-time approximation is valuable for 
quantitative comparison with experiments. 
In any case, given that the errors associated to different approximate solutions to the BTE are not obvious to determine a priori, and that the numerical solution to the full linearized BTE is comparable in terms of computational cost, we consider it a useful in general.

We now discuss the comparison with experimental work, which concerns only the TMDs  reported in Table \ref{tab:mobilities}. As already discussed, the positions of the valleys with respect to the Fermi level play a very important role, and since our bands are computed within DFT without SOC, we do not claim quantitative agreement with experiments. In practice, one could measure both higher or lower mobilities. For example, one could measure a lower mobility if the Q valley is lower, and a higher one if spin conservation restricts available scattered states. 
Nevertheless, experimental mobilities for TMDs are relatively close to our predictions, whereas one would expect additional extrinsic scattering processes to give significantly lower values. This can be explained in several ways.
First, since mobility goes as $1/n$ and intervalley scattering is activated as the Fermi level increases and crosses more valleys, we can expect the high-density intrinsic mobilities computed here to be lower than experiments performed at lower densitites. Second, one could conclude that the importance of extrinsic contributions, like remote phonons, might have been overestimated in the past. Conversely, extrinsic mechanisms leading to an enhanced mobility, such as the phonon mode quenching mentioned in Ref. \onlinecite{Radisavljevic2013}, could be more effective than expected. Undoubtedly, the complexity of electron and phonon dynamics in TMDs calls for further work before reaching numerical agreement between simulations and experimental measurements.

\section{Conclusions}

We have developed an automated procedure to determine the transport properties of 2D 
materials from first-principles. We aim for the highest accuracy achievable within the 
framework of density-functional perturbation theory, with as few assumptions and 
simplifications as possible. 
The method includes several strengths and improvements with respect to existing 
approaches. Electron-phonon coupling matrix elements are directly computed from density 
functional-perturbation theory in the correct dimensionality framework and with the 
correct electrostatics of field-effect doping. The linearized Boltzmann transport 
equation is solved numerically in full, 
beyond relaxation-time 
approximation or any other closed-form expressions for the scattering time. 
The implementation of this entire transport workflow within the AiiDA infrastructure 
provides great flexibility to improve or adapt the method to different applications, 
as well as the data storage and 
provenance necessary to build and disseminate databases.
Here, we studied in detail a small test set of six systems (electron-doped MoS$_2$, WS$_2$, WSe$_2$, arsenene and phosphorene as well as hole-doped phosphorene) 
presenting different characteristics. 
Our results point out the crucial role of intervalley scattering, band anisotropy and 
doping to the transport properties of 2D materials, in turn underscoring the importance 
of an accurate treatment of these aspects in first-principles simulations. 
Hole-doped phosphorene is found to yield the highest mobility, thanks to its mono-valley and anisotropic nature. 
Electron-doped arsenene shows a lower mobility than could be expected, due to the importance of intervalley scattering. The transport properties of electron-doped TMDs are found to be very sensitive to the relative positions of the K and Q valleys: these quantities are still subject of study at the highest levels of theory and experimentally. While this work is based on DFT band structures, more accurate predictions would be reached by including spin-orbit interactions, GW corrections and by fitting at least the most important features of the valleys to experimental data.

\section*{Acknowledgements:}
This work has been in part supported by NCCR MARVEL (N.M. and D.C.). Calculations were performed on the 
Marconi - KNL supercomputer in Cineca under PRACE project PRA15\_3963. D.C.\ acknowledges support from the ‘EPFL Fellows’ fellowship programme co-funded by Marie Sklodowska-Curie, Horizon 2020 grant agreement no. 665667. M.G.\ acknowledges support from the Swiss National Science Foundation through the Ambizione career programme.
\newpage

\onecolumngrid
	 
\newpage


\renewcommand\theequation{A\arabic{equation}}
\renewcommand\thefigure{A\arabic{figure}}
\setcounter{equation}{0}
\setcounter{figure}{0}
\setcounter{table}{0}
\setcounter{section}{0}

\section*{Appendix}

\section{Closed Form of BTE}
\label{app:closed_form}
Re-writing Eq. \ref{eq:BTE} as
\begin{align} 
1 &= \sum_{\bok'} P_{\bok\bok'} \frac{1-f^0(\bok')}{1-f^0(\bok)} \times \left\{\tau(\bok)-  \tau(\bok')  \frac{\bo{v}(\bok') \cdot \bo{u}_{\bo{E}} }{\bo{v}(\bok)\cdot\bo{u}_{\bo{E}}} 
\right\}, 
\end{align}
we see that a closed form can be obtained by assuming $\tau(\bok) \approx \tau(\bok')$:
\begin{align} 
\frac{1}{\tau(\bok)} &= \sum_{\bok'} P_{\bok\bok'} \frac{1-f^0(\bok')}{1-f^0(\bok)} \times \left\{1-  \frac{\bo{v}(\bok') \cdot \bo{u}_{\bo{E}} }{\bo{v}(\bok)\cdot\bo{u}_{\bo{E}}} 
\right\}, 
\end{align}
To obtain the closed form used in this work to initialize the iterative solution to the BTE, we replace $\bo{u}_{\bo{E}}$ by the direction of $\bok$ with respect to the bottom of the corresponding valley, as explained in the main text.
One may also replace $\bo{u}_{\bo{E}}$ with $\bo{v}(\bok)$ to obtain the so-called momentum relaxation-time approximation (mRTA) for the scattering time:
\begin{align}
\frac{1}{\tau_{\rm \footnotesize mRTA}(\bok)} &= \sum_{\bok'} P_{\bok\bok'} \frac{1-f^0(\bok')}{1-f^0(\bok)} \times \left\{1-  \frac{\bo{v}(\bok') \cdot \bo{v}(\bok) }{\bo{v}(\bok)^2} 
\right\},
\end{align}
which gives very similar results.
If the second term in brackets on the right-hand side is neglected, we finally obtain the scattering time within the energy relaxation-time approximation (eRTA):
\begin{align}
\frac{1}{\tau_{\rm\footnotesize eRTA}(\bok)} &= \sum_{\bok'} P_{\bok\bok'} \frac{1-f^0(\bok')}{1-f^0(\bok)} 
\end{align}

\section{Alternative algebraic solution to the BTE}
\label{app:matrix_sol}

We give here an other method to solve the BTE, relying on a direct algebraic solution of the system.
Writing the BTE for the $N_i$ irreducible states yields $N_i$ equations and $N_f$ unknown variables $\tau(\bok')$. To solve the corresponding system, we first need to fold it back on the irreducible states, using the fact that $\tau(\bok') \approx \tau(\tilde{\bok}')$ where $\tilde{\bok}' \in \mathcal{I}$ is the symmetry equivalent of $\bok'\in \mathcal{F}$:
\begin{align}
\begin{split}
& \left( 1-f^0(\tilde{\bok})\right) \bo{v}(\tilde{\bok}) \cdot \bo{u}_{\bo{E}} =   \\
& \sum_{\bok'} P_{\tilde{\bok}\bok'} \left( 1-f^0(\bok')\right) \bo{v}(\tilde{\bok}) \cdot \bo{u}_{\bo{E}} \times 
\tau(\tilde{\bok}) \\
& - 
\sum_{\tilde{\bok}'} \left( \sum_{\bok'\equiv\tilde{\bok}'} P_{\tilde{\bok}\bok'} \left( 1-f^0(\bok')\right) \times  \bo{v}(\bok') \cdot \bo{u}_{\bo{E}}  \right) \tau(\tilde{\bok}').
\end{split}
\end{align}
From there we use the same techniques mentionned in the main text to have a well-behaved set of $N_i$ equations with $N_i$ unknowns. 
Written as a matrix-vector product, it reads: 
\begin{align}
\forall \tilde{\bok} \in \mathcal{I}, \ F_{\tilde{\bok}} = \sum_{\tilde{\bok}'} \mathcal{S}_{\tilde{\bok}, \tilde{\bok}'} \tau_{\tilde{\bok}'},
\end{align}
where 
\begin{align}\label{eq:ScatMat}
F_{\tilde{\bok}} &= \left( 1-f^0(\tilde{\bok})\right) \bo{v}(\tilde{\bok}) \cdot \bo{u}_{\tilde{\bok}} \\ 
\begin{split}
\mathcal{S}_{\tilde{\bok}, \tilde{\bok}'}  & =  \left(\sum_{\bok''} P_{\tilde{\bok}\bok''} \left( 1-f^0(\bok'')\right) \bo{v}(\tilde{\bok}) \cdot \bo{u}_{\tilde{\bok}} \right) 
\delta_{\tilde{\bok}, \tilde{\bok}'} \\
& \ -  \left( \sum_{\bok'\equiv\tilde{\bok}'} P_{\tilde{\bok}\bok'} \left( 1-f^0(\bok')\right) \times  \bo{v}(\bok') \cdot \bo{u}_{\tilde{\bok}}  \right) 
\end{split}
\end{align}
Solving the BTE then boils down to a matrix inversion, and this gives us the values of $\tau$ on the $\mathcal{I}$ grid. It is then rotated according to the symmetries of the band structure, and interpolated linearly on the fine grid of the pockets $\mathcal{F}$. 
In this approach, the $\delta$ functions in Eq. \ref{eq:Pkkp} are replaced by Gaussians, and a convergence test is performed on the associated broadening. This way, one associates a weight to each contribution from each $\bok$ point, and the contributions can then be organized as a matrix.
In the triangles method, one does not associate a contribution to each $\bok$ point of the grid. For the integral to be computed, the integrand must be known a priori. This is not the case here since the integrals involve the scattering times at all $\bok$. 
The approach discussed here and the one in the main text give the same results within a $5\%$ error, provided the broadening parameter is chosen properly.

\section{Bands structures and phonon dispersions}
\label{app:bands_phonons}

\begin{figure*}[h]
\includegraphics[width=0.32\textwidth]{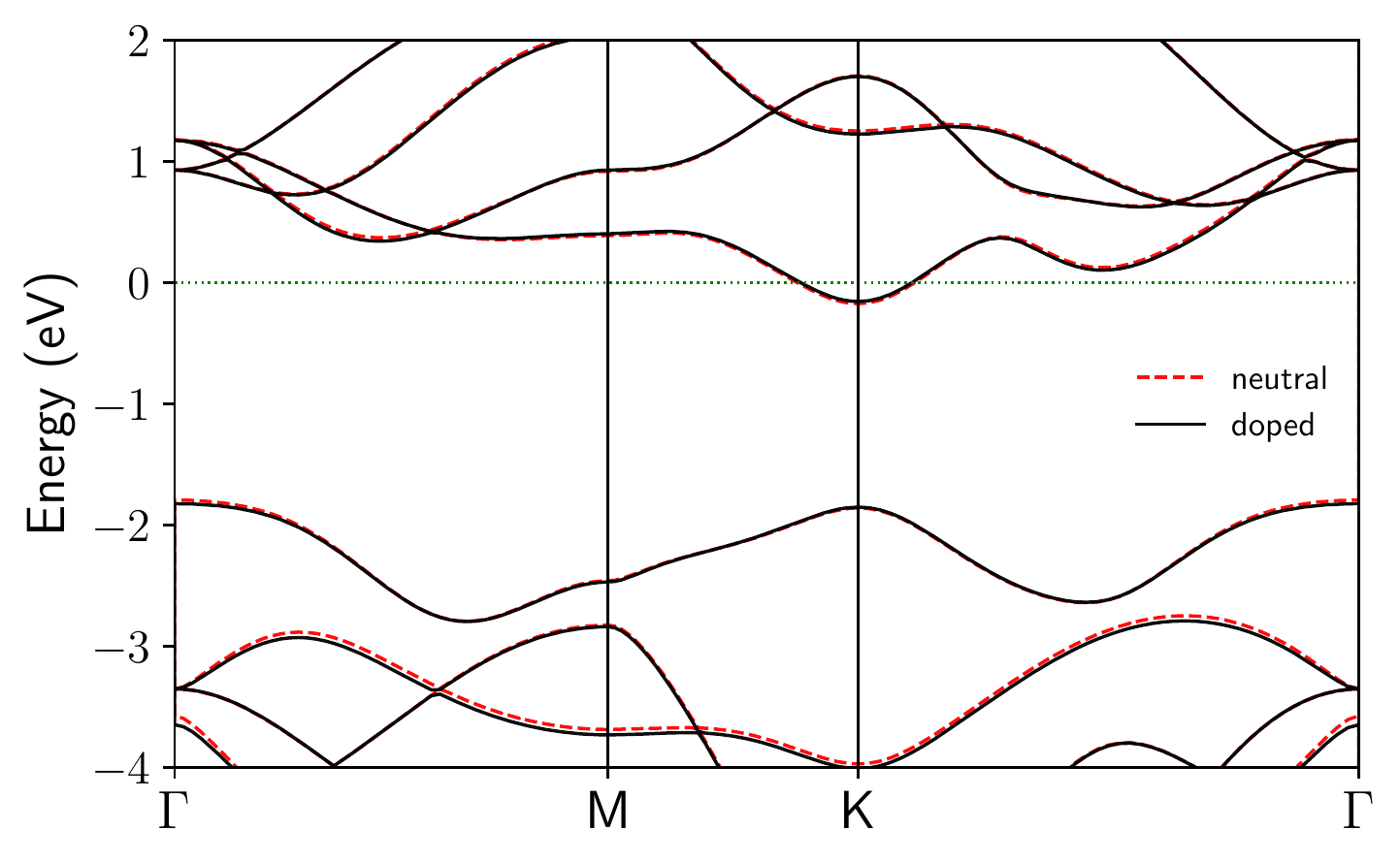}
\includegraphics[width=0.32\textwidth]{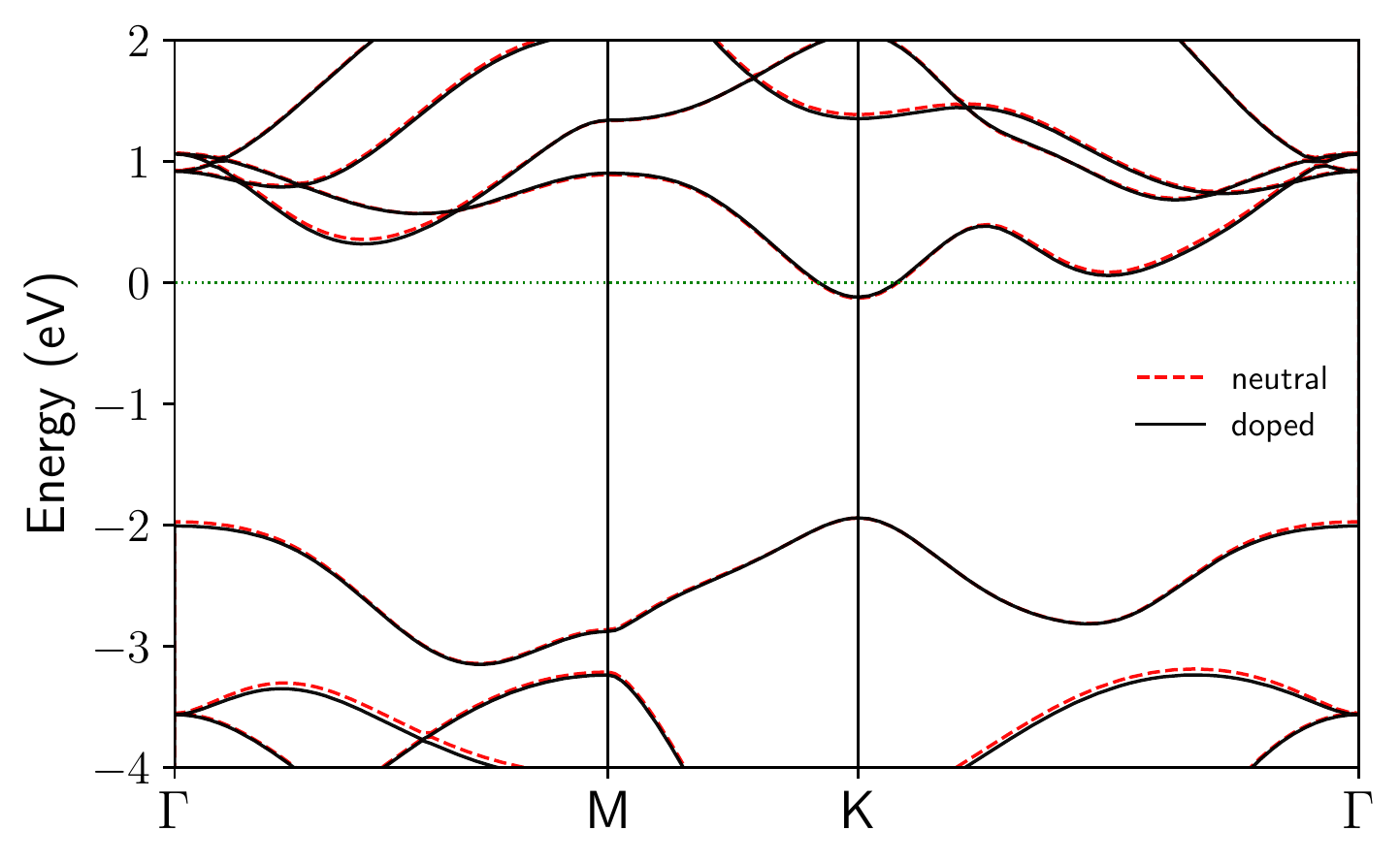}
\includegraphics[width=0.32\textwidth]{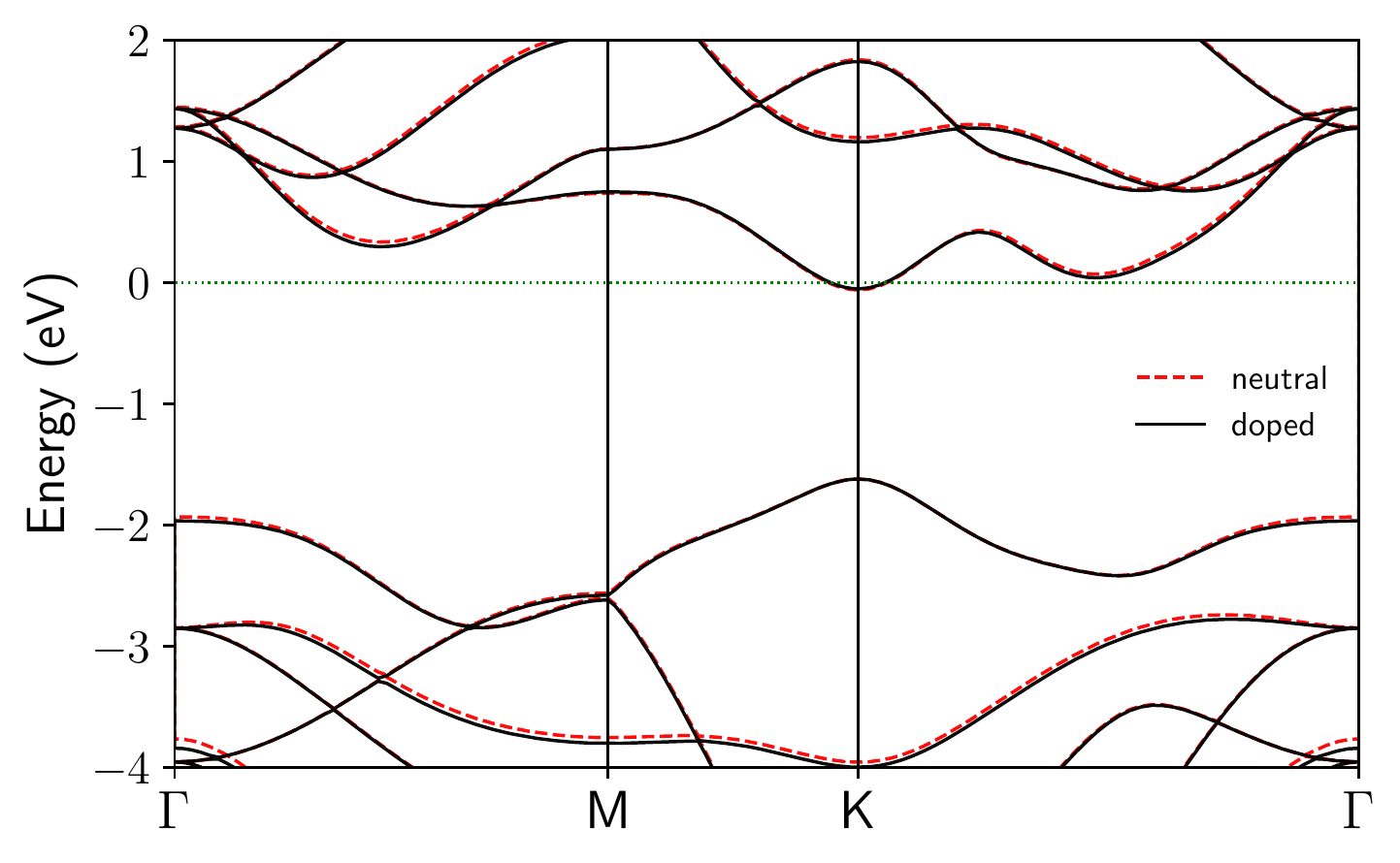}
\includegraphics[width=0.32\textwidth]{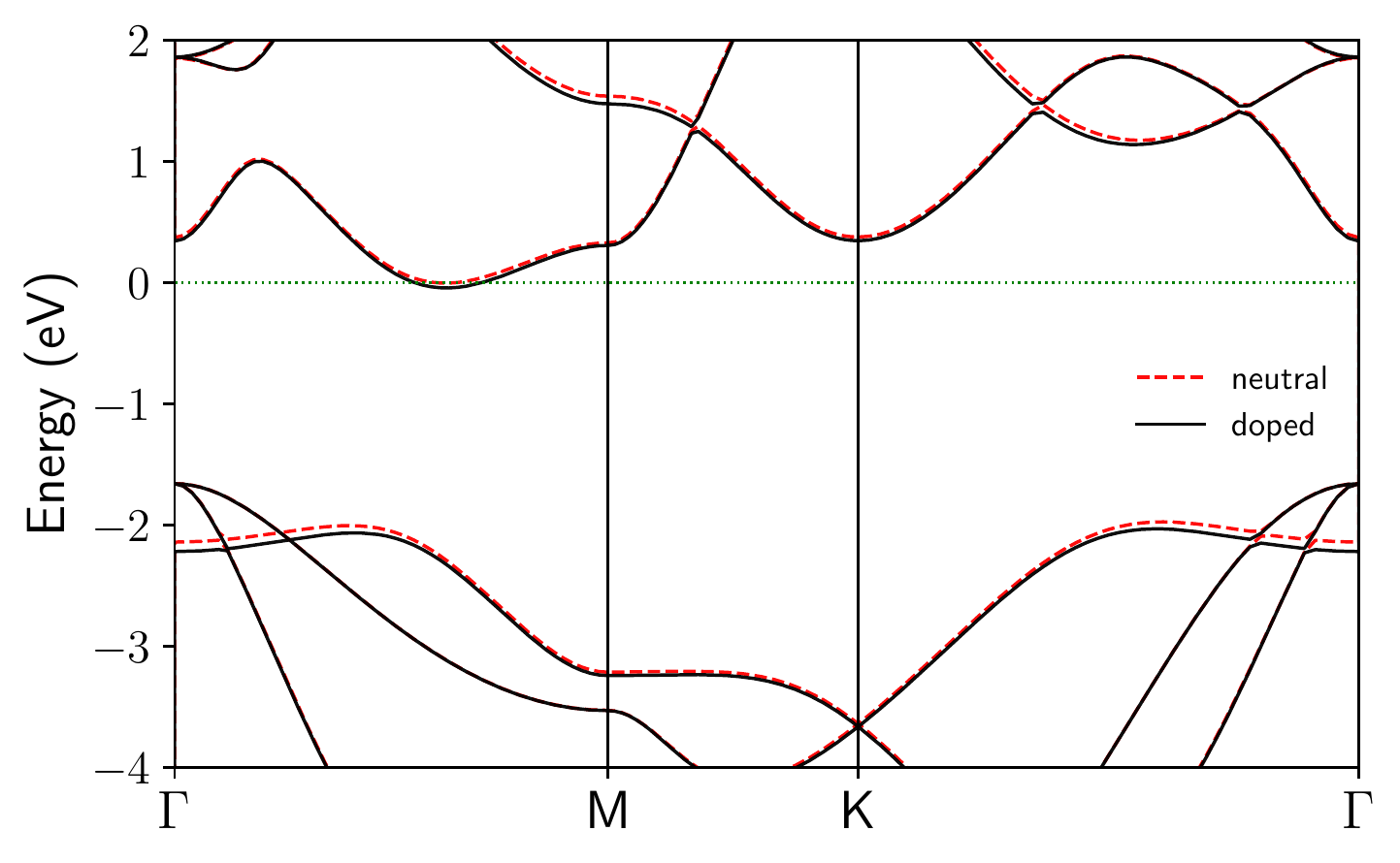}
\includegraphics[width=0.32\textwidth]{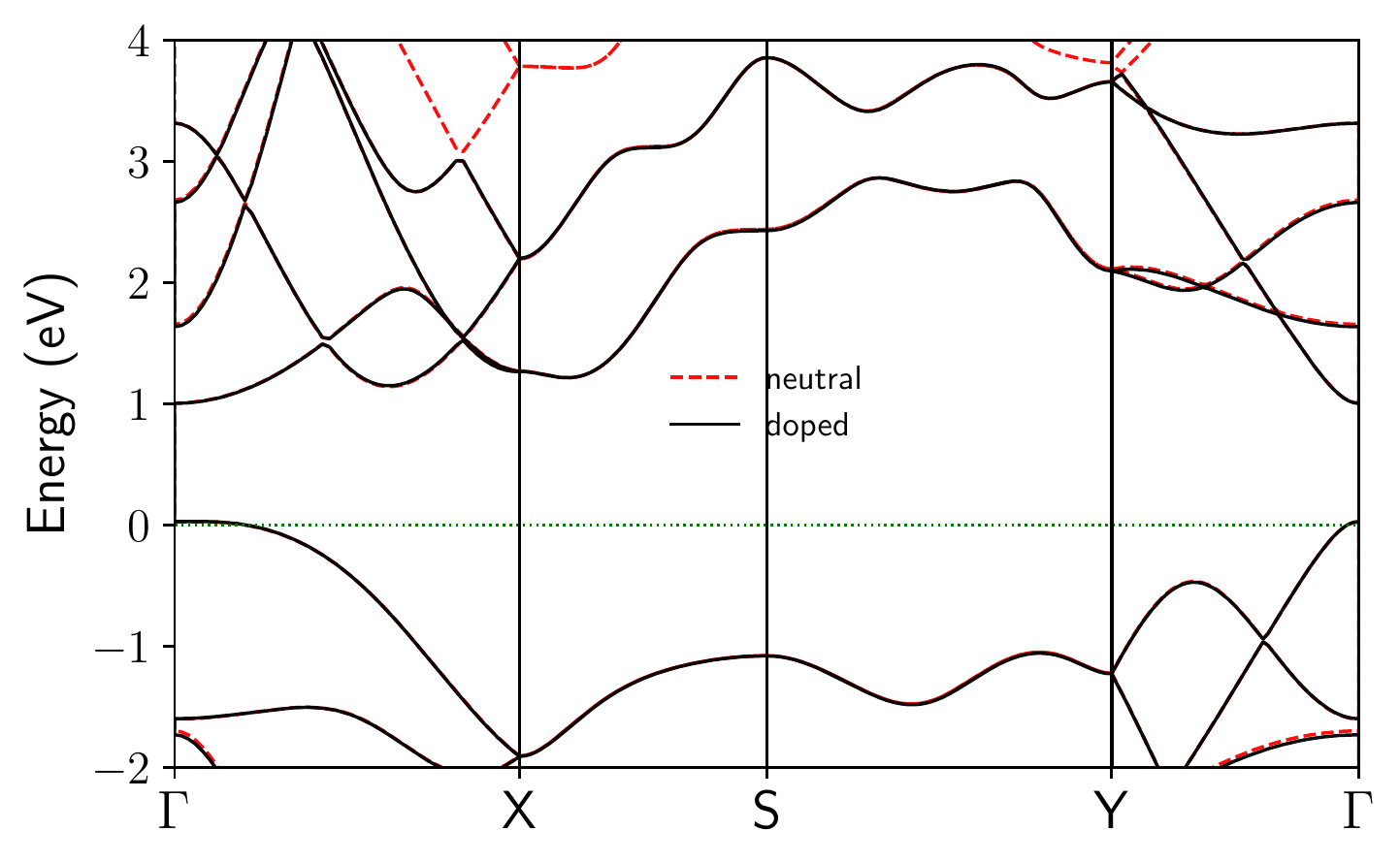}
\includegraphics[width=0.32\textwidth]{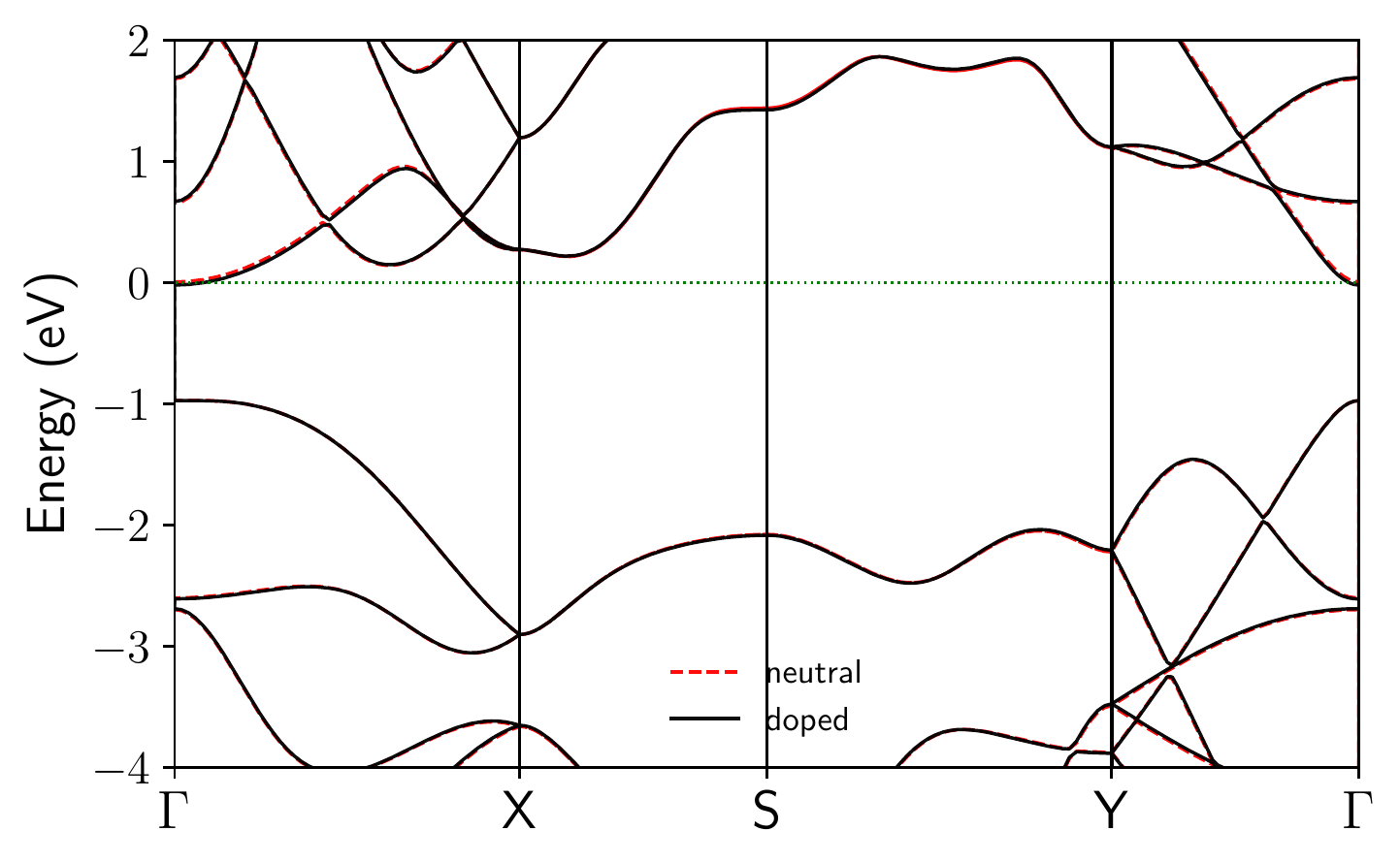}
\caption{From left to right, top to bottom: band structure along high symmetry path for electron-doped MoS\2 , WS\2, and WSe\2, As, P\4, and hole-doped P\4 (black solid lines) compared with the undoped case (red dashed lines). In each case the two band structures are aligned with respect to the top of the valence band. The zero is at the Fermi level of the doped case. }
\label{fig:bands}
\end{figure*}

\begin{figure*}[h]
\includegraphics[width=0.32\textwidth]{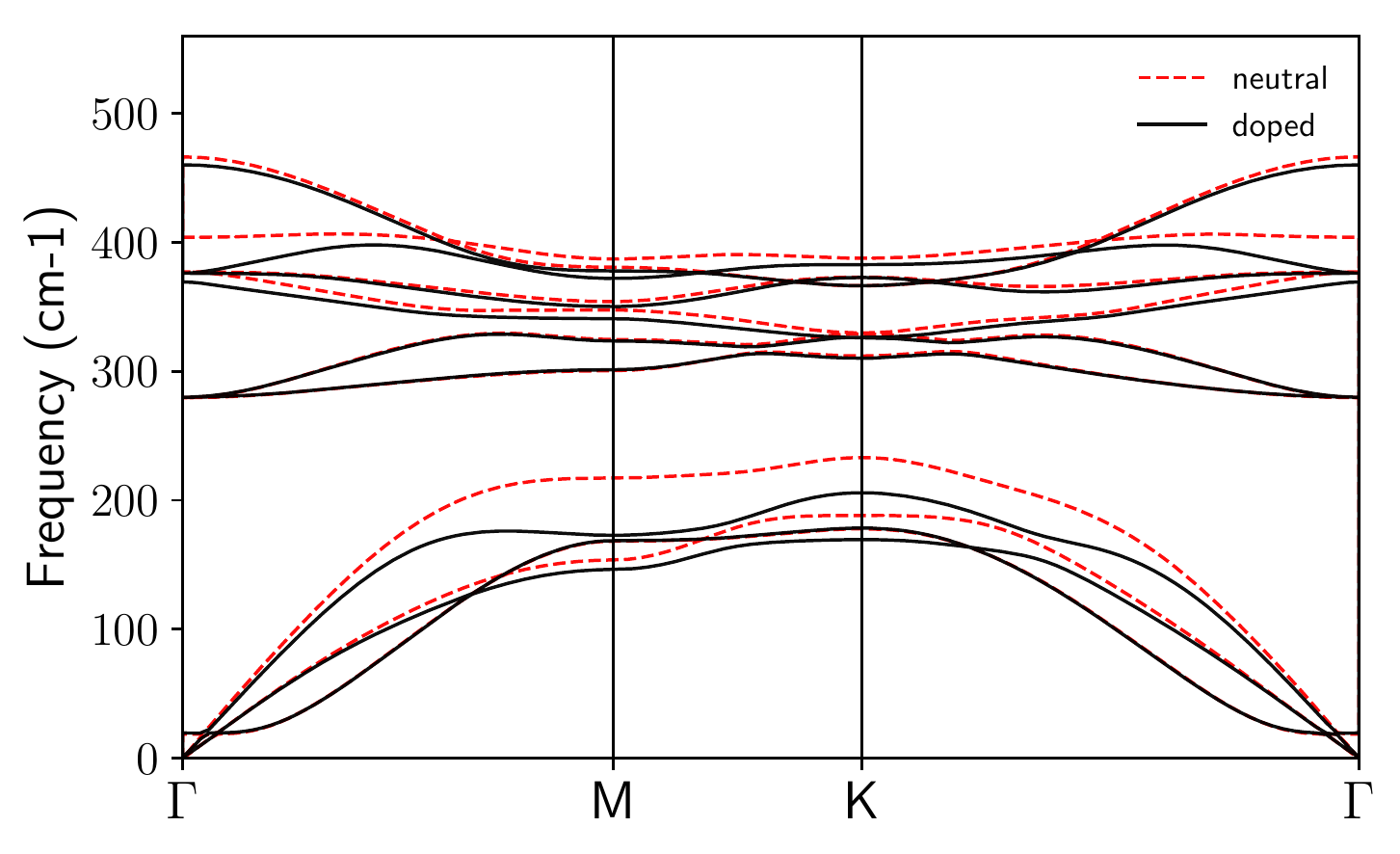}
\includegraphics[width=0.32\textwidth]{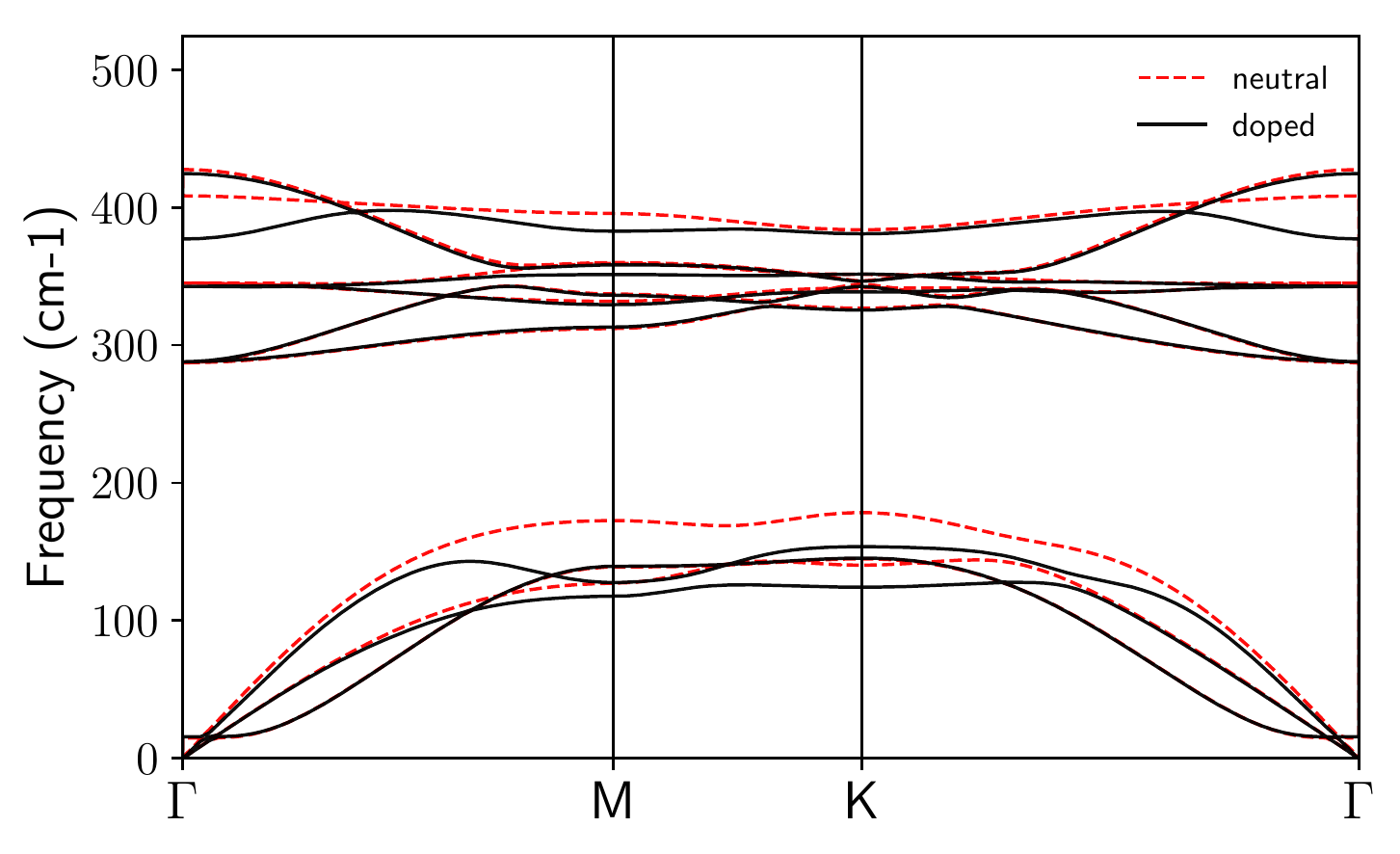}
\includegraphics[width=0.32\textwidth]{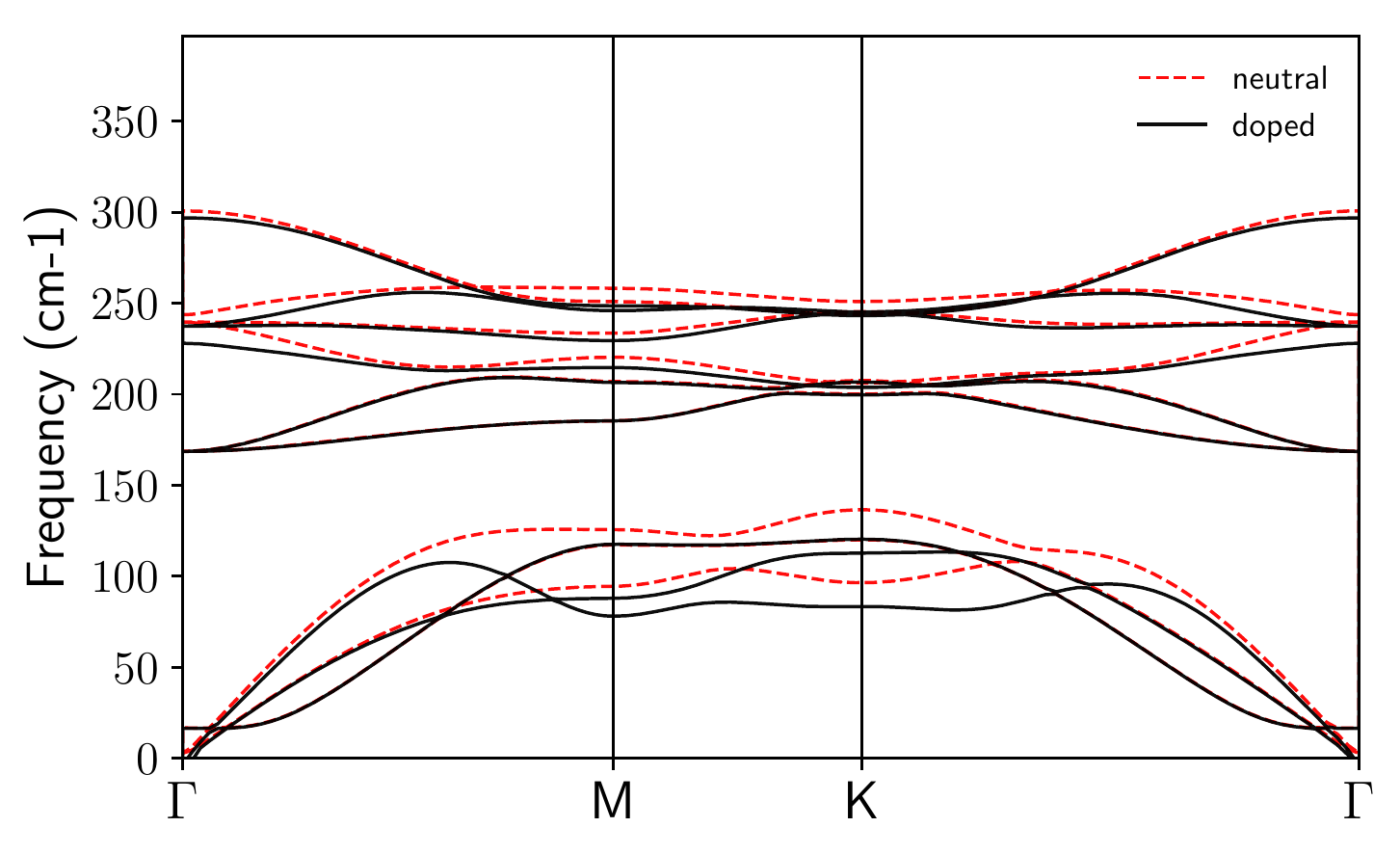}
\includegraphics[width=0.32\textwidth]{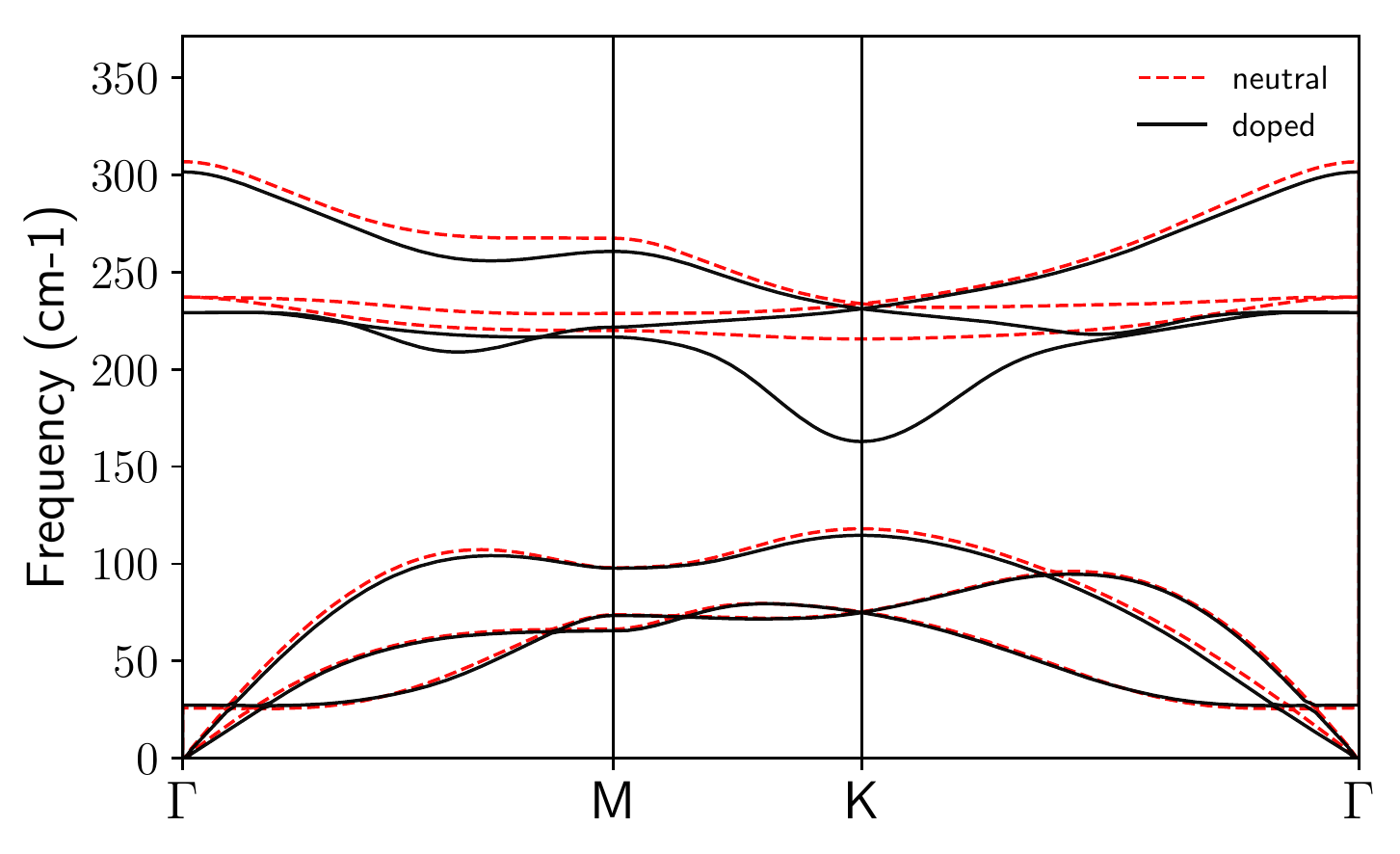}
\includegraphics[width=0.32\textwidth]{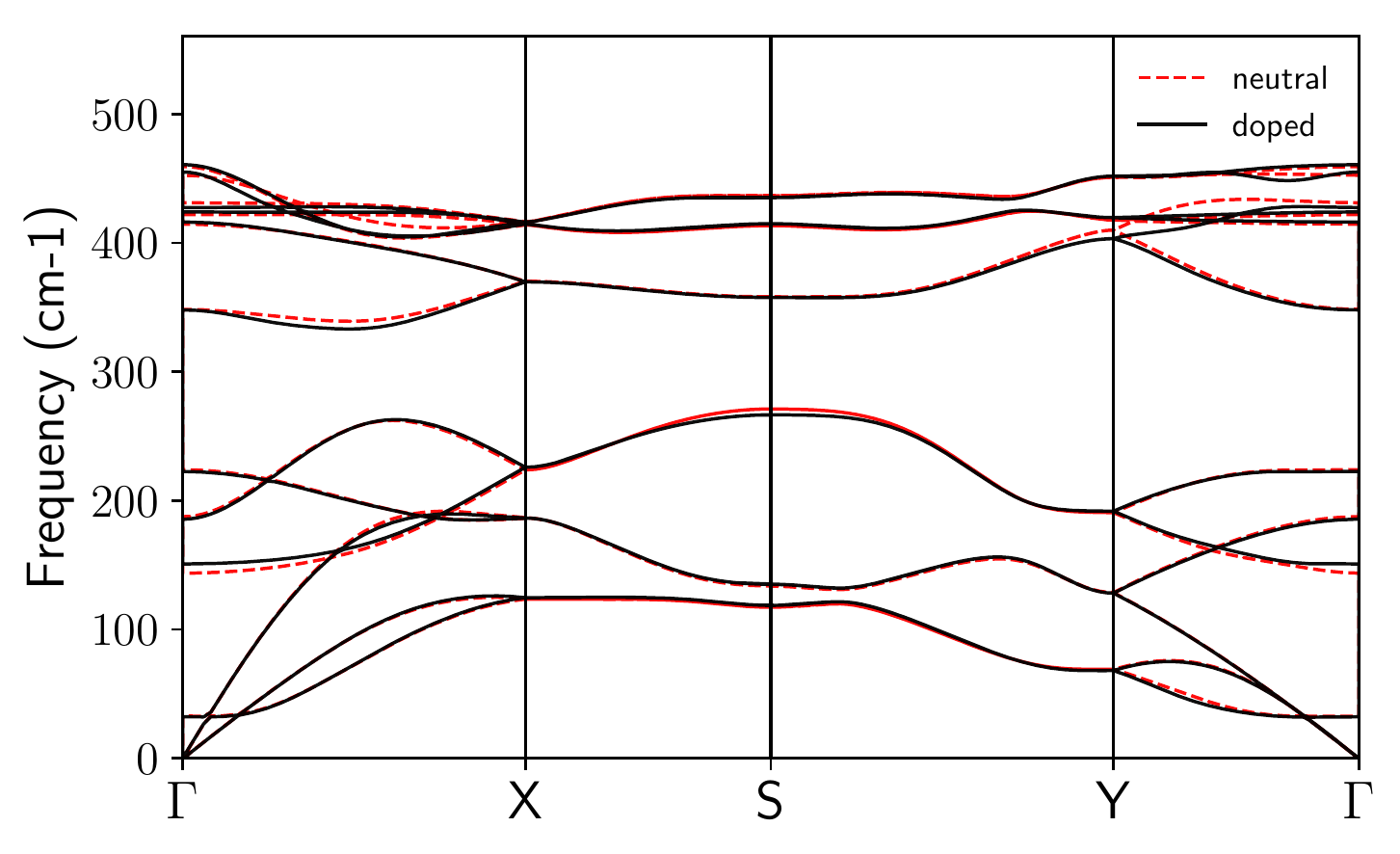}
\includegraphics[width=0.32\textwidth]{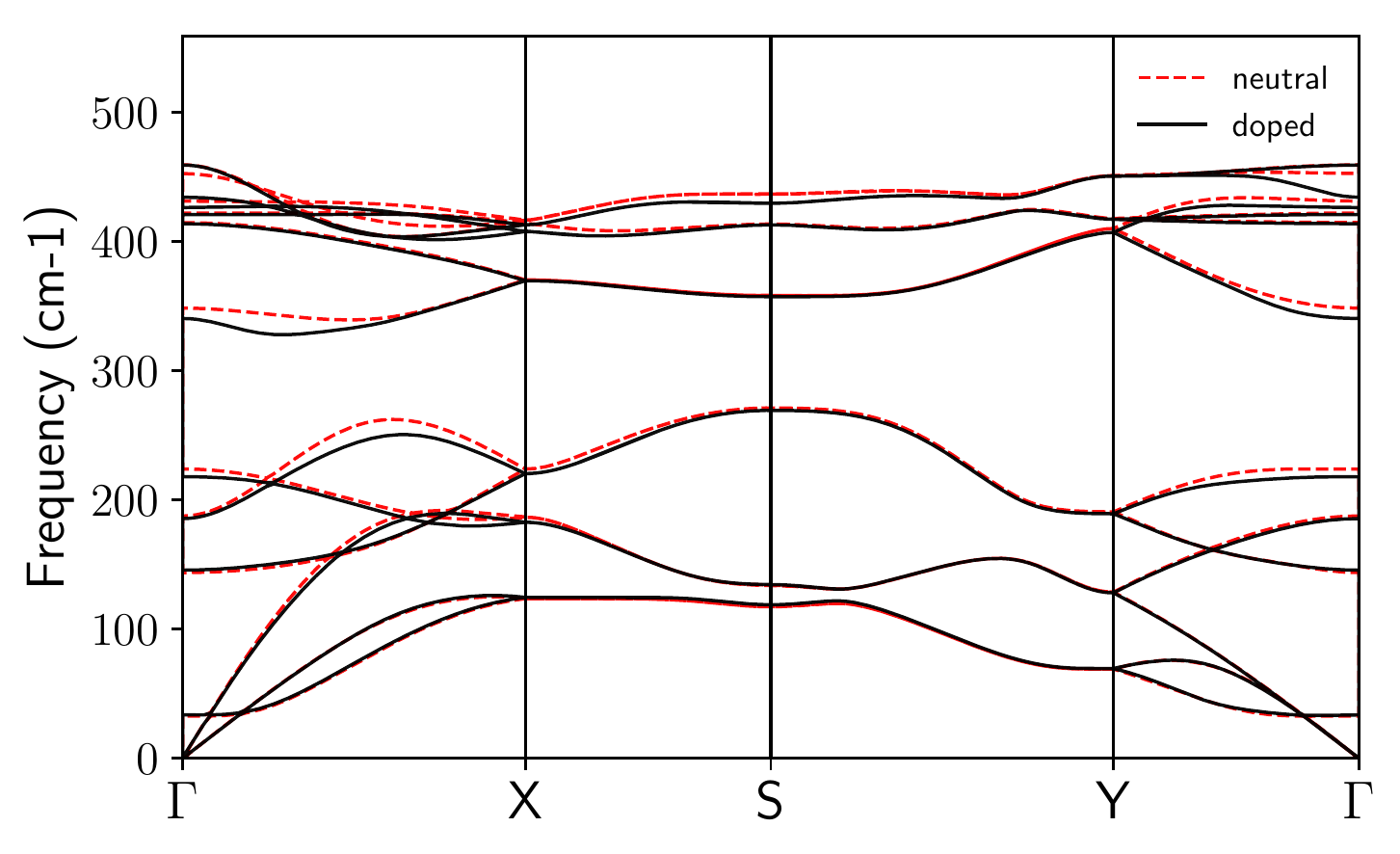}
\caption{From left to right, top to bottom: phonon dispersion along high symmetry path for electron-doped MoS\2 , WS\2, and WSe\2, As, P\4, and hole-doped P\4 (black solid lines) compared with the undoped case (red dashed lines). Phonon softenings can be observed for phonons corresponding to possible electronic transitions with strong electron-phonon coupling. For example, in arsenene, the softenning at K corresponds to the strong intervalley coupling ($g \approx 140$ meV) with mode number $4$ observed in Fig. \ref{fig:EPC_As}.}
\label{fig:phonons}
\end{figure*}

For the materials studied here, we report the band structures in Fig. \ref{fig:bands}, phonon dispersions in Fig. \ref{fig:phonons} and some interesting quantities related to the band structures in Table \ref{tab:effmass}.

\begin{table*}[h!]
\caption{Extracted quantitative characteristics of the band structures. The first column shows effective masses at the band edges relevant for transport, obtained from finite-differences differentiation. If the effective mass is anisotripic the lighter effective mass is indicated as transport effective mass (T) while the mass in the perpendicular direction is referred to as longitudinal mass (L). The second column shows the absolute energy difference between the bottom of the conduction band (top of the valence band) and the Fermi level. The third column shows the absolute energy difference between the bottom of the K valley and the bottom of the Q valley in TMDs.}
\begin{tabular}{ c c c c }
\hline
Material (e or h doped) & Effective masses (m$_0$)  & $\Delta(E_{Fermi} - E_{max,min}) (meV) $ &  $\Delta(E_{K} - E_{Q}) (meV)$  \\ 
\hline
\hline
MoS2(e) & m$^{*}_{K}$=0.417 & 153.3 & 257.4 \\
WS2(e) & m$^{*}_{K}$=0.314 & 118.1 & 178.0\\
WSe2(e) & m$^{*}_{K}$=0.330 & 50.2 & 92.0\\
As(e) &  m$^{*}_{Q}$(T)=0.145,m$^{*}_{Q}$(L)=0.517 & 42.8 & - \\
P (h) & m$^{*}_{\Gamma}$(T)=0.172,m$^{*}_{\Gamma}$(L)=8.872 & 28.9  & -\\
P (e) & m$^{*}_{\Gamma}$(T)=0.139,m$^{*}_{\Gamma}$(L)=1.237  & 18.9 & - \\
\hline
\hline
\end{tabular}
\label{tab:effmass}
\end{table*}

\section{additional EPC plots:}
\label{app:EPC}
Here we show additional plots for the EPC matrix elements of MoS$_2$, WS$_2$, WSe$_2$ and phosphorene.
\begin{figure*}[h]
\includegraphics[width=0.85\textwidth]{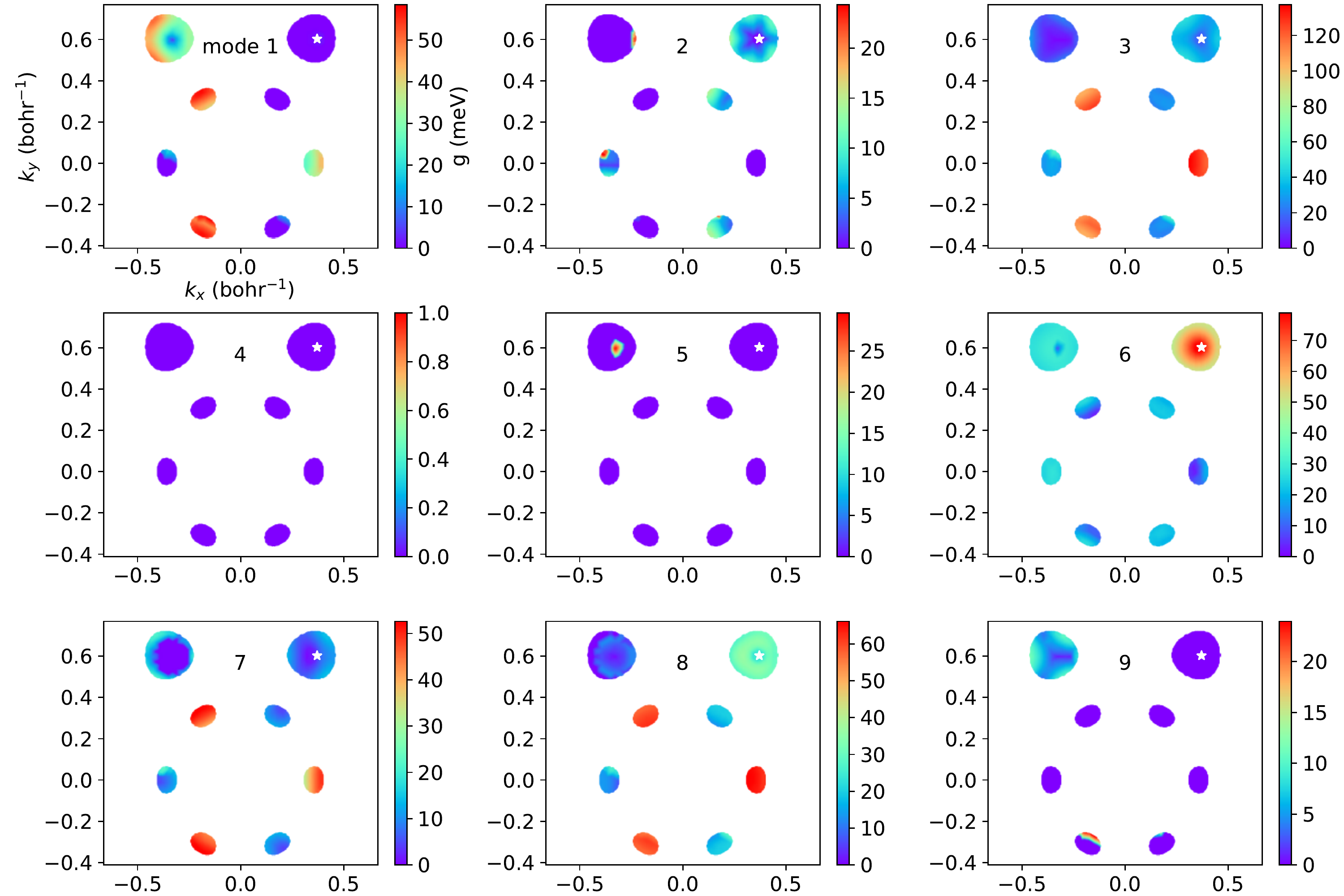}
\caption{Interpolated $g_{\bok \bok'}$ for electron-doped MoS$_2$. The initial state considered is indicated by a white star. The rest of the points are the possible final states in the finely sampled pockets and the color of the point indicates the strength of the electron-phonon coupling matrix element. The index of the phonon mode indicated at the top of each subplot refers to a purely energetic ordering of the phonon modes associated with each transition.}
\label{fig:EPC_MoS2}
\end{figure*}

\begin{figure*}[h]
\includegraphics[width=0.85\textwidth]{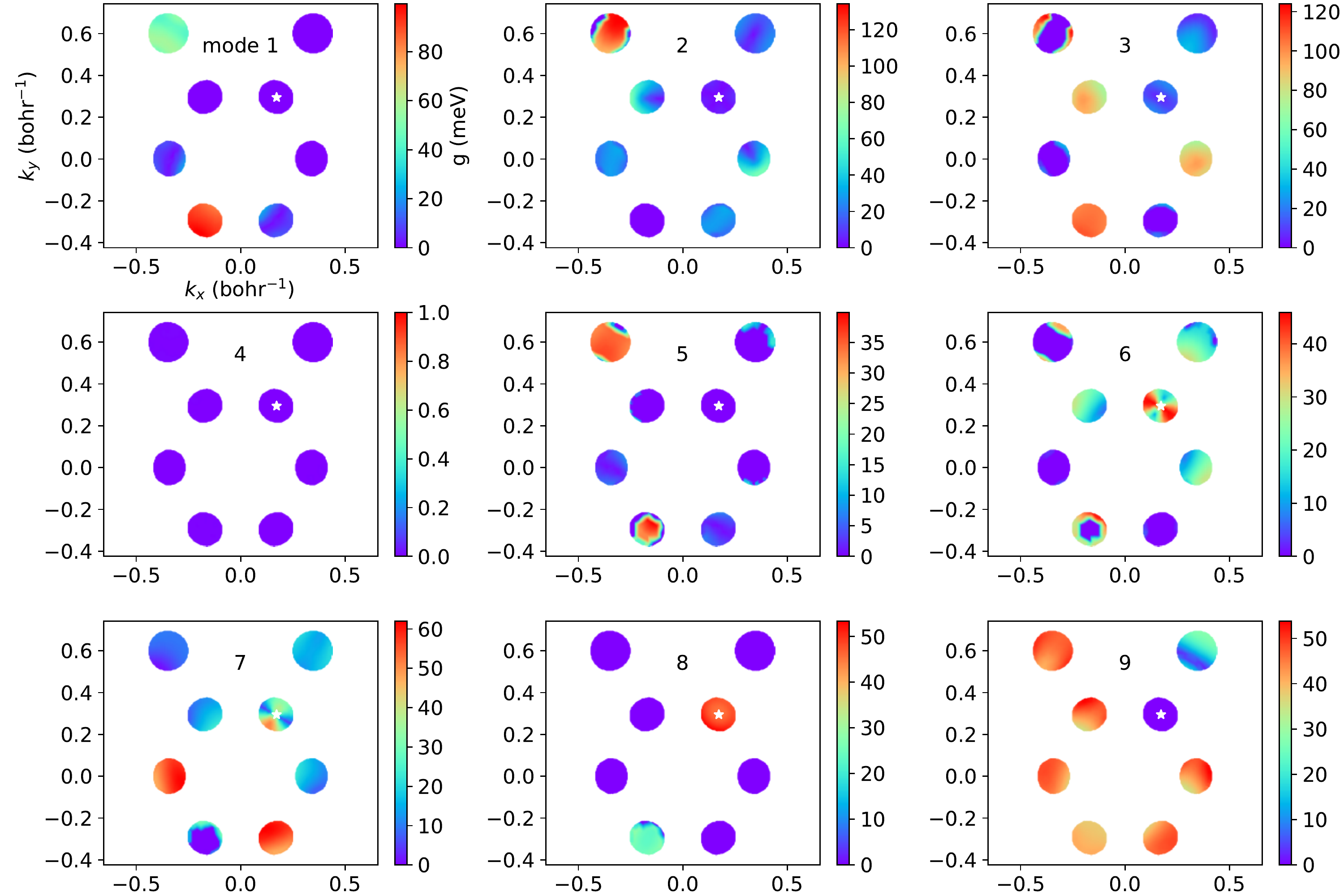}
\caption{Interpolated $g_{\bok \bok'}$ for electron-doped WS$_2$, with the initial state at the bottom of the Q valley.}
\label{fig:EPC_WS2}
\end{figure*}

\begin{figure*}[h]
\includegraphics[width=0.85\textwidth]{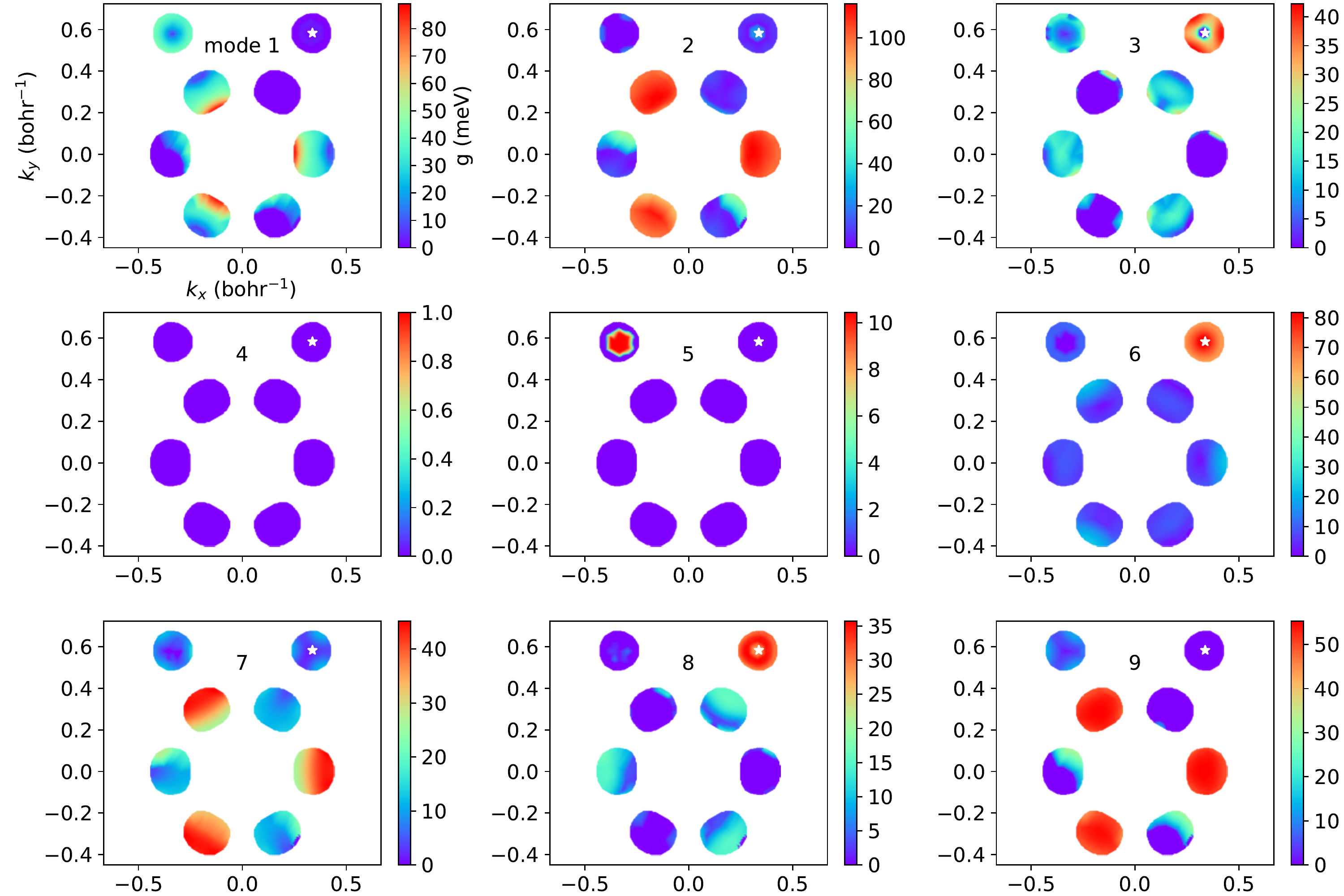}
\caption{Interpolated $g_{\bok \bok'}$ for electron-doped WSe$_2$, with the initial state at the bottom of the K valley.}
\label{fig:EPC_WSe2}
\end{figure*}

\begin{figure*}[h]
\includegraphics[width=0.85\textwidth]{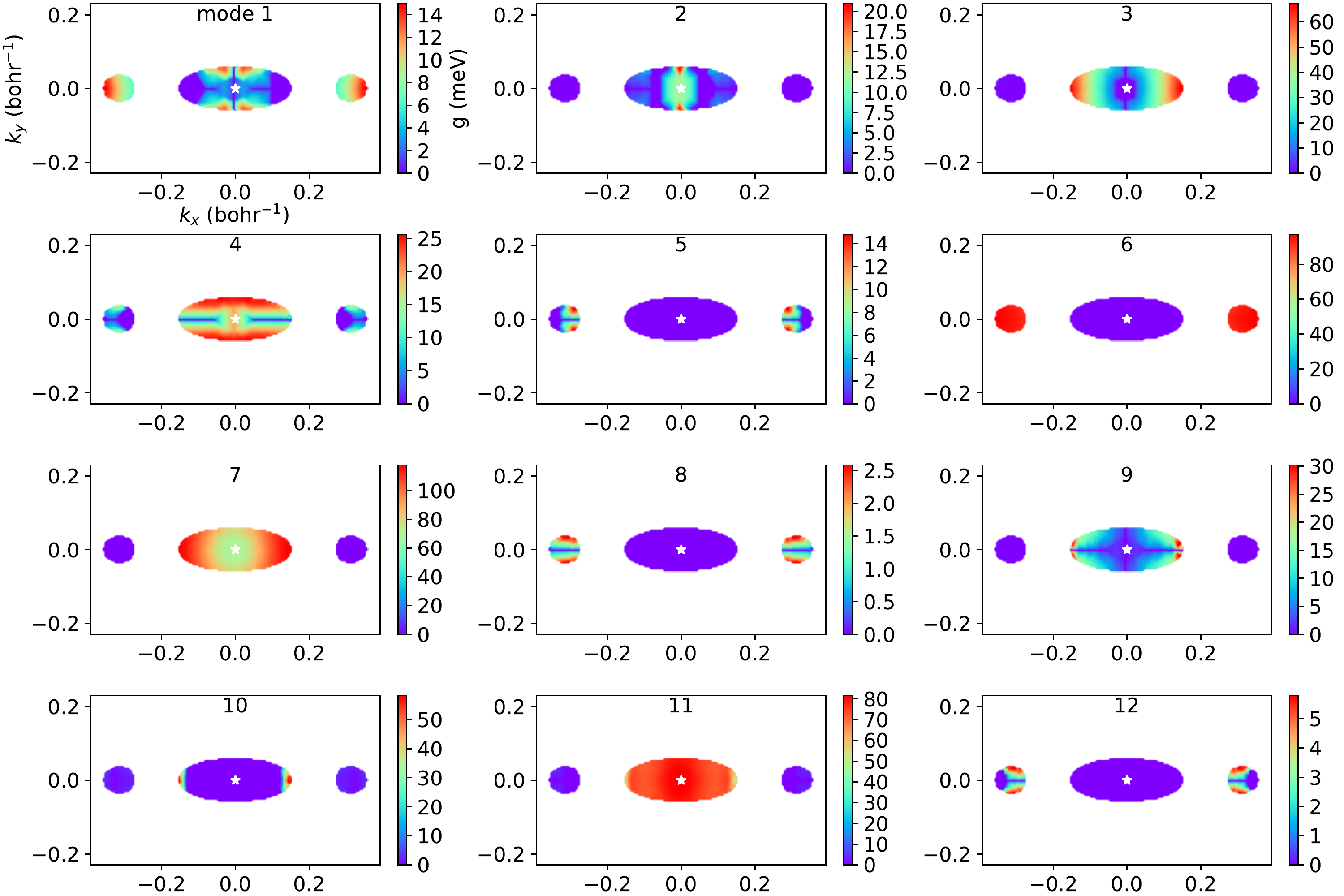}
\caption{Interpolated $g_{\bok \bok'}$ for electron-doped phosphorene, with the initial state at the bottom of the $\Gamma$ valley.}
\label{fig:EPC_P4e}
\end{figure*}

\newpage

\twocolumngrid
\bibliographystyle{myapsrev}
\bibliography{bib}

\begin{thebibliography}{100}%
\makeatletter
\providecommand \@ifxundefined [1]{%
 \ifx #1\undefined \expandafter \@firstoftwo
 \else \expandafter \@secondoftwo
\fi
}%
\providecommand \@ifnum [1]{%
 \ifnum #1\expandafter \@firstoftwo
 \else \expandafter \@secondoftwo
\fi
}%
\providecommand \enquote [1]{``#1''}%
\providecommand \bibnamefont  [1]{#1}%
\providecommand \bibfnamefont [1]{#1}%
\providecommand \citenamefont [1]{#1}%
\providecommand\href[0]{\@sanitize\@href}%
\providecommand\@href[1]{\endgroup\@@startlink{#1}\endgroup\@@href}%
\providecommand\@@href[1]{#1\@@endlink}%
\providecommand \@sanitize [0]{\begingroup\catcode`\&12\catcode`\#12\relax}%
\@ifxundefined \pdfoutput {\@firstoftwo}{%
 \@ifnum{\z@=\pdfoutput}{\@firstoftwo}{\@secondoftwo}%
}{%
 \providecommand\@@startlink[1]{\leavevmode\special{html:<a href="#1">}}%
 \providecommand\@@endlink[0]{\special{html:</a>}}%
}{%
 \providecommand\@@startlink[1]{%
  \leavevmode
  \pdfstartlink
   attr{/Border[0 0 1 ]/H/I/C[0 1 1]}%
   user{/Subtype/Link/A<</Type/Action/S/URI/URI(#1)>>}%
  \relax
 }%
 \providecommand\@@endlink[0]{\pdfendlink}%
}%
\providecommand \url  [0]{\begingroup\@sanitize \@url }%
\providecommand \@url [1]{\endgroup\@href {#1}{\urlprefix}}%
\providecommand \urlprefix [0]{URL }%
\providecommand \Eprint[0]{\href }%
\@ifxundefined \urlstyle {%
  \providecommand \doi [1]{doi:\discretionary{}{}{}#1}%
}{%
  \providecommand \doi [0]{doi:\discretionary{}{}{}\begingroup
  \urlstyle{rm}\Url }%
}%
\providecommand \doibase [0]{http://dx.doi.org/}%
\providecommand \Doi[1]{\href{\doibase#1}}%
\providecommand \bibAnnote [3]{%
  \BibitemShut{#1}%
  \begin{quotation}\noindent
    \textsc{Key:}\ #2\\\textsc{Annotation:}\ #3%
  \end{quotation}%
}%
\providecommand \bibAnnoteFile [2]{%
  \IfFileExists{#2}{\bibAnnote {#1} {#2} {\input{#2}}}{}%
}%
\providecommand \typeout [0]{\immediate \write \m@ne }%
\providecommand \selectlanguage [0]{\@gobble}%
\providecommand \bibinfo [0]{\@secondoftwo}%
\providecommand \bibfield [0]{\@secondoftwo}%
\providecommand \translation [1]{[#1]}%
\providecommand \BibitemOpen[0]{}%
\providecommand \bibitemStop [0]{}%
\providecommand \bibitemNoStop [0]{.\EOS\space}%
\providecommand \EOS [0]{\spacefactor3000\relax}%
\providecommand \BibitemShut [1]{\csname bibitem#1\endcsname}%
\bibitem{Efetov2010}%
  \BibitemOpen
  \bibfield{author}{%
  \bibinfo {author} {\bibfnamefont{D.~K.}\ \bibnamefont{Efetov}}\ and\ \bibinfo
  {author} {\bibfnamefont{P.}~\bibnamefont{Kim}},\ }%
  \Doi{10.1103/PhysRevLett.105.256805}{\emph{\bibinfo {title} {{Controlling
  Electron-Phonon Interactions in Graphene at Ultrahigh Carrier Densities}}}},\
  \bibfield{journal}{%
  \bibinfo {journal} {Physical Review Letters}\ }%
  \textbf{\bibinfo {volume} {105}},\ \bibinfo {pages} {256805} (\bibinfo {year}
  {2010}).%
  \bibAnnoteFile{Stop}{Efetov2010}%
\bibitem{Radisavljevic2011}%
  \BibitemOpen
  \bibfield{author}{%
  \bibinfo {author} {\bibfnamefont{B.}~\bibnamefont{Radisavljevic}}, \bibinfo
  {author} {\bibfnamefont{A.}~\bibnamefont{Radenovic}}, \bibinfo {author}
  {\bibfnamefont{J.}~\bibnamefont{Brivio}}, \bibinfo {author}
  {\bibfnamefont{V.}~\bibnamefont{Giacometti}},\ and\ \bibinfo {author}
  {\bibfnamefont{A.}~\bibnamefont{Kis}},\ }%
  \Doi{10.1038/nnano.2010.279}{\emph{\bibinfo {title} {{Single-layer MoS2
  transistors}}}},\ \bibfield{journal}{%
  \bibinfo {journal} {Nature Nanotechnology}\ }%
  \textbf{\bibinfo {volume} {6}},\ \bibinfo {pages} {147} (\bibinfo {year}
  {2011}).%
  \bibAnnoteFile{Stop}{Radisavljevic2011}%
\bibitem{Lebegue2013}%
  \BibitemOpen
  \bibfield{author}{%
  \bibinfo {author} {\bibfnamefont{S.}~\bibnamefont{Leb\`egue}}, \bibinfo
  {author} {\bibfnamefont{T.}~\bibnamefont{Bj\"orkman}}, \bibinfo {author}
  {\bibfnamefont{M.}~\bibnamefont{Klintenberg}}, \bibinfo {author}
  {\bibfnamefont{R.~M.}\ \bibnamefont{Nieminen}},\ and\ \bibinfo {author}
  {\bibfnamefont{O.}~\bibnamefont{Eriksson}},\ }%
  \Doi{10.1103/PhysRevX.3.031002}{\emph{\bibinfo {title} {Two-Dimensional
  Materials from Data Filtering and \textit{Ab Initio} Calculations}}},\
  \bibfield{journal}{%
  \bibinfo {journal} {Phys. Rev. X}\ }%
  \textbf{\bibinfo {volume} {3}},\ \bibinfo {pages} {031002} (\bibinfo {year}
  {2013}).%
  \bibAnnoteFile{Stop}{Lebegue2013}%
\bibitem{Hanlon2015}%
  \BibitemOpen
  \bibfield{author}{%
  \bibinfo {author} {\bibfnamefont{D.}~\bibnamefont{Hanlon}}, \bibinfo {author}
  {\bibfnamefont{C.}~\bibnamefont{Backes}}, \bibinfo {author}
  {\bibfnamefont{E.}~\bibnamefont{Doherty}}, \bibinfo {author}
  {\bibfnamefont{C.~S.}\ \bibnamefont{Cucinotta}}, \bibinfo {author}
  {\bibfnamefont{N.~C.}\ \bibnamefont{Berner}}, \bibinfo {author}
  {\bibfnamefont{C.}~\bibnamefont{Boland}}, \bibinfo {author}
  {\bibfnamefont{K.}~\bibnamefont{Lee}}, \bibinfo {author}
  {\bibfnamefont{A.}~\bibnamefont{Harvey}}, \bibinfo {author}
  {\bibfnamefont{P.}~\bibnamefont{Lynch}}, \bibinfo {author}
  {\bibfnamefont{Z.}~\bibnamefont{Gholamvand}}, \bibinfo {author}
  {\bibfnamefont{S.}~\bibnamefont{Zhang}}, \bibinfo {author}
  {\bibfnamefont{K.}~\bibnamefont{Wang}}, \bibinfo {author}
  {\bibfnamefont{G.}~\bibnamefont{Moynihan}}, \bibinfo {author}
  {\bibfnamefont{A.}~\bibnamefont{Pokle}}, \bibinfo {author}
  {\bibfnamefont{Q.~M.}\ \bibnamefont{Ramasse}}, \bibinfo {author}
  {\bibfnamefont{N.}~\bibnamefont{McEvoy}}, \bibinfo {author}
  {\bibfnamefont{W.~J.}\ \bibnamefont{Blau}}, \bibinfo {author}
  {\bibfnamefont{J.}~\bibnamefont{Wang}}, \bibinfo {author}
  {\bibfnamefont{G.}~\bibnamefont{Abellan}}, \bibinfo {author}
  {\bibfnamefont{F.}~\bibnamefont{Hauke}}, \bibinfo {author}
  {\bibfnamefont{A.}~\bibnamefont{Hirsch}}, \bibinfo {author}
  {\bibfnamefont{S.}~\bibnamefont{Sanvito}}, \bibinfo {author}
  {\bibfnamefont{D.~D.}\ \bibnamefont{O'Regan}}, \bibinfo {author}
  {\bibfnamefont{G.~S.}\ \bibnamefont{Duesberg}}, \bibinfo {author}
  {\bibfnamefont{V.}~\bibnamefont{Nicolosi}},\ and\ \bibinfo {author}
  {\bibfnamefont{J.~N.}\ \bibnamefont{Coleman}},\ }%
  \Doi{10.1038/ncomms9563}{\emph{\bibinfo {title} {Liquid exfoliation of
  solvent-stabilized few-layer black phosphorus for applications beyond
  electronics}}},\ \bibfield{journal}{%
  \bibinfo {journal} {Nature Communications}\ }%
  \textbf{\bibinfo {volume} {6}},\ \bibinfo {pages} {8563} (\bibinfo {year}
  {2015}).%
  \bibAnnoteFile{Stop}{Hanlon2015}%
\bibitem{Yasaei2015}%
  \BibitemOpen
  \bibfield{author}{%
  \bibinfo {author} {\bibfnamefont{P.}~\bibnamefont{Yasaei}}, \bibinfo {author}
  {\bibfnamefont{B.}~\bibnamefont{Kumar}}, \bibinfo {author}
  {\bibfnamefont{T.}~\bibnamefont{Foroozan}}, \bibinfo {author}
  {\bibfnamefont{C.}~\bibnamefont{Wang}}, \bibinfo {author}
  {\bibfnamefont{M.}~\bibnamefont{Asadi}}, \bibinfo {author}
  {\bibfnamefont{D.}~\bibnamefont{Tuschel}}, \bibinfo {author}
  {\bibfnamefont{J.~E.}\ \bibnamefont{Indacochea}}, \bibinfo {author}
  {\bibfnamefont{R.~F.}\ \bibnamefont{Klie}},\ and\ \bibinfo {author}
  {\bibfnamefont{A.}~\bibnamefont{Salehi-Khojin}},\ }%
  \Doi{10.1002/adma.201405150}{\emph{\bibinfo {title} {High-Quality Black
  Phosphorus Atomic Layers by Liquid-Phase Exfoliation}}},\
  \bibfield{journal}{%
  \bibinfo {journal} {Advanced Materials}\ }%
  \textbf{\bibinfo {volume} {27}},\ \bibinfo {pages} {1887} (\bibinfo {year}
  {2015}).%
  \bibAnnoteFile{Stop}{Yasaei2015}%
\bibitem{Gibaja2016}%
  \BibitemOpen
  \bibfield{author}{%
  \bibinfo {author} {\bibfnamefont{C.}~\bibnamefont{Gibaja}}, \bibinfo {author}
  {\bibfnamefont{D.}~\bibnamefont{Rodriguez-San-Miguel}}, \bibinfo {author}
  {\bibfnamefont{P.}~\bibnamefont{Ares}}, \bibinfo {author}
  {\bibfnamefont{J.}~\bibnamefont{G{\'o}mez-Herrero}}, \bibinfo {author}
  {\bibfnamefont{M.}~\bibnamefont{Varela}}, \bibinfo {author}
  {\bibfnamefont{R.}~\bibnamefont{Gillen}}, \bibinfo {author}
  {\bibfnamefont{J.}~\bibnamefont{Maultzsch}}, \bibinfo {author}
  {\bibfnamefont{F.}~\bibnamefont{Hauke}}, \bibinfo {author}
  {\bibfnamefont{A.}~\bibnamefont{Hirsch}}, \bibinfo {author}
  {\bibfnamefont{G.}~\bibnamefont{Abell{\'a}n}},\ and\ \bibinfo {author}
  {\bibfnamefont{F.}~\bibnamefont{Zamora}},\ }%
  \Doi{10.1002/anie.201605298}{\emph{\bibinfo {title} {Few-Layer Antimonene by
  Liquid-Phase Exfoliation}}},\ \bibfield{journal}{%
  \bibinfo {journal} {Angewandte Chemie International Edition}\ }%
  \textbf{\bibinfo {volume} {55}},\ \bibinfo {pages} {14345} (\bibinfo {year}
  {2016}).%
  \bibAnnoteFile{Stop}{Gibaja2016}%
\bibitem{Ares2016}%
  \BibitemOpen
  \bibfield{author}{%
  \bibinfo {author} {\bibfnamefont{P.}~\bibnamefont{Ares}}, \bibinfo {author}
  {\bibfnamefont{F.}~\bibnamefont{Aguilar-Galindo}}, \bibinfo {author}
  {\bibfnamefont{D.}~\bibnamefont{Rodr{\'\i}guez-San-Miguel}}, \bibinfo
  {author} {\bibfnamefont{D.~A.}\ \bibnamefont{Aldave}}, \bibinfo {author}
  {\bibfnamefont{S.}~\bibnamefont{D{\'\i}az-Tendero}}, \bibinfo {author}
  {\bibfnamefont{M.}~\bibnamefont{Alcam{\'\i}}}, \bibinfo {author}
  {\bibfnamefont{F.}~\bibnamefont{Mart{\'\i}n}}, \bibinfo {author}
  {\bibfnamefont{J.}~\bibnamefont{G{\'o}mez-Herrero}},\ and\ \bibinfo {author}
  {\bibfnamefont{F.}~\bibnamefont{Zamora}},\ }%
  \Doi{10.1002/adma.201602128}{\emph{\bibinfo {title} {Mechanical Isolation of
  Highly Stable Antimonene under Ambient Conditions}}},\ \bibfield{journal}{%
  \bibinfo {journal} {Advanced Materials}\ }%
  \textbf{\bibinfo {volume} {28}},\ \bibinfo {pages} {6332} (\bibinfo {year}
  {2016}).%
  \bibAnnoteFile{Stop}{Ares2016}%
\bibitem{Bandurin2016}%
  \BibitemOpen
  \bibfield{author}{%
  \bibinfo {author} {\bibfnamefont{D.~A.}\ \bibnamefont{Bandurin}}, \bibinfo
  {author} {\bibfnamefont{A.~V.}\ \bibnamefont{Tyurnina}}, \bibinfo {author}
  {\bibfnamefont{G.~L.}\ \bibnamefont{Yu}}, \bibinfo {author}
  {\bibfnamefont{A.}~\bibnamefont{Mishchenko}}, \bibinfo {author}
  {\bibfnamefont{V.}~\bibnamefont{Z{\'o}lyomi}}, \bibinfo {author}
  {\bibfnamefont{S.~V.}\ \bibnamefont{Morozov}}, \bibinfo {author}
  {\bibfnamefont{R.~K.}\ \bibnamefont{Kumar}}, \bibinfo {author}
  {\bibfnamefont{R.~V.}\ \bibnamefont{Gorbachev}}, \bibinfo {author}
  {\bibfnamefont{Z.~R.}\ \bibnamefont{Kudrynskyi}}, \bibinfo {author}
  {\bibfnamefont{S.}~\bibnamefont{Pezzini}}, \bibinfo {author}
  {\bibfnamefont{Z.~D.}\ \bibnamefont{Kovalyuk}}, \bibinfo {author}
  {\bibfnamefont{U.}~\bibnamefont{Zeitler}}, \bibinfo {author}
  {\bibfnamefont{K.~S.}\ \bibnamefont{Novoselov}}, \bibinfo {author}
  {\bibfnamefont{A.}~\bibnamefont{Patan{\`e}}}, \bibinfo {author}
  {\bibfnamefont{L.}~\bibnamefont{Eaves}}, \bibinfo {author}
  {\bibfnamefont{I.~V.}\ \bibnamefont{Grigorieva}}, \bibinfo {author}
  {\bibfnamefont{V.~I.}\ \bibnamefont{Fal'ko}}, \bibinfo {author}
  {\bibfnamefont{A.~K.}\ \bibnamefont{Geim}},\ and\ \bibinfo {author}
  {\bibfnamefont{Y.}~\bibnamefont{Cao}},\ }%
  \href{http://dx.doi.org/10.1038/nnano.2016.242}{\emph{\bibinfo {title} {High
  electron mobility, quantum Hall effect and anomalous optical response in
  atomically thin InSe}}},\ \bibfield{journal}{%
  \bibinfo {journal} {Nature Nanotechnology}\ }%
  \textbf{\bibinfo {volume} {12}},\ \bibinfo {pages} {223 EP } (\bibinfo {year}
  {2016}).%
  \bibAnnoteFile{Stop}{Bandurin2016}%
\bibitem{Ashton2017}%
  \BibitemOpen
  \bibfield{author}{%
  \bibinfo {author} {\bibfnamefont{M.}~\bibnamefont{Ashton}}, \bibinfo {author}
  {\bibfnamefont{J.}~\bibnamefont{Paul}}, \bibinfo {author}
  {\bibfnamefont{S.~B.}\ \bibnamefont{Sinnott}},\ and\ \bibinfo {author}
  {\bibfnamefont{R.~G.}\ \bibnamefont{Hennig}},\ }%
  \Doi{10.1103/PhysRevLett.118.106101}{\emph{\bibinfo {title} {Topology-Scaling
  Identification of Layered Solids and Stable Exfoliated {2D} Materials}}},\
  \bibfield{journal}{%
  \bibinfo {journal} {Phys. Rev. Lett.}\ }%
  \textbf{\bibinfo {volume} {118}},\ \bibinfo {pages} {106101} (\bibinfo {year}
  {2017}).%
  \bibAnnoteFile{Stop}{Ashton2017}%
\bibitem{Cheon2017}%
  \BibitemOpen
  \bibfield{author}{%
  \bibinfo {author} {\bibfnamefont{G.}~\bibnamefont{Cheon}}, \bibinfo {author}
  {\bibfnamefont{K.-A.~N.}\ \bibnamefont{Duerloo}}, \bibinfo {author}
  {\bibfnamefont{A.~D.}\ \bibnamefont{Sendek}}, \bibinfo {author}
  {\bibfnamefont{C.}~\bibnamefont{Porter}}, \bibinfo {author}
  {\bibfnamefont{Y.}~\bibnamefont{Chen}},\ and\ \bibinfo {author}
  {\bibfnamefont{E.~J.}\ \bibnamefont{Reed}},\ }%
  \Doi{10.1021/acs.nanolett.6b05229}{\emph{\bibinfo {title} {Data Mining for
  New Two- and One-Dimensional Weakly Bonded Solids and Lattice-Commensurate
  Heterostructures}}},\ \bibfield{journal}{%
  \bibinfo {journal} {Nano Letters}\ }%
  \textbf{\bibinfo {volume} {17}},\ \bibinfo {pages} {1915} (\bibinfo {year}
  {2017}).%
  \bibAnnoteFile{Stop}{Cheon2017}%
\bibitem{Mounet2018}%
  \BibitemOpen
  \bibfield{author}{%
  \bibinfo {author} {\bibfnamefont{N.}~\bibnamefont{Mounet}}, \bibinfo {author}
  {\bibfnamefont{M.}~\bibnamefont{Gibertini}}, \bibinfo {author}
  {\bibfnamefont{P.}~\bibnamefont{Schwaller}}, \bibinfo {author}
  {\bibfnamefont{D.}~\bibnamefont{Campi}}, \bibinfo {author}
  {\bibfnamefont{A.}~\bibnamefont{Merkys}}, \bibinfo {author}
  {\bibfnamefont{A.}~\bibnamefont{Marrazzo}}, \bibinfo {author}
  {\bibnamefont{Thib}}, \bibinfo {author} {\bibfnamefont{I.~E.}\
  \bibnamefont{Castelli}}, \bibinfo {author}
  {\bibfnamefont{A.}~\bibnamefont{Cepellotti}}, \bibinfo {author}
  {\bibfnamefont{G.}~\bibnamefont{Pizzi}},\ and\ \bibinfo {author}
  {\bibfnamefont{N.}~\bibnamefont{Marzari}},\ }%
  \href{https://www.nature.com/articles/s41565-017-0035-5}{\emph{\bibinfo
  {title} {Two-dimensional materials from high-throughput computational
  exfoliation of experimentally known compounds}}},\ \bibfield{journal}{%
  \bibinfo {journal} {Nature Nanotechnology}\ }%
  \textbf{\bibinfo {volume} {13}},\ \bibinfo {pages} {246} (\bibinfo {year}
  {2018}).%
  \bibAnnoteFile{Stop}{Mounet2018}%
\bibitem{Wang2012}%
  \BibitemOpen
  \bibfield{author}{%
  \bibinfo {author} {\bibfnamefont{Q.~H.}\ \bibnamefont{Wang}}, \bibinfo
  {author} {\bibfnamefont{K.}~\bibnamefont{Kalantar-Zadeh}}, \bibinfo {author}
  {\bibfnamefont{A.}~\bibnamefont{Kis}}, \bibinfo {author}
  {\bibfnamefont{J.~N.}\ \bibnamefont{Coleman}},\ and\ \bibinfo {author}
  {\bibfnamefont{M.~S.}\ \bibnamefont{Strano}},\ }%
  \Doi{10.1038/nnano.2012.193}{\emph{\bibinfo {title} {Electronics and
  optoelectronics of two-dimensional transition metal dichalcogenides}}},\
  \bibfield{journal}{%
  \bibinfo {journal} {Nat. Nanotech.}\ }%
  \textbf{\bibinfo {volume} {7}},\ \bibinfo {pages} {699} (\bibinfo {year}
  {2012}).%
  \bibAnnoteFile{Stop}{Wang2012}%
\bibitem{Chhowalla2016}%
  \BibitemOpen
  \bibfield{author}{%
  \bibinfo {author} {\bibfnamefont{M.}~\bibnamefont{Chhowalla}}, \bibinfo
  {author} {\bibfnamefont{D.}~\bibnamefont{Jena}},\ and\ \bibinfo {author}
  {\bibfnamefont{H.}~\bibnamefont{Zhang}},\ }%
  \href{http://dx.doi.org/10.1038/natrevmats.2016.52}{\emph{\bibinfo {title}
  {Two-dimensional semiconductors for transistors}}},\ \bibfield{journal}{%
  \bibinfo {journal} {Nature Reviews Materials}\ }%
  \textbf{\bibinfo {volume} {1}},\ \bibinfo {pages} {16052} (\bibinfo {year}
  {2016}).%
  \bibAnnoteFile{Stop}{Chhowalla2016}%
\bibitem{Ziman}%
  \BibitemOpen
  \bibfield{author}{%
  \bibinfo {author} {\bibfnamefont{J.~M.}\ \bibnamefont{Ziman}},\ }%
  \emph{\bibinfo {title} {Electrons and phonons: the theory of transport
  phenomena in solids}}\ (\bibinfo {publisher} {Clarendon},\ \bibinfo {address}
  {Oxford},\ \bibinfo {year} {1960}).%
  \bibAnnoteFile{Stop}{Ziman}%
\bibitem{Allen1978}%
  \BibitemOpen
  \bibfield{author}{%
  \bibinfo {author} {\bibfnamefont{P.~B.}\ \bibnamefont{Allen}},\ }%
  \href{http://prb.aps.org/abstract/PRB/v17/i10/p3725{\_}1}{\emph{\bibinfo
  {title} {{New Method for Solving Boltzmann's equation for electrons in
  metals}}}},\ \bibfield{journal}{%
  \bibinfo {journal} {Physical Review B}\ }%
  \textbf{\bibinfo {volume} {17}},\ \bibinfo {pages} {3725} (\bibinfo {year}
  {1978}).%
  \bibAnnoteFile{Stop}{Allen1978}%
\bibitem{Grimvall}%
  \BibitemOpen
  \bibfield{author}{%
  \bibinfo {author} {\bibfnamefont{G.}~\bibnamefont{Grimvall}},\ }%
  \emph{\bibinfo {title} {The electron-phonon interaction in metals}}\
  (\bibinfo {publisher} {North-Holland},\ \bibinfo {address} {Amsterdam},\
  \bibinfo {year} {1981}).%
  \bibAnnoteFile{Stop}{Grimvall}%
\bibitem{Giustino2017}%
  \BibitemOpen
  \bibfield{author}{%
  \bibinfo {author} {\bibfnamefont{F.}~\bibnamefont{Giustino}},\ }%
  \Doi{10.1103/RevModPhys.89.015003}{\emph{\bibinfo {title} {{Electron-phonon
  interactions from first principles}}}},\ \bibfield{journal}{%
  \bibinfo {journal} {Reviews of Modern Physics}\ }%
  \textbf{\bibinfo {volume} {89}},\ \bibinfo {pages} {015003} (\bibinfo {year}
  {2017}).%
  \bibAnnoteFile{Stop}{Giustino2017}%
\bibitem{Savrasov1994}%
  \BibitemOpen
  \bibfield{author}{%
  \bibinfo {author} {\bibfnamefont{S.~Y.}\ \bibnamefont{Savrasov}}, \bibinfo
  {author} {\bibfnamefont{D.~Y.}\ \bibnamefont{Savrasov}},\ and\ \bibinfo
  {author} {\bibfnamefont{O.~K.}\ \bibnamefont{Andersen}},\ }%
  \Doi{10.1103/PhysRevLett.72.372}{\emph{\bibinfo {title} {Linear-response
  calculations of electron-phonon interactions}}},\ \bibfield{journal}{%
  \bibinfo {journal} {Phys. Rev. Lett.}\ }%
  \textbf{\bibinfo {volume} {72}},\ \bibinfo {pages} {372} (\bibinfo {year}
  {1994}).%
  \bibAnnoteFile{Stop}{Savrasov1994}%
\bibitem{Liu1996}%
  \BibitemOpen
  \bibfield{author}{%
  \bibinfo {author} {\bibfnamefont{A.~Y.}\ \bibnamefont{Liu}}\ and\ \bibinfo
  {author} {\bibfnamefont{A.~A.}\ \bibnamefont{Quong}},\ }%
  \Doi{10.1103/PhysRevB.53.R7575}{\emph{\bibinfo {title} {Linear-response
  calculation of electron-phonon coupling parameters}}},\ \bibfield{journal}{%
  \bibinfo {journal} {Phys. Rev. B}\ }%
  \textbf{\bibinfo {volume} {53}},\ \bibinfo {pages} {R7575} (\bibinfo {year}
  {1996}).%
  \bibAnnoteFile{Stop}{Liu1996}%
\bibitem{Mauri1996}%
  \BibitemOpen
  \bibfield{author}{%
  \bibinfo {author} {\bibfnamefont{F.}~\bibnamefont{Mauri}}, \bibinfo {author}
  {\bibfnamefont{O.}~\bibnamefont{Zakharov}}, \bibinfo {author}
  {\bibfnamefont{S.}~\bibnamefont{de~Gironcoli}}, \bibinfo {author}
  {\bibfnamefont{S.~G.}\ \bibnamefont{Louie}},\ and\ \bibinfo {author}
  {\bibfnamefont{M.~L.}\ \bibnamefont{Cohen}},\ }%
  \Doi{10.1103/PhysRevLett.77.1151}{\emph{\bibinfo {title} {Phonon Softening
  and Superconductivity in Tellurium under Pressure}}},\ \bibfield{journal}{%
  \bibinfo {journal} {Phys. Rev. Lett.}\ }%
  \textbf{\bibinfo {volume} {77}},\ \bibinfo {pages} {1151} (\bibinfo {year}
  {1996}).%
  \bibAnnoteFile{Stop}{Mauri1996}%
\bibitem{Bauer1998}%
  \BibitemOpen
  \bibfield{author}{%
  \bibinfo {author} {\bibfnamefont{R.}~\bibnamefont{Bauer}}, \bibinfo {author}
  {\bibfnamefont{A.}~\bibnamefont{Schmid}}, \bibinfo {author}
  {\bibfnamefont{P.}~\bibnamefont{Pavone}},\ and\ \bibinfo {author}
  {\bibfnamefont{D.}~\bibnamefont{Strauch}},\ }%
  \Doi{10.1103/PhysRevB.57.11276}{\emph{\bibinfo {title} {{Electron-phonon
  coupling in the metallic elements Al, Au, Na, and Nb: A first-principles
  study}}}},\ \bibfield{journal}{%
  \bibinfo {journal} {Phys. Rev. B}\ }%
  \textbf{\bibinfo {volume} {57}},\ \bibinfo {pages} {11276} (\bibinfo {year}
  {1998}).%
  \bibAnnoteFile{Stop}{Bauer1998}%
\bibitem{Baroni2001}%
  \BibitemOpen
  \bibfield{author}{%
  \bibinfo {author} {\bibfnamefont{S.}~\bibnamefont{Baroni}}, \bibinfo {author}
  {\bibfnamefont{S.}~\bibnamefont{de~Gironcoli}}, \bibinfo {author}
  {\bibfnamefont{A.}~\bibnamefont{Dal~Corso}},\ and\ \bibinfo {author}
  {\bibfnamefont{P.}~\bibnamefont{Giannozzi}},\ }%
  \Doi{10.1103/RevModPhys.73.515}{\emph{\bibinfo {title} {Phonons and related
  crystal properties from density-functional perturbation theory}}},\
  \bibfield{journal}{%
  \bibinfo {journal} {Rev. Mod. Phys.}\ }%
  \textbf{\bibinfo {volume} {73}},\ \bibinfo {pages} {515} (\bibinfo {year}
  {2001}).%
  \bibAnnoteFile{Stop}{Baroni2001}%
\bibitem{Dacorogna1985}%
  \BibitemOpen
  \bibfield{author}{%
  \bibinfo {author} {\bibfnamefont{M.~M.}\ \bibnamefont{Dacorogna}}, \bibinfo
  {author} {\bibfnamefont{M.~L.}\ \bibnamefont{Cohen}},\ and\ \bibinfo {author}
  {\bibfnamefont{P.~K.}\ \bibnamefont{Lam}},\ }%
  \Doi{10.1103/PhysRevLett.55.837}{\emph{\bibinfo {title} {Self-Consistent
  Calculation of the q Dependence of the Electron-Phonon Coupling in
  Aluminum}}},\ \bibfield{journal}{%
  \bibinfo {journal} {Phys. Rev. Lett.}\ }%
  \textbf{\bibinfo {volume} {55}},\ \bibinfo {pages} {837} (\bibinfo {year}
  {1985}).%
  \bibAnnoteFile{Stop}{Dacorogna1985}%
\bibitem{Gunst2016a}%
  \BibitemOpen
  \bibfield{author}{%
  \bibinfo {author} {\bibfnamefont{T.}~\bibnamefont{Gunst}}, \bibinfo {author}
  {\bibfnamefont{T.}~\bibnamefont{Markussen}}, \bibinfo {author}
  {\bibfnamefont{K.}~\bibnamefont{Stokbro}},\ and\ \bibinfo {author}
  {\bibfnamefont{M.}~\bibnamefont{Brandbyge}},\ }%
  \Doi{10.1103/PhysRevB.93.035414}{\emph{\bibinfo {title} {{First-principles
  method for electron-phonon coupling and electron mobility: Applications to
  two-dimensional materials}}}},\ \bibfield{journal}{%
  \bibinfo {journal} {Physical Review B}\ }%
  \textbf{\bibinfo {volume} {93}},\ \bibinfo {pages} {035414} (\bibinfo {year}
  {2016}).%
  \bibAnnoteFile{Stop}{Gunst2016a}%
\bibitem{Sohier2016}%
  \BibitemOpen
  \bibfield{author}{%
  \bibinfo {author} {\bibfnamefont{T.}~\bibnamefont{Sohier}}, \bibinfo {author}
  {\bibfnamefont{M.}~\bibnamefont{Calandra}},\ and\ \bibinfo {author}
  {\bibfnamefont{F.}~\bibnamefont{Mauri}},\ }%
  \Doi{10.1103/PhysRevB.94.085415}{\emph{\bibinfo {title} {{Two-dimensional
  Fr{\"{o}}hlich interaction in transition-metal dichalcogenide monolayers:
  Theoretical modeling and first-principles calculations}}}},\
  \bibfield{journal}{%
  \bibinfo {journal} {Physical Review B}\ }%
  \textbf{\bibinfo {volume} {94}},\ \bibinfo {pages} {085415} (\bibinfo {year}
  {2016}).%
  \bibAnnoteFile{Stop}{Sohier2016}%
\bibitem{Kozinsky2006}%
  \BibitemOpen
  \bibfield{author}{%
  \bibinfo {author} {\bibfnamefont{B.}~\bibnamefont{Kozinsky}}\ and\ \bibinfo
  {author} {\bibfnamefont{N.}~\bibnamefont{Marzari}},\ }%
  \Doi{10.1103/PhysRevLett.96.166801}{\emph{\bibinfo {title} {{Static
  Dielectric Properties of Carbon Nanotubes from First Principles}}}},\
  \bibfield{journal}{%
  \bibinfo {journal} {Physical Review Letters}\ }%
  \textbf{\bibinfo {volume} {96}},\ \bibinfo {pages} {166801} (\bibinfo {year}
  {2006}).%
  \bibAnnoteFile{Stop}{Kozinsky2006}%
\bibitem{Sohier2015}%
  \BibitemOpen
  \bibfield{author}{%
  \bibinfo {author} {\bibfnamefont{T.}~\bibnamefont{Sohier}}, \bibinfo {author}
  {\bibfnamefont{M.}~\bibnamefont{Calandra}},\ and\ \bibinfo {author}
  {\bibfnamefont{F.}~\bibnamefont{Mauri}},\ }%
  \Doi{10.1103/PhysRevB.91.165428}{\emph{\bibinfo {title} {Density-functional
  calculation of static screening in two-dimensional materials: The
  long-wavelength dielectric function of graphene}}},\ \bibfield{journal}{%
  \bibinfo {journal} {Phys. Rev. B}\ }%
  \textbf{\bibinfo {volume} {91}},\ \bibinfo {pages} {165428} (\bibinfo {year}
  {2015}).%
  \bibAnnoteFile{Stop}{Sohier2015}%
\bibitem{Sohier2017nl}%
  \BibitemOpen
  \bibfield{author}{%
  \bibinfo {author} {\bibfnamefont{T.}~\bibnamefont{Sohier}}, \bibinfo {author}
  {\bibfnamefont{M.}~\bibnamefont{Gibertini}}, \bibinfo {author}
  {\bibfnamefont{M.}~\bibnamefont{Calandra}}, \bibinfo {author}
  {\bibfnamefont{F.}~\bibnamefont{Mauri}},\ and\ \bibinfo {author}
  {\bibfnamefont{N.}~\bibnamefont{Marzari}},\ }%
  \Doi{10.1021/acs.nanolett.7b01090}{\emph{\bibinfo {title} {Breakdown of
  Optical Phonons' Splitting in Two-Dimensional Materials}}},\
  \bibfield{journal}{%
  \bibinfo {journal} {Nano Letters}\ }%
  \textbf{\bibinfo {volume} {17}},\ \bibinfo {pages} {3758} (\bibinfo {year}
  {2017}).%
  \bibAnnoteFile{Stop}{Sohier2017nl}%
\bibitem{Sohier2017}%
  \BibitemOpen
  \bibfield{author}{%
  \bibinfo {author} {\bibfnamefont{T.}~\bibnamefont{Sohier}}, \bibinfo {author}
  {\bibfnamefont{M.}~\bibnamefont{Calandra}},\ and\ \bibinfo {author}
  {\bibfnamefont{F.}~\bibnamefont{Mauri}},\ }%
  \Doi{10.1103/PhysRevB.96.075448}{\emph{\bibinfo {title} {{Density functional
  perturbation theory for gated two-dimensional heterostructures: Theoretical
  developments and application to flexural phonons in graphene}}}},\
  \bibfield{journal}{%
  \bibinfo {journal} {Physical Review B}\ }%
  \textbf{\bibinfo {volume} {96}},\ \bibinfo {pages} {075448} (\bibinfo {year}
  {2017}).%
  \bibAnnoteFile{Stop}{Sohier2017}%
\bibitem{Lee2005}%
  \BibitemOpen
  \bibfield{author}{%
  \bibinfo {author} {\bibfnamefont{Y.~S.}\ \bibnamefont{Lee}}, \bibinfo
  {author} {\bibfnamefont{M.}~\bibnamefont{{Buongiorno Nardelli}}},\ and\
  \bibinfo {author} {\bibfnamefont{N.}~\bibnamefont{Marzari}},\ }%
  \Doi{10.1103/PhysRevLett.95.076804}{\emph{\bibinfo {title} {{Band structure
  and quantum conductance of nanostructures from maximally localized Wannier
  functions: The case of functionalized carbon nanotubes}}}},\
  \bibfield{journal}{%
  \bibinfo {journal} {Physical Review Letters}\ }%
  \textbf{\bibinfo {volume} {95}},\ \bibinfo {pages} {076804} (\bibinfo {year}
  {2005}).%
  \bibAnnoteFile{Stop}{Lee2005}%
\bibitem{Yates2007}%
  \BibitemOpen
  \bibfield{author}{%
  \bibinfo {author} {\bibfnamefont{J.~R.}\ \bibnamefont{Yates}}, \bibinfo
  {author} {\bibfnamefont{X.}~\bibnamefont{Wang}}, \bibinfo {author}
  {\bibfnamefont{D.}~\bibnamefont{Vanderbilt}},\ and\ \bibinfo {author}
  {\bibfnamefont{I.}~\bibnamefont{Souza}},\ }%
  \Doi{10.1103/PhysRevB.75.195121}{\emph{\bibinfo {title} {{Spectral and Fermi
  surface properties from Wannier interpolation}}}},\ \bibfield{journal}{%
  \bibinfo {journal} {Physical Review B}\ }%
  \textbf{\bibinfo {volume} {75}},\ \bibinfo {pages} {195121} (\bibinfo {year}
  {2007}).%
  \bibAnnoteFile{Stop}{Yates2007}%
\bibitem{Giustino2007}%
  \BibitemOpen
  \bibfield{author}{%
  \bibinfo {author} {\bibfnamefont{F.}~\bibnamefont{Giustino}}, \bibinfo
  {author} {\bibfnamefont{M.~L.}\ \bibnamefont{Cohen}},\ and\ \bibinfo {author}
  {\bibfnamefont{S.~G.}\ \bibnamefont{Louie}},\ }%
  \Doi{10.1103/PhysRevB.76.165108}{\emph{\bibinfo {title} {{Electron-phonon
  interaction using Wannier functions}}}},\ \bibfield{journal}{%
  \bibinfo {journal} {Phys. Rev. B}\ }%
  \textbf{\bibinfo {volume} {76}},\ \bibinfo {pages} {165108} (\bibinfo {year}
  {2007}).%
  \bibAnnoteFile{Stop}{Giustino2007}%
\bibitem{Calandra2010}%
  \BibitemOpen
  \bibfield{author}{%
  \bibinfo {author} {\bibfnamefont{M.}~\bibnamefont{Calandra}}, \bibinfo
  {author} {\bibfnamefont{G.}~\bibnamefont{Profeta}},\ and\ \bibinfo {author}
  {\bibfnamefont{F.}~\bibnamefont{Mauri}},\ }%
  \Doi{10.1103/PhysRevB.82.165111}{\emph{\bibinfo {title} {Adiabatic and
  nonadiabatic phonon dispersion in a Wannier function approach}}},\
  \bibfield{journal}{%
  \bibinfo {journal} {Phys. Rev. B}\ }%
  \textbf{\bibinfo {volume} {82}},\ \bibinfo {pages} {165111} (\bibinfo {year}
  {2010}).%
  \bibAnnoteFile{Stop}{Calandra2010}%
\bibitem{Marzari2012}%
  \BibitemOpen
  \bibfield{author}{%
  \bibinfo {author} {\bibfnamefont{N.}~\bibnamefont{Marzari}}, \bibinfo
  {author} {\bibfnamefont{A.~A.}\ \bibnamefont{Mostofi}}, \bibinfo {author}
  {\bibfnamefont{J.~R.}\ \bibnamefont{Yates}}, \bibinfo {author}
  {\bibfnamefont{I.}~\bibnamefont{Souza}},\ and\ \bibinfo {author}
  {\bibfnamefont{D.}~\bibnamefont{Vanderbilt}},\ }%
  \Doi{10.1103/RevModPhys.84.1419}{\emph{\bibinfo {title} {Maximally localized
  Wannier functions: Theory and applications}}},\ \bibfield{journal}{%
  \bibinfo {journal} {Rev. Mod. Phys.}\ }%
  \textbf{\bibinfo {volume} {84}},\ \bibinfo {pages} {1419} (\bibinfo {year}
  {2012}).%
  \bibAnnoteFile{Stop}{Marzari2012}%
\bibitem{Ponce2016}%
  \BibitemOpen
  \bibfield{author}{%
  \bibinfo {author} {\bibfnamefont{S.}~\bibnamefont{Ponc{\'{e}}}}, \bibinfo
  {author} {\bibfnamefont{E.}~\bibnamefont{Margine}}, \bibinfo {author}
  {\bibfnamefont{C.}~\bibnamefont{Verdi}},\ and\ \bibinfo {author}
  {\bibfnamefont{F.}~\bibnamefont{Giustino}},\ }%
  \Doi{10.1016/J.CPC.2016.07.028}{\emph{\bibinfo {title} {{EPW:
  Electron--phonon coupling, transport and superconducting properties using
  maximally localized Wannier functions}}}},\ \bibfield{journal}{%
  \bibinfo {journal} {Computer Physics Communications}\ }%
  \textbf{\bibinfo {volume} {209}},\ \bibinfo {pages} {116} (\bibinfo {year}
  {2016}).%
  \bibAnnoteFile{Stop}{Ponce2016}%
\bibitem{Sjakste2015}%
  \BibitemOpen
  \bibfield{author}{%
  \bibinfo {author} {\bibfnamefont{J.}~\bibnamefont{Sjakste}}, \bibinfo
  {author} {\bibfnamefont{N.}~\bibnamefont{Vast}}, \bibinfo {author}
  {\bibfnamefont{M.}~\bibnamefont{Calandra}},\ and\ \bibinfo {author}
  {\bibfnamefont{F.}~\bibnamefont{Mauri}},\ }%
  \Doi{10.1103/PhysRevB.92.054307}{\emph{\bibinfo {title} {Wannier
  interpolation of the electron-phonon matrix elements in polar semiconductors:
  Polar-optical coupling in GaAs}}},\ \bibfield{journal}{%
  \bibinfo {journal} {Phys. Rev. B}\ }%
  \textbf{\bibinfo {volume} {92}},\ \bibinfo {pages} {054307} (\bibinfo {year}
  {2015}).%
  \bibAnnoteFile{Stop}{Sjakste2015}%
\bibitem{Verdi2015}%
  \BibitemOpen
  \bibfield{author}{%
  \bibinfo {author} {\bibfnamefont{C.}~\bibnamefont{Verdi}}\ and\ \bibinfo
  {author} {\bibfnamefont{F.}~\bibnamefont{Giustino}},\ }%
  \Doi{10.1103/PhysRevLett.115.176401}{\emph{\bibinfo {title} {Fr\"ohlich
  Electron-Phonon Vertex from First Principles}}},\ \bibfield{journal}{%
  \bibinfo {journal} {Phys. Rev. Lett.}\ }%
  \textbf{\bibinfo {volume} {115}},\ \bibinfo {pages} {176401} (\bibinfo {year}
  {2015}).%
  \bibAnnoteFile{Stop}{Verdi2015}%
\bibitem{Savrasov1996}%
  \BibitemOpen
  \bibfield{author}{%
  \bibinfo {author} {\bibfnamefont{S.~Y.}\ \bibnamefont{Savrasov}}\ and\
  \bibinfo {author} {\bibfnamefont{D.~Y.}\ \bibnamefont{Savrasov}},\ }%
  \Doi{10.1103/PhysRevB.54.16487}{\emph{\bibinfo {title} {Electron-phonon
  interactions and related physical properties of metals from linear-response
  theory}}},\ \bibfield{journal}{%
  \bibinfo {journal} {Phys. Rev. B}\ }%
  \textbf{\bibinfo {volume} {54}},\ \bibinfo {pages} {16487} (\bibinfo {year}
  {1996}).%
  \bibAnnoteFile{Stop}{Savrasov1996}%
\bibitem{Restrepo2009}%
  \BibitemOpen
  \bibfield{author}{%
  \bibinfo {author} {\bibfnamefont{O.~D.}\ \bibnamefont{Restrepo}}, \bibinfo
  {author} {\bibfnamefont{K.}~\bibnamefont{Varga}},\ and\ \bibinfo {author}
  {\bibfnamefont{S.~T.}\ \bibnamefont{Pantelides}},\ }%
  \Doi{10.1063/1.3147189}{\emph{\bibinfo {title} {First-principles calculations
  of electron mobilities in silicon: Phonon and Coulomb scattering}}},\
  \bibfield{journal}{%
  \bibinfo {journal} {Applied Physics Letters}\ }%
  \textbf{\bibinfo {volume} {94}},\ \bibinfo {pages} {212103} (\bibinfo {year}
  {2009}).%
  \bibAnnoteFile{Stop}{Restrepo2009}%
\bibitem{Shishir2009}%
  \BibitemOpen
  \bibfield{author}{%
  \bibinfo {author} {\bibfnamefont{R.~S.}\ \bibnamefont{Shishir}}\ and\
  \bibinfo {author} {\bibfnamefont{D.~K.}\ \bibnamefont{Ferry}},\ }%
  \Doi{10.1088/0953-8984/21/23/232204}{\emph{\bibinfo {title} {{Intrinsic
  mobility in graphene.}}}},\ \bibfield{journal}{%
  \bibinfo {journal} {Journal of physics. Condensed matter : an Institute of
  Physics journal}\ }%
  \textbf{\bibinfo {volume} {21}},\ \bibinfo {pages} {232204} (\bibinfo {year}
  {2009}).%
  \bibAnnoteFile{Stop}{Shishir2009}%
\bibitem{Borysenko2010}%
  \BibitemOpen
  \bibfield{author}{%
  \bibinfo {author} {\bibfnamefont{K.~M.}\ \bibnamefont{Borysenko}}, \bibinfo
  {author} {\bibfnamefont{J.~T.}\ \bibnamefont{Mullen}}, \bibinfo {author}
  {\bibfnamefont{E.~A.}\ \bibnamefont{Barry}}, \bibinfo {author}
  {\bibfnamefont{S.}~\bibnamefont{Paul}}, \bibinfo {author}
  {\bibfnamefont{Y.~G.}\ \bibnamefont{Semenov}}, \bibinfo {author}
  {\bibfnamefont{J.~M.}\ \bibnamefont{Zavada}}, \bibinfo {author}
  {\bibfnamefont{M.}~\bibnamefont{{Buongiorno Nardelli}}},\ and\ \bibinfo
  {author} {\bibfnamefont{K.~W.}\ \bibnamefont{Kim}},\ }%
  \Doi{10.1103/PhysRevB.81.121412}{\emph{\bibinfo {title} {{First-principles
  analysis of electron-phonon interactions in graphene}}}},\
  \bibfield{journal}{%
  \bibinfo {journal} {Physical Review B}\ }%
  \textbf{\bibinfo {volume} {81}},\ \bibinfo {pages} {121412} (\bibinfo {year}
  {2010}).%
  \bibAnnoteFile{Stop}{Borysenko2010}%
\bibitem{Jin2014}%
  \BibitemOpen
  \bibfield{author}{%
  \bibinfo {author} {\bibfnamefont{Z.}~\bibnamefont{Jin}}, \bibinfo {author}
  {\bibfnamefont{X.}~\bibnamefont{Li}}, \bibinfo {author}
  {\bibfnamefont{J.~T.}\ \bibnamefont{Mullen}},\ and\ \bibinfo {author}
  {\bibfnamefont{K.~W.}\ \bibnamefont{Kim}},\ }%
  \Doi{10.1103/PhysRevB.90.045422}{\emph{\bibinfo {title} {{Intrinsic transport
  properties of electrons and holes in monolayer transition-metal
  dichalcogenides}}}},\ \bibfield{journal}{%
  \bibinfo {journal} {Physical Review B}\ }%
  \textbf{\bibinfo {volume} {90}},\ \bibinfo {pages} {045422} (\bibinfo {year}
  {2014}).%
  \bibAnnoteFile{Stop}{Jin2014}%
\bibitem{Restrepo2014}%
  \BibitemOpen
  \bibfield{author}{%
  \bibinfo {author} {\bibfnamefont{O.~D.}\ \bibnamefont{Restrepo}}, \bibinfo
  {author} {\bibfnamefont{K.~E.}\ \bibnamefont{Krymowski}}, \bibinfo {author}
  {\bibfnamefont{J.}~\bibnamefont{Goldberger}},\ and\ \bibinfo {author}
  {\bibfnamefont{W.}~\bibnamefont{Windl}},\ }%
  \Doi{10.1088/1367-2630/16/10/105009}{\emph{\bibinfo {title} {{A first
  principles method to simulate electron mobilities in 2D materials}}}},\
  \bibfield{journal}{%
  \bibinfo {journal} {New Journal of Physics}\ }%
  \textbf{\bibinfo {volume} {16}} (\bibinfo {year} {2014}).%
  \bibAnnoteFile{Stop}{Restrepo2014}%
\bibitem{Park2014}%
  \BibitemOpen
  \bibfield{author}{%
  \bibinfo {author} {\bibfnamefont{C.-H.}\ \bibnamefont{Park}}, \bibinfo
  {author} {\bibfnamefont{N.}~\bibnamefont{Bonini}}, \bibinfo {author}
  {\bibfnamefont{T.}~\bibnamefont{Sohier}}, \bibinfo {author}
  {\bibfnamefont{G.}~\bibnamefont{Samsonidze}}, \bibinfo {author}
  {\bibfnamefont{B.}~\bibnamefont{Kozinsky}}, \bibinfo {author}
  {\bibfnamefont{M.}~\bibnamefont{Calandra}}, \bibinfo {author}
  {\bibfnamefont{F.}~\bibnamefont{Mauri}},\ and\ \bibinfo {author}
  {\bibfnamefont{N.}~\bibnamefont{Marzari}},\ }%
  \Doi{10.1021/nl402696q}{\emph{\bibinfo {title} {{Electron-Phonon Interactions
  and the Intrinsic Electrical Resistivity of Graphene.}}}},\
  \bibfield{journal}{%
  \bibinfo {journal} {Nano letters}\ }%
  \textbf{\bibinfo {volume} {14}},\ \bibinfo {pages} {1113} (\bibinfo {year}
  {2014}).%
  \bibAnnoteFile{Stop}{Park2014}%
\bibitem{Jin2016}%
  \BibitemOpen
  \bibfield{author}{%
  \bibinfo {author} {\bibfnamefont{Z.}~\bibnamefont{Jin}}, \bibinfo {author}
  {\bibfnamefont{J.~T.}\ \bibnamefont{Mullen}},\ and\ \bibinfo {author}
  {\bibfnamefont{K.~W.}\ \bibnamefont{Kim}},\ }%
  \Doi{10.1063/1.4960526}{\emph{\bibinfo {title} {{Highly anisotropic
  electronic transport properties of monolayer and bilayer phosphorene from
  first principles}}}},\ \bibfield{journal}{%
  \bibinfo {journal} {Applied Physics Letters}\ }%
  \textbf{\bibinfo {volume} {109}},\ \bibinfo {pages} {053108} (\bibinfo {year}
  {2016}).%
  \bibAnnoteFile{Stop}{Jin2016}%
\bibitem{Rudenko2016}%
  \BibitemOpen
  \bibfield{author}{%
  \bibinfo {author} {\bibfnamefont{A.~N.}\ \bibnamefont{Rudenko}}, \bibinfo
  {author} {\bibfnamefont{S.}~\bibnamefont{Brener}},\ and\ \bibinfo {author}
  {\bibfnamefont{M.~I.}\ \bibnamefont{Katsnelson}},\ }%
  \Doi{10.1103/PhysRevLett.116.246401}{\emph{\bibinfo {title} {{Intrinsic
  Charge Carrier Mobility in Single-Layer Black Phosphorus}}}},\
  \bibfield{journal}{%
  \bibinfo {journal} {Physical Review Letters}\ }%
  \textbf{\bibinfo {volume} {116}},\ \bibinfo {pages} {246401} (\bibinfo {year}
  {2016}).%
  \bibAnnoteFile{Stop}{Rudenko2016}%
\bibitem{Trushkov2017}%
  \BibitemOpen
  \bibfield{author}{%
  \bibinfo {author} {\bibfnamefont{Y.}~\bibnamefont{Trushkov}}\ and\ \bibinfo
  {author} {\bibfnamefont{V.}~\bibnamefont{Perebeinos}},\ }%
  \Doi{10.1103/PhysRevB.95.075436}{\emph{\bibinfo {title} {{Phonon-limited
  carrier mobility in monolayer black phosphorus}}}},\ \bibfield{journal}{%
  \bibinfo {journal} {Physical Review B}\ }%
  \textbf{\bibinfo {volume} {95}},\ \bibinfo {pages} {075436} (\bibinfo {year}
  {2017}).%
  \bibAnnoteFile{Stop}{Trushkov2017}%
\bibitem{Gaddemane2018}%
  \BibitemOpen
  \bibfield{author}{%
  \bibinfo {author} {\bibfnamefont{G.}~\bibnamefont{Gaddemane}}, \bibinfo
  {author} {\bibfnamefont{W.~G.}\ \bibnamefont{Vandenberghe}}, \bibinfo
  {author} {\bibfnamefont{M.~L.}\ \bibnamefont{Van~de Put}}, \bibinfo {author}
  {\bibfnamefont{S.}~\bibnamefont{Chen}}, \bibinfo {author}
  {\bibfnamefont{S.}~\bibnamefont{Tiwari}}, \bibinfo {author}
  {\bibfnamefont{E.}~\bibnamefont{Chen}},\ and\ \bibinfo {author}
  {\bibfnamefont{M.~V.}\ \bibnamefont{Fischetti}},\ }%
  \Doi{10.1103/PhysRevB.98.115416}{\emph{\bibinfo {title} {Theoretical studies
  of electronic transport in monolayer and bilayer phosphorene: A critical
  overview}}},\ \bibfield{journal}{%
  \bibinfo {journal} {Phys. Rev. B}\ }%
  \textbf{\bibinfo {volume} {98}},\ \bibinfo {pages} {115416} (\bibinfo {year}
  {2018}).%
  \bibAnnoteFile{Stop}{Gaddemane2018}%
\bibitem{Ponce2018}%
  \BibitemOpen
  \bibfield{author}{%
  \bibinfo {author} {\bibfnamefont{S.}~\bibnamefont{Ponc{\'{e}}}}, \bibinfo
  {author} {\bibfnamefont{E.~R.}\ \bibnamefont{Margine}},\ and\ \bibinfo
  {author} {\bibfnamefont{F.}~\bibnamefont{Giustino}},\ }%
  \Doi{10.1103/PhysRevB.97.121201}{\emph{\bibinfo {title} {{Towards predictive
  many-body calculations of phonon-limited carrier mobilities in
  semiconductors}}}},\ \bibfield{journal}{%
  \bibinfo {journal} {Physical Review B}\ }%
  \textbf{\bibinfo {volume} {97}} (\bibinfo {year} {2018}).%
  \bibAnnoteFile{Stop}{Ponce2018}%
\bibitem{Rode1970a}%
  \BibitemOpen
  \bibfield{author}{%
  \bibinfo {author} {\bibfnamefont{D.~L.}\ \bibnamefont{Rode}},\ }%
  \Doi{10.1103/PhysRevB.2.1012}{\emph{\bibinfo {title} {Electron Mobility in
  Direct-Gap Polar Semiconductors}}},\ \bibfield{journal}{%
  \bibinfo {journal} {Phys. Rev. B}\ }%
  \textbf{\bibinfo {volume} {2}},\ \bibinfo {pages} {1012} (\bibinfo {year}
  {1970}).%
  \bibAnnoteFile{Stop}{Rode1970a}%
\bibitem{Rode1970b}%
  \BibitemOpen
  \bibfield{author}{%
  \bibinfo {author} {\bibfnamefont{D.~L.}\ \bibnamefont{Rode}},\ }%
  \Doi{10.1103/PhysRevB.2.4036}{\emph{\bibinfo {title} {Electron Mobility in
  II-VI Semiconductors}}},\ \bibfield{journal}{%
  \bibinfo {journal} {Phys. Rev. B}\ }%
  \textbf{\bibinfo {volume} {2}},\ \bibinfo {pages} {4036} (\bibinfo {year}
  {1970}).%
  \bibAnnoteFile{Stop}{Rode1970b}%
\bibitem{Kaasbjerg2012a}%
  \BibitemOpen
  \bibfield{author}{%
  \bibinfo {author} {\bibfnamefont{K.}~\bibnamefont{Kaasbjerg}}, \bibinfo
  {author} {\bibfnamefont{K.~S.}\ \bibnamefont{Thygesen}},\ and\ \bibinfo
  {author} {\bibfnamefont{K.~W.}\ \bibnamefont{Jacobsen}},\ }%
  \Doi{10.1103/PhysRevB.85.115317}{\emph{\bibinfo {title} {{Phonon-limited
  mobility in n-type single-layer MoS 2 from first principles}}}},\
  \bibfield{journal}{%
  \bibinfo {journal} {Physical Review B}\ }%
  \textbf{\bibinfo {volume} {85}},\ \bibinfo {pages} {115317} (\bibinfo {year}
  {2012}).%
  \bibAnnoteFile{Stop}{Kaasbjerg2012a}%
\bibitem{Kaasbjerg2013}%
  \BibitemOpen
  \bibfield{author}{%
  \bibinfo {author} {\bibfnamefont{K.}~\bibnamefont{Kaasbjerg}}, \bibinfo
  {author} {\bibfnamefont{K.~S.}\ \bibnamefont{Thygesen}},\ and\ \bibinfo
  {author} {\bibfnamefont{A.-p.~P.}\ \bibnamefont{Jauho}},\ }%
  \Doi{10.1103/PhysRevB.87.235312}{\emph{\bibinfo {title} {{Acoustic phonon
  limited mobility in two-dimensional semiconductors: Deformation potential and
  piezoelectric scattering in monolayer MoS 2 from first principles}}}},\
  \bibfield{journal}{%
  \bibinfo {journal} {Physical Review B}\ }%
  \textbf{\bibinfo {volume} {87}},\ \bibinfo {pages} {235312} (\bibinfo {year}
  {2013}).%
  \bibAnnoteFile{Stop}{Kaasbjerg2013}%
\bibitem{Li2015}%
  \BibitemOpen
  \bibfield{author}{%
  \bibinfo {author} {\bibfnamefont{W.}~\bibnamefont{Li}},\ }%
  \Doi{10.1103/PhysRevB.92.075405}{\emph{\bibinfo {title} {{Electrical
  transport limited by electron-phonon coupling from Boltzmann transport
  equation: An ab initio study of Si, Al, and MoS 2}}}},\ \bibfield{journal}{%
  \bibinfo {journal} {Physical Review B}\ }%
  \textbf{\bibinfo {volume} {92}},\ \bibinfo {pages} {075405} (\bibinfo {year}
  {2015}).%
  \bibAnnoteFile{Stop}{Li2015}%
\bibitem{Ma2018}%
  \BibitemOpen
  \bibfield{author}{%
  \bibinfo {author} {\bibfnamefont{J.}~\bibnamefont{Ma}}, \bibinfo {author}
  {\bibfnamefont{A.~S.}\ \bibnamefont{Nissimagoudar}},\ and\ \bibinfo {author}
  {\bibfnamefont{W.}~\bibnamefont{Li}},\ }%
  \Doi{10.1103/PhysRevB.97.045201}{\emph{\bibinfo {title} {First-principles
  study of electron and hole mobilities of Si and GaAs}}},\
  \bibfield{journal}{%
  \bibinfo {journal} {Phys. Rev. B}\ }%
  \textbf{\bibinfo {volume} {97}},\ \bibinfo {pages} {045201} (\bibinfo {year}
  {2018}).%
  \bibAnnoteFile{Stop}{Ma2018}%
\bibitem{Fiorentini2016}%
  \BibitemOpen
  \bibfield{author}{%
  \bibinfo {author} {\bibfnamefont{M.}~\bibnamefont{Fiorentini}}\ and\ \bibinfo
  {author} {\bibfnamefont{N.}~\bibnamefont{Bonini}},\ }%
  \Doi{10.1103/PhysRevB.94.085204}{\emph{\bibinfo {title} {Thermoelectric
  coefficients of $n$-doped silicon from first principles via the solution of
  the Boltzmann transport equation}}},\ \bibfield{journal}{%
  \bibinfo {journal} {Phys. Rev. B}\ }%
  \textbf{\bibinfo {volume} {94}},\ \bibinfo {pages} {085204} (\bibinfo {year}
  {2016}).%
  \bibAnnoteFile{Stop}{Fiorentini2016}%
\bibitem{Fugallo2013}%
  \BibitemOpen
  \bibfield{author}{%
  \bibinfo {author} {\bibfnamefont{G.}~\bibnamefont{Fugallo}}, \bibinfo
  {author} {\bibfnamefont{M.}~\bibnamefont{Lazzeri}}, \bibinfo {author}
  {\bibfnamefont{L.}~\bibnamefont{Paulatto}},\ and\ \bibinfo {author}
  {\bibfnamefont{F.}~\bibnamefont{Mauri}},\ }%
  \Doi{10.1103/PhysRevB.88.045430}{\emph{\bibinfo {title} {Ab initio
  variational approach for evaluating lattice thermal conductivity}}},\
  \bibfield{journal}{%
  \bibinfo {journal} {Phys. Rev. B}\ }%
  \textbf{\bibinfo {volume} {88}},\ \bibinfo {pages} {045430} (\bibinfo {year}
  {2013}).%
  \bibAnnoteFile{Stop}{Fugallo2013}%
\bibitem{Fugallo2014}%
  \BibitemOpen
  \bibfield{author}{%
  \bibinfo {author} {\bibfnamefont{G.}~\bibnamefont{Fugallo}}, \bibinfo
  {author} {\bibfnamefont{A.}~\bibnamefont{Cepellotti}}, \bibinfo {author}
  {\bibfnamefont{L.}~\bibnamefont{Paulatto}}, \bibinfo {author}
  {\bibfnamefont{M.}~\bibnamefont{Lazzeri}}, \bibinfo {author}
  {\bibfnamefont{N.}~\bibnamefont{Marzari}},\ and\ \bibinfo {author}
  {\bibfnamefont{F.}~\bibnamefont{Mauri}},\ }%
  \Doi{10.1021/nl502059f}{\emph{\bibinfo {title} {Thermal Conductivity of
  Graphene and Graphite: Collective Excitations and Mean Free Paths}}},\
  \bibfield{journal}{%
  \bibinfo {journal} {Nano Letters}\ }%
  \textbf{\bibinfo {volume} {14}},\ \bibinfo {pages} {6109} (\bibinfo {year}
  {2014}).%
  \bibAnnoteFile{Stop}{Fugallo2014}%
\bibitem{Cepellotti2015}%
  \BibitemOpen
  \bibfield{author}{%
  \bibinfo {author} {\bibfnamefont{A.}~\bibnamefont{Cepellotti}}, \bibinfo
  {author} {\bibfnamefont{G.}~\bibnamefont{Fugallo}}, \bibinfo {author}
  {\bibfnamefont{L.}~\bibnamefont{Paulatto}}, \bibinfo {author}
  {\bibfnamefont{M.}~\bibnamefont{Lazzeri}}, \bibinfo {author}
  {\bibfnamefont{F.}~\bibnamefont{Mauri}},\ and\ \bibinfo {author}
  {\bibfnamefont{N.}~\bibnamefont{Marzari}},\ }%
  \href{http://dx.doi.org/10.1038/ncomms7400}{\emph{\bibinfo {title} {Phonon
  hydrodynamics in two-dimensional materials}}},\ \bibfield{journal}{%
  \bibinfo {journal} {Nature Communications}\ }%
  \textbf{\bibinfo {volume} {6}},\ \bibinfo {pages} {6400} (\bibinfo {year}
  {2015}).%
  \bibAnnoteFile{Stop}{Cepellotti2015}%
\bibitem{Sohier2014a}%
  \BibitemOpen
  \bibfield{author}{%
  \bibinfo {author} {\bibfnamefont{T.}~\bibnamefont{Sohier}}, \bibinfo {author}
  {\bibfnamefont{M.}~\bibnamefont{Calandra}}, \bibinfo {author}
  {\bibfnamefont{C.-H.}\ \bibnamefont{Park}}, \bibinfo {author}
  {\bibfnamefont{N.}~\bibnamefont{Bonini}}, \bibinfo {author}
  {\bibfnamefont{N.}~\bibnamefont{Marzari}},\ and\ \bibinfo {author}
  {\bibfnamefont{F.}~\bibnamefont{Mauri}},\ }%
  \Doi{10.1103/PhysRevB.90.125414}{\emph{\bibinfo {title} {{Phonon-limited
  resistivity of graphene by first-principles calculations: Electron-phonon
  interactions, strain-induced gauge field, and Boltzmann equation}}}},\
  \bibfield{journal}{%
  \bibinfo {journal} {Physical Review B}\ }%
  \textbf{\bibinfo {volume} {90}},\ \bibinfo {pages} {125414} (\bibinfo {year}
  {2014}).%
  \bibAnnoteFile{Stop}{Sohier2014a}%
\bibitem{Ashraff1987}%
  \BibitemOpen
  \bibfield{author}{%
  \bibinfo {author} {\bibfnamefont{J.~A.}\ \bibnamefont{Ashraff}}\ and\
  \bibinfo {author} {\bibfnamefont{P.~D.}\ \bibnamefont{Loly}},\ }%
  \href{http://stacks.iop.org/0022-3719/20/i=29/a=017}{\emph{\bibinfo {title}
  {The triangular linear analytic method for two-dimensional spectral
  functions}}},\ \bibfield{journal}{%
  \bibinfo {journal} {Journal of Physics C: Solid State Physics}\ }%
  \textbf{\bibinfo {volume} {20}},\ \bibinfo {pages} {4823} (\bibinfo {year}
  {1987}).%
  \bibAnnoteFile{Stop}{Ashraff1987}%
\bibitem{Pedersen2008}%
  \BibitemOpen
  \bibfield{author}{%
  \bibinfo {author} {\bibfnamefont{T.~G.}\ \bibnamefont{Pedersen}}, \bibinfo
  {author} {\bibfnamefont{C.}~\bibnamefont{Flindt}}, \bibinfo {author}
  {\bibfnamefont{J.}~\bibnamefont{Pedersen}}, \bibinfo {author}
  {\bibfnamefont{A.-P.}\ \bibnamefont{Jauho}}, \bibinfo {author}
  {\bibfnamefont{N.~A.}\ \bibnamefont{Mortensen}},\ and\ \bibinfo {author}
  {\bibfnamefont{K.}~\bibnamefont{Pedersen}},\ }%
  \Doi{10.1103/PhysRevB.77.245431}{\emph{\bibinfo {title} {Optical properties
  of graphene antidot lattices}}},\ \bibfield{journal}{%
  \bibinfo {journal} {Phys. Rev. B}\ }%
  \textbf{\bibinfo {volume} {77}},\ \bibinfo {pages} {245431} (\bibinfo {year}
  {2008}).%
  \bibAnnoteFile{Stop}{Pedersen2008}%
\bibitem{Ando2006}%
  \BibitemOpen
  \bibfield{author}{%
  \bibinfo {author} {\bibfnamefont{T.}~\bibnamefont{Ando}},\ }%
  \Doi{10.1143/JPSJ.75.074716}{\emph{\bibinfo {title} {{Screening Effect and
  Impurity Scattering in Monolayer Graphene}}}},\ \bibfield{journal}{%
  \bibinfo {journal} {Journal of the Physics Society of Japan}\ }%
  \textbf{\bibinfo {volume} {75}},\ \bibinfo {pages} {074716} (\bibinfo {year}
  {2006}).%
  \bibAnnoteFile{Stop}{Ando2006}%
\bibitem{Hwang2007}%
  \BibitemOpen
  \bibfield{author}{%
  \bibinfo {author} {\bibfnamefont{E.~H.}\ \bibnamefont{Hwang}}, \bibinfo
  {author} {\bibfnamefont{S.}~\bibnamefont{Adam}},\ and\ \bibinfo {author}
  {\bibfnamefont{S.}~\bibnamefont{{Das Sarma}}},\ }%
  \Doi{10.1103/PhysRevLett.98.186806}{\emph{\bibinfo {title} {{Carrier
  Transport in Two-Dimensional Graphene Layers}}}},\ \bibfield{journal}{%
  \bibinfo {journal} {Physical Review Letters}\ }%
  \textbf{\bibinfo {volume} {98}},\ \bibinfo {pages} {186806} (\bibinfo {year}
  {2007}).%
  \bibAnnoteFile{Stop}{Hwang2007}%
\bibitem{Manes2007}%
  \BibitemOpen
  \bibfield{author}{%
  \bibinfo {author} {\bibfnamefont{J.~L.}\ \bibnamefont{Manes}},\ }%
  \Doi{10.1103/PhysRevB.76.045430}{\emph{\bibinfo {title} {{Symmetry-based
  approach to electron-phonon interactions in graphene}}}},\
  \bibfield{journal}{%
  \bibinfo {journal} {Physical Review B}\ }%
  \textbf{\bibinfo {volume} {76}},\ \bibinfo {pages} {045430} (\bibinfo {year}
  {2007}).%
  \bibAnnoteFile{Stop}{Manes2007}%
\bibitem{Samsonidze2007}%
  \BibitemOpen
  \bibfield{author}{%
  \bibinfo {author} {\bibfnamefont{G.~G.}\ \bibnamefont{Samsonidze}}, \bibinfo
  {author} {\bibfnamefont{E.~B.}\ \bibnamefont{Barros}}, \bibinfo {author}
  {\bibfnamefont{R.}~\bibnamefont{Saito}}, \bibinfo {author}
  {\bibfnamefont{J.}~\bibnamefont{Jiang}}, \bibinfo {author}
  {\bibfnamefont{G.}~\bibnamefont{Dresselhaus}},\ and\ \bibinfo {author}
  {\bibfnamefont{M.~S.}\ \bibnamefont{Dresselhaus}},\ }%
  \Doi{10.1103/PhysRevB.75.155420}{\emph{\bibinfo {title} {{Electron-phonon
  coupling mechanism in two-dimensional graphite and single-wall carbon
  nanotubes}}}},\ \bibfield{journal}{%
  \bibinfo {journal} {Physical Review B}\ }%
  \textbf{\bibinfo {volume} {75}},\ \bibinfo {pages} {155420} (\bibinfo {year}
  {2007}).%
  \bibAnnoteFile{Stop}{Samsonidze2007}%
\bibitem{Stauber2007}%
  \BibitemOpen
  \bibfield{author}{%
  \bibinfo {author} {\bibfnamefont{T.}~\bibnamefont{Stauber}}, \bibinfo
  {author} {\bibfnamefont{N.}~\bibnamefont{Peres}},\ and\ \bibinfo {author}
  {\bibfnamefont{F.}~\bibnamefont{Guinea}},\ }%
  \Doi{10.1103/PhysRevB.76.205423}{\emph{\bibinfo {title} {{Electronic
  transport in graphene: A semiclassical approach including midgap states}}}},\
  \bibfield{journal}{%
  \bibinfo {journal} {Physical Review B}\ }%
  \textbf{\bibinfo {volume} {76}},\ \bibinfo {pages} {205423} (\bibinfo {year}
  {2007}).%
  \bibAnnoteFile{Stop}{Stauber2007}%
\bibitem{Fratini2008}%
  \BibitemOpen
  \bibfield{author}{%
  \bibinfo {author} {\bibfnamefont{S.}~\bibnamefont{Fratini}}\ and\ \bibinfo
  {author} {\bibfnamefont{F.}~\bibnamefont{Guinea}},\ }%
  \href{http://prb.aps.org/abstract/PRB/v77/i19/e195415}{\emph{\bibinfo {title}
  {{Substrate-limited electron dynamics in graphene}}}},\ \bibfield{journal}{%
  \bibinfo {journal} {Physical Review B}\ }%
  \textbf{\bibinfo {volume} {77}},\ \bibinfo {pages} {195415} (\bibinfo {year}
  {2008}).%
  \bibAnnoteFile{Stop}{Fratini2008}%
\bibitem{Hwang2008a}%
  \BibitemOpen
  \bibfield{author}{%
  \bibinfo {author} {\bibfnamefont{E.~H.}\ \bibnamefont{Hwang}}\ and\ \bibinfo
  {author} {\bibfnamefont{S.}~\bibnamefont{{Das Sarma}}},\ }%
  \href{http://prb.aps.org/abstract/PRB/v77/i11/e115449}{\emph{\bibinfo {title}
  {{Acoustic phonon scattering limited carrier mobility in two-dimensional
  extrinsic graphene}}}},\ \bibfield{journal}{%
  \bibinfo {journal} {Physical Review B}\ }%
  \textbf{\bibinfo {volume} {77}},\ \bibinfo {pages} {115449} (\bibinfo {year}
  {2008}).%
  \bibAnnoteFile{Stop}{Hwang2008a}%
\bibitem{Hwang2009}%
  \BibitemOpen
  \bibfield{author}{%
  \bibinfo {author} {\bibfnamefont{E.~H.}\ \bibnamefont{Hwang}}, \bibinfo
  {author} {\bibfnamefont{S.~D.}\ \bibnamefont{Sarma}},\ and\ \bibinfo {author}
  {\bibfnamefont{S.}~\bibnamefont{{Das Sarma}}},\ }%
  \Doi{10.1103/PhysRevB.79.165404}{\emph{\bibinfo {title} {{Screening-induced
  temperature-dependent transport in two-dimensional graphene}}}},\
  \bibfield{journal}{%
  \bibinfo {journal} {Physical Review B}\ }%
  \textbf{\bibinfo {volume} {79}},\ \bibinfo {pages} {165404} (\bibinfo {year}
  {2009}).%
  \bibAnnoteFile{Stop}{Hwang2009}%
\bibitem{Mariani2010}%
  \BibitemOpen
  \bibfield{author}{%
  \bibinfo {author} {\bibfnamefont{E.}~\bibnamefont{Mariani}}\ and\ \bibinfo
  {author} {\bibfnamefont{F.}~\bibnamefont{von Oppen}},\ }%
  \Doi{10.1103/PhysRevB.82.195403}{\emph{\bibinfo {title}
  {{Temperature-dependent resistivity of suspended graphene}}}},\
  \bibfield{journal}{%
  \bibinfo {journal} {Physical Review B}\ }%
  \textbf{\bibinfo {volume} {82}},\ \bibinfo {pages} {195403} (\bibinfo {year}
  {2010}).%
  \bibAnnoteFile{Stop}{Mariani2010}%
\bibitem{DasSarma2011}%
  \BibitemOpen
  \bibfield{author}{%
  \bibinfo {author} {\bibfnamefont{S.}~\bibnamefont{{Das Sarma}}}, \bibinfo
  {author} {\bibfnamefont{S.}~\bibnamefont{Adam}}, \bibinfo {author}
  {\bibfnamefont{E.~H.}\ \bibnamefont{Hwang}},\ and\ \bibinfo {author}
  {\bibfnamefont{E.}~\bibnamefont{Rossi}},\ }%
  \Doi{10.1103/RevModPhys.83.407}{\emph{\bibinfo {title} {{Electronic transport
  in two-dimensional graphene}}}},\ \bibfield{journal}{%
  \bibinfo {journal} {Reviews of Modern Physics}\ }%
  \textbf{\bibinfo {volume} {83}},\ \bibinfo {pages} {407} (\bibinfo {year}
  {2011}).%
  \bibAnnoteFile{Stop}{DasSarma2011}%
\bibitem{Kaasbjerg2012}%
  \BibitemOpen
  \bibfield{author}{%
  \bibinfo {author} {\bibfnamefont{K.}~\bibnamefont{Kaasbjerg}}, \bibinfo
  {author} {\bibfnamefont{K.~K.}\ \bibnamefont{Thygesen}},\ and\ \bibinfo
  {author} {\bibfnamefont{K.~K.}\ \bibnamefont{Jacobsen}},\ }%
  \Doi{10.1103/PhysRevB.85.165440}{\emph{\bibinfo {title} {{Unraveling the
  acoustic electron-phonon interaction in graphene}}}},\ \bibfield{journal}{%
  \bibinfo {journal} {Physical Review B}\ }%
  \textbf{\bibinfo {volume} {85}},\ \bibinfo {pages} {165440} (\bibinfo {year}
  {2012}).%
  \bibAnnoteFile{Stop}{Kaasbjerg2012}%
\bibitem{Hwang}%
  \BibitemOpen
  \bibfield{author}{%
  \bibinfo {author} {\bibfnamefont{S.}~\bibnamefont{{Das Sarma}}}\ and\
  \bibinfo {author} {\bibfnamefont{E.~H.}\ \bibnamefont{Hwang}},\ }%
  \Doi{10.1103/PhysRevB.87.035415}{\emph{\bibinfo {title} {{Density-dependent
  electrical conductivity in suspended graphene: Approaching the Dirac point in
  transport}}}},\ \bibfield{journal}{%
  \bibinfo {journal} {Physical Review B}\ }%
  \textbf{\bibinfo {volume} {87}},\ \bibinfo {pages} {035415} (\bibinfo {year}
  {2013}).%
  \bibAnnoteFile{Stop}{Hwang}%
\bibitem{Gunst2017}%
  \BibitemOpen
  \bibfield{author}{%
  \bibinfo {author} {\bibfnamefont{T.}~\bibnamefont{Gunst}}, \bibinfo {author}
  {\bibfnamefont{K.}~\bibnamefont{Kaasbjerg}},\ and\ \bibinfo {author}
  {\bibfnamefont{M.}~\bibnamefont{Brandbyge}},\ }%
  \Doi{10.1103/PhysRevLett.118.046601}{\emph{\bibinfo {title} {{Flexural-Phonon
  Scattering Induced by Electrostatic Gating in Graphene}}}},\
  \bibfield{journal}{%
  \bibinfo {journal} {Physical Review Letters}\ }%
  \textbf{\bibinfo {volume} {118}},\ \bibinfo {pages} {046601} (\bibinfo {year}
  {2017}).%
  \bibAnnoteFile{Stop}{Gunst2017}%
\bibitem{Qiao2014}%
  \BibitemOpen
  \bibfield{author}{%
  \bibinfo {author} {\bibfnamefont{J.}~\bibnamefont{Qiao}}, \bibinfo {author}
  {\bibfnamefont{X.}~\bibnamefont{Kong}}, \bibinfo {author}
  {\bibfnamefont{Z.-X.}\ \bibnamefont{Hu}}, \bibinfo {author}
  {\bibfnamefont{F.}~\bibnamefont{Yang}},\ and\ \bibinfo {author}
  {\bibfnamefont{W.}~\bibnamefont{Ji}},\ }%
  \Doi{10.1038/ncomms5475}{\emph{\bibinfo {title} {{High-mobility transport
  anisotropy and linear dichroism in few-layer black phosphorus}}}},\
  \bibfield{journal}{%
  \bibinfo {journal} {Nature Communications}\ }%
  \textbf{\bibinfo {volume} {5}},\ \bibinfo {pages} {4475} (\bibinfo {year}
  {2014}).%
  \bibAnnoteFile{Stop}{Qiao2014}%
\bibitem{Liao2015}%
  \BibitemOpen
  \bibfield{author}{%
  \bibinfo {author} {\bibfnamefont{B.}~\bibnamefont{Liao}}, \bibinfo {author}
  {\bibfnamefont{J.}~\bibnamefont{Zhou}}, \bibinfo {author}
  {\bibfnamefont{B.}~\bibnamefont{Qiu}}, \bibinfo {author}
  {\bibfnamefont{M.~S.}\ \bibnamefont{Dresselhaus}},\ and\ \bibinfo {author}
  {\bibfnamefont{G.}~\bibnamefont{Chen}},\ }%
  \Doi{10.1103/PhysRevB.91.235419}{\emph{\bibinfo {title} {{Ab initio study of
  electron-phonon interaction in phosphorene}}}},\ \bibfield{journal}{%
  \bibinfo {journal} {Physical Review B}\ }%
  \textbf{\bibinfo {volume} {91}},\ \bibinfo {pages} {235419} (\bibinfo {year}
  {2015}).%
  \bibAnnoteFile{Stop}{Liao2015}%
\bibitem{Wang2015}%
  \BibitemOpen
  \bibfield{author}{%
  \bibinfo {author} {\bibfnamefont{Y.}~\bibnamefont{Wang}}\ and\ \bibinfo
  {author} {\bibfnamefont{Y.}~\bibnamefont{Ding}},\ }%
  \Doi{10.1186/s11671-015-0955-7}{\emph{\bibinfo {title} {Electronic Structure
  and Carrier Mobilities of Arsenene and Antimonene Nanoribbons: A
  First-Principle Study}}},\ \bibfield{journal}{%
  \bibinfo {journal} {Nanoscale Research Letters}\ }%
  \textbf{\bibinfo {volume} {10}},\ \bibinfo {pages} {254} (\bibinfo {year}
  {2015}).%
  \bibAnnoteFile{Stop}{Wang2015}%
\bibitem{Pizzi2016nc}%
  \BibitemOpen
  \bibfield{author}{%
  \bibinfo {author} {\bibfnamefont{G.}~\bibnamefont{Pizzi}}, \bibinfo {author}
  {\bibfnamefont{M.}~\bibnamefont{Gibertini}}, \bibinfo {author}
  {\bibfnamefont{E.}~\bibnamefont{Dib}}, \bibinfo {author}
  {\bibfnamefont{N.}~\bibnamefont{Marzari}}, \bibinfo {author}
  {\bibfnamefont{G.}~\bibnamefont{Iannaccone}},\ and\ \bibinfo {author}
  {\bibfnamefont{G.}~\bibnamefont{Fiori}},\ }%
  \href{http://dx.doi.org/10.1038/ncomms12585}{\emph{\bibinfo {title}
  {Performance of arsenene and antimonene double-gate MOSFETs from first
  principles}}},\ \bibfield{journal}{%
  \bibinfo {journal} {Nature Communications}\ }%
  \textbf{\bibinfo {volume} {7}},\ \bibinfo {pages} {1258} (\bibinfo {year}
  {2016}).%
  \bibAnnoteFile{Stop}{Pizzi2016nc}%
\bibitem{Wang2017}%
  \BibitemOpen
  \bibfield{author}{%
  \bibinfo {author} {\bibfnamefont{Y.}~\bibnamefont{Wang}}, \bibinfo {author}
  {\bibfnamefont{P.}~\bibnamefont{Huang}}, \bibinfo {author}
  {\bibfnamefont{M.}~\bibnamefont{Ye}}, \bibinfo {author}
  {\bibfnamefont{R.}~\bibnamefont{Quhe}}, \bibinfo {author}
  {\bibfnamefont{Y.}~\bibnamefont{Pan}}, \bibinfo {author}
  {\bibfnamefont{H.}~\bibnamefont{Zhang}}, \bibinfo {author}
  {\bibfnamefont{H.}~\bibnamefont{Zhong}}, \bibinfo {author}
  {\bibfnamefont{J.}~\bibnamefont{Shi}},\ and\ \bibinfo {author}
  {\bibfnamefont{J.}~\bibnamefont{Lu}},\ }%
  \Doi{10.1021/acs.chemmater.6b04909}{\emph{\bibinfo {title} {Many-body Effect,
  Carrier Mobility, and Device Performance of Hexagonal Arsenene and
  Antimonene}}},\ \bibfield{journal}{%
  \bibinfo {journal} {Chemistry of Materials}\ }%
  \textbf{\bibinfo {volume} {29}},\ \bibinfo {pages} {2191} (\bibinfo {year}
  {2017}).%
  \bibAnnoteFile{Stop}{Wang2017}%
\bibitem{Ashcroft1976}%
  \BibitemOpen
  \bibfield{author}{%
  \bibinfo {author} {\bibfnamefont{N.~W.}\ \bibnamefont{Ashcroft}}\ and\
  \bibinfo {author} {\bibfnamefont{N.~D.}\ \bibnamefont{Mermin}},\ }%
  \emph{\bibinfo {title} {{Solid State Physics}}}\ (\bibinfo {publisher}
  {Brooks Cole},\ \bibinfo {address} {Belmont (USA)},\ \bibinfo {year} {1976})\
  ISBN \bibinfo {isbn} {0030839939}.%
  \bibAnnoteFile{Stop}{Ashcroft1976}%
\bibitem{Nag1980}%
  \BibitemOpen
  \bibfield{author}{%
  \bibinfo {author} {\bibfnamefont{B.}~\bibnamefont{Nag}},\ }%
  \Doi{10.1007/978-3-642-81416-7}{\emph{\bibinfo {title} {{Electron Transport
  in Compound Semiconductors}}}},\ \bibfield{journal}{%
  \bibinfo {journal} {Springer Series in Solid-State Sciences},\ }%
  \bibinfo {series} {Springer Series in Solid-State Sciences}\ \textbf{\bibinfo
  {volume} {11}} (\bibinfo {year} {1980}).%
  \bibAnnoteFile{Stop}{Nag1980}%
\bibitem{gllb1}%
  \BibitemOpen
  \bibfield{author}{%
  \bibinfo {author} {\bibfnamefont{O.}~\bibnamefont{Gritsenko}}, \bibinfo
  {author} {\bibfnamefont{R.}~\bibnamefont{van Leeuwen}}, \bibinfo {author}
  {\bibfnamefont{E.}~\bibnamefont{van Lenthe}},\ and\ \bibinfo {author}
  {\bibfnamefont{E.}~\bibnamefont{Baerends}},\ }%
  \emph{\bibinfo {title} {{Self-consistent approximation to the Kohn-Sham
  exchange potential}}},\ \bibfield{journal}{%
  \bibinfo {journal} {Physical Review A}\ }%
  \textbf{\bibinfo {volume} {51}},\ \bibinfo {pages} {1944} (\bibinfo {year}
  {1995}).%
  \bibAnnoteFile{Stop}{gllb1}%
\bibitem{gllb2}%
  \BibitemOpen
  \bibfield{author}{%
  \bibinfo {author} {\bibfnamefont{M.}~\bibnamefont{Kuisma}}, \bibinfo {author}
  {\bibfnamefont{J.}~\bibnamefont{Ojanen}}, \bibinfo {author}
  {\bibfnamefont{J.}~\bibnamefont{Enkovaara}},\ and\ \bibinfo {author}
  {\bibfnamefont{T.}~\bibnamefont{Rantala}},\ }%
  \Doi{10.1103/PhysRevB.82.115106}{\emph{\bibinfo {title} {{Kohn-Sham potential
  with discontinuity for band gap materials}}}},\ \bibfield{journal}{%
  \bibinfo {journal} {Physical Review B}\ }%
  \textbf{\bibinfo {volume} {82}},\ \bibinfo {pages} {115106} (\bibinfo {year}
  {2010}).%
  \bibAnnoteFile{Stop}{gllb2}%
\bibitem{Cohen1966}%
  \BibitemOpen
  \bibfield{author}{%
  \bibinfo {author} {\bibfnamefont{M.~L.}\ \bibnamefont{Cohen}}\ and\ \bibinfo
  {author} {\bibfnamefont{T.~K.}\ \bibnamefont{Bergstresser}},\ }%
  \Doi{10.1103/PhysRev.141.789}{\emph{\bibinfo {title} {{Band Structures and
  Pseudopotential Form Factors for Fourteen Semiconductors of the Diamond and
  Zinc-blende Structures}}}},\ \bibfield{journal}{%
  \bibinfo {journal} {Physical Review}\ }%
  \textbf{\bibinfo {volume} {141}},\ \bibinfo {pages} {789} (\bibinfo {year}
  {1966}).%
  \bibAnnoteFile{Stop}{Cohen1966}%
\bibitem{Wang1995}%
  \BibitemOpen
  \bibfield{author}{%
  \bibinfo {author} {\bibfnamefont{L.-W.}\ \bibnamefont{Wang}}\ and\ \bibinfo
  {author} {\bibfnamefont{A.}~\bibnamefont{Zunger}},\ }%
  \Doi{10.1103/PhysRevB.51.17398}{\emph{\bibinfo {title}
  {{Local-density-derived semiempirical pseudopotentials}}}},\
  \bibfield{journal}{%
  \bibinfo {journal} {Physical Review B}\ }%
  \textbf{\bibinfo {volume} {51}},\ \bibinfo {pages} {17398} (\bibinfo {year}
  {1995}).%
  \bibAnnoteFile{Stop}{Wang1995}%
\bibitem{Giannozzi2017}%
  \BibitemOpen
  \bibfield{author}{%
  \bibinfo {author} {\bibfnamefont{P.}~\bibnamefont{Giannozzi}}, \bibinfo
  {author} {\bibfnamefont{O.}~\bibnamefont{Andreussi}}, \bibinfo {author}
  {\bibfnamefont{T.}~\bibnamefont{Brumme}}, \bibinfo {author}
  {\bibfnamefont{O.}~\bibnamefont{Bunau}}, \bibinfo {author}
  {\bibfnamefont{M.}~\bibnamefont{{Buongiorno Nardelli}}}, \bibinfo {author}
  {\bibfnamefont{M.}~\bibnamefont{Calandra}}, \bibinfo {author}
  {\bibfnamefont{R.}~\bibnamefont{Car}}, \bibinfo {author}
  {\bibfnamefont{C.}~\bibnamefont{Cavazzoni}}, \bibinfo {author}
  {\bibfnamefont{D.}~\bibnamefont{Ceresoli}}, \bibinfo {author}
  {\bibfnamefont{M.}~\bibnamefont{Cococcioni}}, \bibinfo {author}
  {\bibfnamefont{N.}~\bibnamefont{Colonna}}, \bibinfo {author}
  {\bibfnamefont{I.}~\bibnamefont{Carnimeo}}, \bibinfo {author}
  {\bibfnamefont{A.}~\bibnamefont{{Dal Corso}}}, \bibinfo {author}
  {\bibfnamefont{S.}~\bibnamefont{de~Gironcoli}}, \bibinfo {author}
  {\bibfnamefont{P.}~\bibnamefont{Delugas}}, \bibinfo {author}
  {\bibfnamefont{R.~A.}\ \bibnamefont{DiStasio}}, \bibinfo {author}
  {\bibfnamefont{A.}~\bibnamefont{Ferretti}}, \bibinfo {author}
  {\bibfnamefont{A.}~\bibnamefont{Floris}}, \bibinfo {author}
  {\bibfnamefont{G.}~\bibnamefont{Fratesi}}, \bibinfo {author}
  {\bibfnamefont{G.}~\bibnamefont{Fugallo}}, \bibinfo {author}
  {\bibfnamefont{R.}~\bibnamefont{Gebauer}}, \bibinfo {author}
  {\bibfnamefont{U.}~\bibnamefont{Gerstmann}}, \bibinfo {author}
  {\bibfnamefont{F.}~\bibnamefont{Giustino}}, \bibinfo {author}
  {\bibfnamefont{T.}~\bibnamefont{Gorni}}, \bibinfo {author}
  {\bibfnamefont{J.}~\bibnamefont{Jia}}, \bibinfo {author}
  {\bibfnamefont{M.}~\bibnamefont{Kawamura}}, \bibinfo {author}
  {\bibfnamefont{H.-Y.}\ \bibnamefont{Ko}}, \bibinfo {author}
  {\bibfnamefont{A.}~\bibnamefont{Kokalj}}, \bibinfo {author}
  {\bibfnamefont{E.}~\bibnamefont{K{\"{u}}{\c{c}}{\"{u}}kbenli}}, \bibinfo
  {author} {\bibfnamefont{M.}~\bibnamefont{Lazzeri}}, \bibinfo {author}
  {\bibfnamefont{M.}~\bibnamefont{Marsili}}, \bibinfo {author}
  {\bibfnamefont{N.}~\bibnamefont{Marzari}}, \bibinfo {author}
  {\bibfnamefont{F.}~\bibnamefont{Mauri}}, \bibinfo {author}
  {\bibfnamefont{N.~L.}\ \bibnamefont{Nguyen}}, \bibinfo {author}
  {\bibfnamefont{H.-V.}\ \bibnamefont{Nguyen}}, \bibinfo {author}
  {\bibfnamefont{A.}~\bibnamefont{Otero-de-la Roza}}, \bibinfo {author}
  {\bibfnamefont{L.}~\bibnamefont{Paulatto}}, \bibinfo {author}
  {\bibfnamefont{S.}~\bibnamefont{Ponc{\'{e}}}}, \bibinfo {author}
  {\bibfnamefont{D.}~\bibnamefont{Rocca}}, \bibinfo {author}
  {\bibfnamefont{R.}~\bibnamefont{Sabatini}}, \bibinfo {author}
  {\bibfnamefont{B.}~\bibnamefont{Santra}}, \bibinfo {author}
  {\bibfnamefont{M.}~\bibnamefont{Schlipf}}, \bibinfo {author}
  {\bibfnamefont{A.~P.}\ \bibnamefont{Seitsonen}}, \bibinfo {author}
  {\bibfnamefont{A.}~\bibnamefont{Smogunov}}, \bibinfo {author}
  {\bibfnamefont{I.}~\bibnamefont{Timrov}}, \bibinfo {author}
  {\bibfnamefont{T.}~\bibnamefont{Thonhauser}}, \bibinfo {author}
  {\bibfnamefont{P.}~\bibnamefont{Umari}}, \bibinfo {author}
  {\bibfnamefont{N.}~\bibnamefont{Vast}}, \bibinfo {author}
  {\bibfnamefont{X.}~\bibnamefont{Wu}},\ and\ \bibinfo {author}
  {\bibfnamefont{S.}~\bibnamefont{Baroni}},\ }%
  \Doi{10.1088/1361-648X/aa8f79}{\emph{\bibinfo {title} {{Advanced capabilities
  for materials modelling with Quantum ESPRESSO}}}},\ \bibfield{journal}{%
  \bibinfo {journal} {Journal of Physics: Condensed Matter}\ }%
  \textbf{\bibinfo {volume} {29}},\ \bibinfo {pages} {465901} (\bibinfo {year}
  {2017}).%
  \bibAnnoteFile{Stop}{Giannozzi2017}%
\bibitem{Giannozzi2009}%
  \BibitemOpen
  \bibfield{author}{%
  \bibinfo {author} {\bibfnamefont{P.}~\bibnamefont{Giannozzi}}, \bibinfo
  {author} {\bibfnamefont{S.}~\bibnamefont{Baroni}}, \bibinfo {author}
  {\bibfnamefont{N.}~\bibnamefont{Bonini}}, \bibinfo {author}
  {\bibfnamefont{M.}~\bibnamefont{Calandra}}, \bibinfo {author}
  {\bibfnamefont{R.}~\bibnamefont{Car}}, \bibinfo {author}
  {\bibfnamefont{C.}~\bibnamefont{Cavazzoni}}, \bibinfo {author}
  {\bibfnamefont{D.}~\bibnamefont{Ceresoli}}, \bibinfo {author}
  {\bibfnamefont{G.~L.}\ \bibnamefont{Chiarotti}}, \bibinfo {author}
  {\bibfnamefont{M.}~\bibnamefont{Cococcioni}}, \bibinfo {author}
  {\bibfnamefont{I.}~\bibnamefont{Dabo}}, \bibinfo {author}
  {\bibfnamefont{A.}~\bibnamefont{{Dal Corso}}}, \bibinfo {author}
  {\bibfnamefont{S.}~\bibnamefont{de~Gironcoli}}, \bibinfo {author}
  {\bibfnamefont{S.}~\bibnamefont{Fabris}}, \bibinfo {author}
  {\bibfnamefont{G.}~\bibnamefont{Fratesi}}, \bibinfo {author}
  {\bibfnamefont{R.}~\bibnamefont{Gebauer}}, \bibinfo {author}
  {\bibfnamefont{U.}~\bibnamefont{Gerstmann}}, \bibinfo {author}
  {\bibfnamefont{C.}~\bibnamefont{Gougoussis}}, \bibinfo {author}
  {\bibfnamefont{A.}~\bibnamefont{Kokalj}}, \bibinfo {author}
  {\bibfnamefont{M.}~\bibnamefont{Lazzeri}}, \bibinfo {author}
  {\bibfnamefont{L.}~\bibnamefont{Martin-Samos}}, \bibinfo {author}
  {\bibfnamefont{N.}~\bibnamefont{Marzari}}, \bibinfo {author}
  {\bibfnamefont{F.}~\bibnamefont{Mauri}}, \bibinfo {author}
  {\bibfnamefont{R.}~\bibnamefont{Mazzarello}}, \bibinfo {author}
  {\bibfnamefont{S.}~\bibnamefont{Paolini}}, \bibinfo {author}
  {\bibfnamefont{A.}~\bibnamefont{Pasquarello}}, \bibinfo {author}
  {\bibfnamefont{L.}~\bibnamefont{Paulatto}}, \bibinfo {author}
  {\bibfnamefont{C.}~\bibnamefont{Sbraccia}}, \bibinfo {author}
  {\bibfnamefont{S.}~\bibnamefont{Scandolo}}, \bibinfo {author}
  {\bibfnamefont{G.}~\bibnamefont{Sclauzero}}, \bibinfo {author}
  {\bibfnamefont{A.~P.}\ \bibnamefont{Seitsonen}}, \bibinfo {author}
  {\bibfnamefont{A.}~\bibnamefont{Smogunov}}, \bibinfo {author}
  {\bibfnamefont{P.}~\bibnamefont{Umari}},\ and\ \bibinfo {author}
  {\bibfnamefont{R.~M.}\ \bibnamefont{Wentzcovitch}},\ }%
  \Doi{10.1088/0953-8984/21/39/395502}{\emph{\bibinfo {title} {{QUANTUM
  ESPRESSO: a modular and open-source software project for quantum simulations
  of materials.}}}},\ \bibfield{journal}{%
  \bibinfo {journal} {Journal of Physics: Condensed Matter}\ }%
  \textbf{\bibinfo {volume} {21}},\ \bibinfo {pages} {395502} (\bibinfo {year}
  {2009}).%
  \bibAnnoteFile{Stop}{Giannozzi2009}%
\bibitem{Rozzi2006}%
  \BibitemOpen
  \bibfield{author}{%
  \bibinfo {author} {\bibfnamefont{C.~A.}\ \bibnamefont{Rozzi}}, \bibinfo
  {author} {\bibfnamefont{D.}~\bibnamefont{Varsano}}, \bibinfo {author}
  {\bibfnamefont{A.}~\bibnamefont{Marini}}, \bibinfo {author}
  {\bibfnamefont{E.~K.~U.}\ \bibnamefont{Gross}},\ and\ \bibinfo {author}
  {\bibfnamefont{A.}~\bibnamefont{Rubio}},\ }%
  \Doi{10.1103/PhysRevB.73.205119}{\emph{\bibinfo {title} {Exact Coulomb cutoff
  technique for supercell calculations}}},\ \bibfield{journal}{%
  \bibinfo {journal} {Phys. Rev. B}\ }%
  \textbf{\bibinfo {volume} {73}},\ \bibinfo {pages} {205119} (\bibinfo {year}
  {2006}).%
  \bibAnnoteFile{Stop}{Rozzi2006}%
\bibitem{IsmailBeigi2006}%
  \BibitemOpen
  \bibfield{author}{%
  \bibinfo {author} {\bibfnamefont{S.}~\bibnamefont{Ismail-Beigi}},\ }%
  \Doi{10.1103/PhysRevB.73.233103}{\emph{\bibinfo {title} {Truncation of
  periodic image interactions for confined systems}}},\ \bibfield{journal}{%
  \bibinfo {journal} {Phys. Rev. B}\ }%
  \textbf{\bibinfo {volume} {73}},\ \bibinfo {pages} {233103} (\bibinfo {year}
  {2006}).%
  \bibAnnoteFile{Stop}{IsmailBeigi2006}%
\bibitem{Pizzi2016}%
  \BibitemOpen
  \bibfield{author}{%
  \bibinfo {author} {\bibfnamefont{G.}~\bibnamefont{Pizzi}}, \bibinfo {author}
  {\bibfnamefont{A.}~\bibnamefont{Cepellotti}}, \bibinfo {author}
  {\bibfnamefont{R.}~\bibnamefont{Sabatini}}, \bibinfo {author}
  {\bibfnamefont{N.}~\bibnamefont{Marzari}},\ and\ \bibinfo {author}
  {\bibfnamefont{B.}~\bibnamefont{Kozinsky}},\ }%
  \Doi{10.1016/j.commatsci.2015.09.013}{\emph{\bibinfo {title} {{AiiDA:
  automated interactive infrastructure and database for computational
  science}}}},\ \bibfield{journal}{%
  \bibinfo {journal} {Computational Materials Science}\ }%
  \textbf{\bibinfo {volume} {111}},\ \bibinfo {pages} {218} (\bibinfo {year}
  {2016}).%
  \bibAnnoteFile{Stop}{Pizzi2016}%
\bibitem{Blochl1994}%
  \BibitemOpen
  \bibfield{author}{%
  \bibinfo {author} {\bibfnamefont{P.~E.}\ \bibnamefont{Bl\"ochl}}, \bibinfo
  {author} {\bibfnamefont{O.}~\bibnamefont{Jepsen}},\ and\ \bibinfo {author}
  {\bibfnamefont{O.~K.}\ \bibnamefont{Andersen}},\ }%
  \Doi{10.1103/PhysRevB.49.16223}{\emph{\bibinfo {title} {Improved tetrahedron
  method for Brillouin-zone integrations}}},\ \bibfield{journal}{%
  \bibinfo {journal} {Phys. Rev. B}\ }%
  \textbf{\bibinfo {volume} {49}},\ \bibinfo {pages} {16223} (\bibinfo {year}
  {1994}).%
  \bibAnnoteFile{Stop}{Blochl1994}%
\bibitem{Brumme2015}%
  \BibitemOpen
  \bibfield{author}{%
  \bibinfo {author} {\bibfnamefont{T.}~\bibnamefont{Brumme}}, \bibinfo {author}
  {\bibfnamefont{M.}~\bibnamefont{Calandra}},\ and\ \bibinfo {author}
  {\bibfnamefont{F.}~\bibnamefont{Mauri}},\ }%
  \Doi{10.1103/PhysRevB.91.155436}{\emph{\bibinfo {title} {{First-principles
  theory of field-effect doping in transition-metal dichalcogenides: Structural
  properties, electronic structure, Hall coefficient, and electrical
  conductivity}}}},\ \bibfield{journal}{%
  \bibinfo {journal} {Physical Review B}\ }%
  \textbf{\bibinfo {volume} {91}},\ \bibinfo {pages} {155436} (\bibinfo {year}
  {2015}).%
  \bibAnnoteFile{Stop}{Brumme2015}%
\bibitem{Shi2013a}%
  \BibitemOpen
  \bibfield{author}{%
  \bibinfo {author} {\bibfnamefont{H.}~\bibnamefont{Shi}}, \bibinfo {author}
  {\bibfnamefont{H.}~\bibnamefont{Pan}}, \bibinfo {author}
  {\bibfnamefont{Y.-W.}\ \bibnamefont{Zhang}},\ and\ \bibinfo {author}
  {\bibfnamefont{B.~I.}\ \bibnamefont{Yakobson}},\ }%
  \Doi{10.1103/PhysRevB.87.155304}{\emph{\bibinfo {title} {{Quasiparticle band
  structures and optical properties of strained monolayer MoS 2 and WS 2}}}},\
  \bibfield{journal}{%
  \bibinfo {journal} {Physical Review B}\ }%
  \textbf{\bibinfo {volume} {87}},\ \bibinfo {pages} {155304} (\bibinfo {year}
  {2013}).%
  \bibAnnoteFile{Stop}{Shi2013a}%
\bibitem{Maksimov1996}%
  \BibitemOpen
  \bibfield{author}{%
  \bibinfo {author} {\bibfnamefont{E.}~\bibnamefont{Maksimov}}\ and\ \bibinfo
  {author} {\bibfnamefont{S.}~\bibnamefont{Shulga}},\ }%
  \Doi{10.1016/0038-1098(95)00745-8}{\emph{\bibinfo {title} {Nonadiabatic
  effects in optical phonon self-energy}}},\ \bibfield{journal}{%
  \bibinfo {journal} {Solid State Communications}\ }%
  \textbf{\bibinfo {volume} {97}},\ \bibinfo {pages} {553 } (\bibinfo {year}
  {1996}).%
  \bibAnnoteFile{Stop}{Maksimov1996}%
\bibitem{Saitta2008}%
  \BibitemOpen
  \bibfield{author}{%
  \bibinfo {author} {\bibfnamefont{A.~M.}\ \bibnamefont{Saitta}}, \bibinfo
  {author} {\bibfnamefont{M.}~\bibnamefont{Lazzeri}}, \bibinfo {author}
  {\bibfnamefont{M.}~\bibnamefont{Calandra}},\ and\ \bibinfo {author}
  {\bibfnamefont{F.}~\bibnamefont{Mauri}},\ }%
  \Doi{10.1103/PhysRevLett.100.226401}{\emph{\bibinfo {title} {Giant
  Nonadiabatic Effects in Layer Metals: Raman Spectra of Intercalated Graphite
  Explained}}},\ \bibfield{journal}{%
  \bibinfo {journal} {Phys. Rev. Lett.}\ }%
  \textbf{\bibinfo {volume} {100}},\ \bibinfo {pages} {226401} (\bibinfo {year}
  {2008}).%
  \bibAnnoteFile{Stop}{Saitta2008}%
\bibitem{Mahan}%
  \BibitemOpen
  \bibfield{author}{%
  \bibinfo {author} {\bibfnamefont{G.~D.}\ \bibnamefont{Mahan}},\ }%
  \emph{\bibinfo {title} {Electrons and phonons: the theory of transport
  phenomena in solids}},\ \bibinfo {edition} {3rd}\ ed.\ (\bibinfo {publisher}
  {Plenum},\ \bibinfo {address} {New York},\ \bibinfo {year} {2000}).%
  \bibAnnoteFile{Stop}{Mahan}%
\bibitem{Lazzeri2006}%
  \BibitemOpen
  \bibfield{author}{%
  \bibinfo {author} {\bibfnamefont{M.}~\bibnamefont{Lazzeri}}\ and\ \bibinfo
  {author} {\bibfnamefont{F.}~\bibnamefont{Mauri}},\ }%
  \Doi{10.1103/PhysRevLett.97.266407}{\emph{\bibinfo {title} {Nonadiabatic Kohn
  Anomaly in a Doped Graphene Monolayer}}},\ \bibfield{journal}{%
  \bibinfo {journal} {Phys. Rev. Lett.}\ }%
  \textbf{\bibinfo {volume} {97}},\ \bibinfo {pages} {266407} (\bibinfo {year}
  {2006}).%
  \bibAnnoteFile{Stop}{Lazzeri2006}%
\bibitem{Pisana2007}%
  \BibitemOpen
  \bibfield{author}{%
  \bibinfo {author} {\bibfnamefont{S.}~\bibnamefont{Pisana}}, \bibinfo {author}
  {\bibfnamefont{M.}~\bibnamefont{Lazzeri}}, \bibinfo {author}
  {\bibfnamefont{C.}~\bibnamefont{Casiraghi}}, \bibinfo {author}
  {\bibfnamefont{K.~S.}\ \bibnamefont{Novoselov}}, \bibinfo {author}
  {\bibfnamefont{A.~K.}\ \bibnamefont{Geim}}, \bibinfo {author}
  {\bibfnamefont{A.~C.}\ \bibnamefont{Ferrari}},\ and\ \bibinfo {author}
  {\bibfnamefont{F.}~\bibnamefont{Mauri}},\ }%
  \Doi{10.1038/nmat1846}{\emph{\bibinfo {title} {Breakdown of the adiabatic
  Born--Oppenheimer approximation in graphene}}},\ \bibfield{journal}{%
  \bibinfo {journal} {Nature Materials}\ }%
  \textbf{\bibinfo {volume} {6}},\ \bibinfo {pages} {198} (\bibinfo {year}
  {2007}).%
  \bibAnnoteFile{Stop}{Pisana2007}%
\bibitem{Lazzeri2008}%
  \BibitemOpen
  \bibfield{author}{%
  \bibinfo {author} {\bibfnamefont{M.}~\bibnamefont{Lazzeri}}, \bibinfo
  {author} {\bibfnamefont{C.}~\bibnamefont{Attaccalite}}, \bibinfo {author}
  {\bibfnamefont{L.}~\bibnamefont{Wirtz}},\ and\ \bibinfo {author}
  {\bibfnamefont{F.}~\bibnamefont{Mauri}},\ }%
  \Doi{10.1103/PhysRevB.78.081406}{\emph{\bibinfo {title} {Impact of the
  electron-electron correlation on phonon dispersion: Failure of LDA and GGA
  DFT functionals in graphene and graphite}}},\ \bibfield{journal}{%
  \bibinfo {journal} {Phys. Rev. B}\ }%
  \textbf{\bibinfo {volume} {78}},\ \bibinfo {pages} {081406} (\bibinfo {year}
  {2008}).%
  \bibAnnoteFile{Stop}{Lazzeri2008}%
\bibitem{Verdi2017}%
  \BibitemOpen
  \bibfield{author}{%
  \bibinfo {author} {\bibfnamefont{C.}~\bibnamefont{Verdi}}, \bibinfo {author}
  {\bibfnamefont{F.}~\bibnamefont{Caruso}},\ and\ \bibinfo {author}
  {\bibfnamefont{F.}~\bibnamefont{Giustino}},\ }%
  \Doi{10.1038/ncomms15769}{\emph{\bibinfo {title} {Origin of the crossover
  from polarons to Fermi liquids in transition metal oxides}}},\
  \bibfield{journal}{%
  \bibinfo {journal} {Nature Communications}\ }%
  \textbf{\bibinfo {volume} {8}},\ \bibinfo {pages} {15769} (\bibinfo {year}
  {2017}).%
  \bibAnnoteFile{Stop}{Verdi2017}%
\bibitem{Bisri2017}%
  \BibitemOpen
  \bibfield{author}{%
  \bibinfo {author} {\bibfnamefont{S.~Z.}\ \bibnamefont{Bisri}}, \bibinfo
  {author} {\bibfnamefont{S.}~\bibnamefont{Shimizu}}, \bibinfo {author}
  {\bibfnamefont{M.}~\bibnamefont{Nakano}},\ and\ \bibinfo {author}
  {\bibfnamefont{Y.}~\bibnamefont{Iwasa}},\ }%
  \Doi{10.1002/adma.201607054}{\emph{\bibinfo {title} {Endeavor of Iontronics:
  From Fundamentals to Applications of Ion-Controlled Electronics}}},\
  \bibfield{journal}{%
  \bibinfo {journal} {Advanced Materials}\ }%
  \textbf{\bibinfo {volume} {29}},\ \bibinfo {pages} {1607054} (\bibinfo {year}
  {2017}).%
  \bibAnnoteFile{Stop}{Bisri2017}%
\bibitem{Hu1996}%
  \BibitemOpen
  \bibfield{author}{%
  \bibinfo {author} {\bibfnamefont{B.~Y.-K.}\ \bibnamefont{Hu}}\ and\ \bibinfo
  {author} {\bibfnamefont{K.}~\bibnamefont{Flensberg}},\ }%
  \Doi{10.1103/PhysRevB.53.10072}{\emph{\bibinfo {title} {Electron-electron
  scattering in linear transport in two-dimensional systems}}},\
  \bibfield{journal}{%
  \bibinfo {journal} {Phys. Rev. B}\ }%
  \textbf{\bibinfo {volume} {53}},\ \bibinfo {pages} {10072} (\bibinfo {year}
  {1996}).%
  \bibAnnoteFile{Stop}{Hu1996}%
\bibitem{Caruso2016}%
  \BibitemOpen
  \bibfield{author}{%
  \bibinfo {author} {\bibfnamefont{F.}~\bibnamefont{Caruso}}\ and\ \bibinfo
  {author} {\bibfnamefont{F.}~\bibnamefont{Giustino}},\ }%
  \Doi{10.1103/PhysRevB.94.115208}{\emph{\bibinfo {title} {Theory of
  electron-plasmon coupling in semiconductors}}},\ \bibfield{journal}{%
  \bibinfo {journal} {Phys. Rev. B}\ }%
  \textbf{\bibinfo {volume} {94}},\ \bibinfo {pages} {115208} (\bibinfo {year}
  {2016}).%
  \bibAnnoteFile{Stop}{Caruso2016}%
\bibitem{Prandini2018a}%
  \BibitemOpen
  \bibfield{author}{%
  \bibinfo {author} {\bibfnamefont{G.}~\bibnamefont{Prandini}}, \bibinfo
  {author} {\bibfnamefont{A.}~\bibnamefont{Marrazzo}}, \bibinfo {author}
  {\bibfnamefont{I.~E.}\ \bibnamefont{Castelli}}, \bibinfo {author}
  {\bibfnamefont{N.}~\bibnamefont{Mounet}},\ and\ \bibinfo {author}
  {\bibfnamefont{N.}~\bibnamefont{Marzari}},\ }%
  \emph{\bibinfo {title} {{Precision and efficiency in solid-state
  pseudopotential calculations}}},\ \bibfield{journal}{%
  \bibinfo {journal} {http://www.materialscloud.org/sssp/}}%
   (\bibinfo {year} {2018}).%
  \bibAnnoteFile{Stop}{Prandini2018a}%
\bibitem{Hamann2013}%
  \BibitemOpen
  \bibfield{author}{%
  \bibinfo {author} {\bibfnamefont{D.~R.}\ \bibnamefont{Hamann}},\ }%
  \Doi{10.1103/PhysRevB.88.085117}{\emph{\bibinfo {title} {{Optimized
  norm-conserving Vanderbilt pseudopotentials}}}},\ \bibfield{journal}{%
  \bibinfo {journal} {Physical Review B}\ }%
  \textbf{\bibinfo {volume} {88}},\ \bibinfo {pages} {085117} (\bibinfo {year}
  {2013}).%
  \bibAnnoteFile{Stop}{Hamann2013}%
\bibitem{Vanderbilt1990}%
  \BibitemOpen
  \bibfield{author}{%
  \bibinfo {author} {\bibfnamefont{D.}~\bibnamefont{Vanderbilt}},\ }%
  \Doi{10.1103/PhysRevB.41.7892}{\emph{\bibinfo {title} {{Soft self-consistent
  pseudopotentials in a generalized eigenvalue formalism}}}},\
  \bibfield{journal}{%
  \bibinfo {journal} {Physical Review B}\ }%
  \textbf{\bibinfo {volume} {41}},\ \bibinfo {pages} {7892} (\bibinfo {year}
  {1990}).%
  \bibAnnoteFile{Stop}{Vanderbilt1990}%
\bibitem{DalCorso2014}%
  \BibitemOpen
  \bibfield{author}{%
  \bibinfo {author} {\bibfnamefont{A.}~\bibnamefont{{Dal Corso}}},\ }%
  \Doi{10.1016/J.COMMATSCI.2014.07.043}{\emph{\bibinfo {title}
  {{Pseudopotentials periodic table: From H to Pu}}}},\ \bibfield{journal}{%
  \bibinfo {journal} {Computational Materials Science}\ }%
  \textbf{\bibinfo {volume} {95}},\ \bibinfo {pages} {337} (\bibinfo {year}
  {2014}).%
  \bibAnnoteFile{Stop}{DalCorso2014}%
\bibitem{Li2013}%
  \BibitemOpen
  \bibfield{author}{%
  \bibinfo {author} {\bibfnamefont{X.}~\bibnamefont{Li}}, \bibinfo {author}
  {\bibfnamefont{J.~T.}\ \bibnamefont{Mullen}}, \bibinfo {author}
  {\bibfnamefont{Z.}~\bibnamefont{Jin}}, \bibinfo {author}
  {\bibfnamefont{K.~M.}\ \bibnamefont{Borysenko}}, \bibinfo {author}
  {\bibfnamefont{M.}~\bibnamefont{{Buongiorno Nardelli}}},\ and\ \bibinfo
  {author} {\bibfnamefont{K.~W.}\ \bibnamefont{Kim}},\ }%
  \Doi{10.1103/PhysRevB.87.115418}{\emph{\bibinfo {title} {{Intrinsic
  electrical transport properties of monolayer silicene and MoS2from first
  principles}}}},\ \bibfield{journal}{%
  \bibinfo {journal} {Physical Review B}\ }%
  \textbf{\bibinfo {volume} {87}},\ \bibinfo {pages} {115418} (\bibinfo {year}
  {2013}).%
  \bibAnnoteFile{Stop}{Li2013}%
\bibitem{Zhang2014}%
  \BibitemOpen
  \bibfield{author}{%
  \bibinfo {author} {\bibfnamefont{W.}~\bibnamefont{Zhang}}, \bibinfo {author}
  {\bibfnamefont{Z.}~\bibnamefont{Huang}}, \bibinfo {author}
  {\bibfnamefont{W.}~\bibnamefont{Zhang}},\ and\ \bibinfo {author}
  {\bibfnamefont{Y.}~\bibnamefont{Li}},\ }%
  \Doi{10.1007/s12274-014-0532-x}{\emph{\bibinfo {title} {{Two-dimensional
  semiconductors with possible high room temperature mobility }}}},\
  \bibfield{journal}{%
  \bibinfo {journal} {Nano research}\ }%
  \textbf{\bibinfo {volume} {7}},\ \bibinfo {pages} {1731} (\bibinfo {year}
  {2014}).%
  \bibAnnoteFile{Stop}{Zhang2014}%
\bibitem{Yu2016}%
  \BibitemOpen
  \bibfield{author}{%
  \bibinfo {author} {\bibfnamefont{Z.}~\bibnamefont{Yu}}, \bibinfo {author}
  {\bibfnamefont{Z.~Y.}\ \bibnamefont{Ong}}, \bibinfo {author}
  {\bibfnamefont{Y.}~\bibnamefont{Pan}}, \bibinfo {author}
  {\bibfnamefont{Y.}~\bibnamefont{Cui}}, \bibinfo {author}
  {\bibfnamefont{R.}~\bibnamefont{Xin}}, \bibinfo {author}
  {\bibfnamefont{Y.}~\bibnamefont{Shi}}, \bibinfo {author}
  {\bibfnamefont{B.}~\bibnamefont{Wang}}, \bibinfo {author}
  {\bibfnamefont{Y.}~\bibnamefont{Wu}}, \bibinfo {author}
  {\bibfnamefont{T.}~\bibnamefont{Chen}}, \bibinfo {author}
  {\bibfnamefont{Y.~W.}\ \bibnamefont{Zhang}}, \bibinfo {author}
  {\bibfnamefont{G.}~\bibnamefont{Zhang}},\ and\ \bibinfo {author}
  {\bibfnamefont{X.}~\bibnamefont{Wang}},\ }%
  \Doi{10.1002/adma.201503033}{\emph{\bibinfo {title} {{Realization of
  Room-Temperature Phonon-Limited Carrier Transport in Monolayer MoS2by
  Dielectric and Carrier Screening}}}},\ \bibfield{journal}{%
  \bibinfo {journal} {Advanced Materials}\ }%
  \textbf{\bibinfo {volume} {28}},\ \bibinfo {pages} {547} (\bibinfo {year}
  {2016}).%
  \bibAnnoteFile{Stop}{Yu2016}%
\bibitem{Yu2014}%
  \BibitemOpen
  \bibfield{author}{%
  \bibinfo {author} {\bibfnamefont{Z.}~\bibnamefont{Yu}}, \bibinfo {author}
  {\bibfnamefont{Y.}~\bibnamefont{Pan}}, \bibinfo {author}
  {\bibfnamefont{Y.}~\bibnamefont{Shen}}, \bibinfo {author}
  {\bibfnamefont{Z.}~\bibnamefont{Wang}}, \bibinfo {author}
  {\bibfnamefont{Z.-Y.}\ \bibnamefont{Ong}}, \bibinfo {author}
  {\bibfnamefont{T.}~\bibnamefont{Xu}}, \bibinfo {author}
  {\bibfnamefont{R.}~\bibnamefont{Xin}}, \bibinfo {author}
  {\bibfnamefont{L.}~\bibnamefont{Pan}}, \bibinfo {author}
  {\bibfnamefont{B.}~\bibnamefont{Wang}}, \bibinfo {author}
  {\bibfnamefont{L.}~\bibnamefont{Sun}}, \bibinfo {author}
  {\bibfnamefont{J.}~\bibnamefont{Wang}}, \bibinfo {author}
  {\bibfnamefont{G.}~\bibnamefont{Zhang}}, \bibinfo {author}
  {\bibfnamefont{W.~Y.}\ \bibnamefont{Zhang}}, \bibinfo {author}
  {\bibfnamefont{Y.}~\bibnamefont{Shi}},\ and\ \bibinfo {author}
  {\bibfnamefont{X.}~\bibnamefont{Wang}},\ }%
  \Doi{10.1038/ncomms6290}{\emph{\bibinfo {title} {{Towards intrinsic charge
  transport in monolayer molybdenum disulfide by defect and interface
  engineering }}}},\ \bibfield{journal}{%
  \bibinfo {journal} {Nature Communications}\ }%
  \textbf{\bibinfo {volume} {5}},\ \bibinfo {pages} {5290} (\bibinfo {year}
  {2014}).%
  \bibAnnoteFile{Stop}{Yu2014}%
\bibitem{Radisavljevic2013}%
  \BibitemOpen
  \bibfield{author}{%
  \bibinfo {author} {\bibfnamefont{B.}~\bibnamefont{Radisavljevic}}\ and\
  \bibinfo {author} {\bibfnamefont{A.}~\bibnamefont{Kis}},\ }%
  \Doi{10.1038/nmat3687}{\emph{\bibinfo {title} {{Mobility engineering and a
  metal--insulator transition in monolayer MoS2}}}},\ \bibfield{journal}{%
  \bibinfo {journal} {Nature Materials}\ }%
  \textbf{\bibinfo {volume} {12}},\ \bibinfo {pages} {815} (\bibinfo {year}
  {2013}).%
  \bibAnnoteFile{Stop}{Radisavljevic2013}%
\bibitem{Jo2014}%
  \BibitemOpen
  \bibfield{author}{%
  \bibinfo {author} {\bibfnamefont{S.}~\bibnamefont{Jo}}, \bibinfo {author}
  {\bibfnamefont{N.}~\bibnamefont{Ubrig}}, \bibinfo {author}
  {\bibfnamefont{H.}~\bibnamefont{Berger}}, \bibinfo {author}
  {\bibfnamefont{A.~B.}\ \bibnamefont{Kuzmenko}},\ and\ \bibinfo {author}
  {\bibfnamefont{A.~F.}\ \bibnamefont{Morpurgo}},\ }%
  \Doi{10.1021/nl500171v}{\emph{\bibinfo {title} {{Mono- and Bilayer WS2
  Light-Emitting Transistors}}}},\ \bibfield{journal}{%
  \bibinfo {journal} {Nano Letters}\ }%
  \textbf{\bibinfo {volume} {14}},\ \bibinfo {pages} {2019} (\bibinfo {year}
  {2014}).%
  \bibAnnoteFile{Stop}{Jo2014}%
\bibitem{Ovchinnikov2014}%
  \BibitemOpen
  \bibfield{author}{%
  \bibinfo {author} {\bibfnamefont{D.}~\bibnamefont{Ovchinnikov}}, \bibinfo
  {author} {\bibfnamefont{A.}~\bibnamefont{Allain}}, \bibinfo {author}
  {\bibfnamefont{Y.-S.}\ \bibnamefont{Huang}}, \bibinfo {author}
  {\bibfnamefont{D.}~\bibnamefont{Dumcenco}},\ and\ \bibinfo {author}
  {\bibfnamefont{A.}~\bibnamefont{Kis}},\ }%
  \Doi{10.1021/nn502362b}{\emph{\bibinfo {title} {{Electrical Transport
  Properties of Single-Layer WS2}}}},\ \bibfield{journal}{%
  \bibinfo {journal} {ACS Nano}\ }%
  \textbf{\bibinfo {volume} {8}},\ \bibinfo {pages} {8174} (\bibinfo {year}
  {2014}).%
  \bibAnnoteFile{Stop}{Ovchinnikov2014}%
\bibitem{Aji2017}%
  \BibitemOpen
  \bibfield{author}{%
  \bibinfo {author} {\bibfnamefont{A.~S.}\ \bibnamefont{Aji}}, \bibinfo
  {author} {\bibfnamefont{P.}~\bibnamefont{Sol{\'\i}s-Fern{\'a}ndez}}, \bibinfo
  {author} {\bibfnamefont{H.~G.}\ \bibnamefont{Ji}}, \bibinfo {author}
  {\bibfnamefont{K.}~\bibnamefont{Fukuda}},\ and\ \bibinfo {author}
  {\bibfnamefont{H.}~\bibnamefont{Ago}},\ }%
  \Doi{10.1002/adfm.201703448}{\emph{\bibinfo {title} {{High Mobility WS2
  Transistors Realized by MultilayerGraphene Electrodes and Application to High
  Responsivity Flexible Photodetectors}}}},\ \bibfield{journal}{%
  \bibinfo {journal} {Advanced Functional Materials}\ }%
  \textbf{\bibinfo {volume} {27}},\ \bibinfo {pages} {1703448} (\bibinfo {year}
  {2017}).%
  \bibAnnoteFile{Stop}{Aji2017}%
\bibitem{Cui2015}%
  \BibitemOpen
  \bibfield{author}{%
  \bibinfo {author} {\bibfnamefont{Y.}~\bibnamefont{Cui}}, \bibinfo {author}
  {\bibfnamefont{R.}~\bibnamefont{Xin}}, \bibinfo {author}
  {\bibfnamefont{Z.}~\bibnamefont{Yu}}, \bibinfo {author}
  {\bibfnamefont{Y.}~\bibnamefont{Pan}}, \bibinfo {author}
  {\bibfnamefont{Z.-Y.}\ \bibnamefont{Ong}}, \bibinfo {author}
  {\bibfnamefont{X.}~\bibnamefont{Wei}}, \bibinfo {author}
  {\bibfnamefont{J.}~\bibnamefont{Wang}}, \bibinfo {author}
  {\bibfnamefont{H.}~\bibnamefont{Nan}}, \bibinfo {author}
  {\bibfnamefont{Z.}~\bibnamefont{Ni}}, \bibinfo {author}
  {\bibfnamefont{Y.}~\bibnamefont{Wu}}, \bibinfo {author}
  {\bibfnamefont{T.}~\bibnamefont{Chen}}, \bibinfo {author}
  {\bibfnamefont{Y.}~\bibnamefont{Shi}}, \bibinfo {author}
  {\bibfnamefont{B.}~\bibnamefont{Wang}}, \bibinfo {author}
  {\bibfnamefont{G.}~\bibnamefont{Zhang}}, \bibinfo {author}
  {\bibfnamefont{Y.-W.}\ \bibnamefont{Zhang}},\ and\ \bibinfo {author}
  {\bibfnamefont{X.}~\bibnamefont{Wang}},\ }%
  \Doi{10.1002/adma.201502222}{\emph{\bibinfo {title} {{High-Performance
  Monolayer WS2 Field-Effect Transistors on High-k Dielectrics}}}},\
  \bibfield{journal}{%
  \bibinfo {journal} {Advanced Materials}\ }%
  \textbf{\bibinfo {volume} {27}},\ \bibinfo {pages} {5230} (\bibinfo {year}
  {2015}).%
  \bibAnnoteFile{Stop}{Cui2015}%
\bibitem{Iqbal2015}%
  \BibitemOpen
  \bibfield{author}{%
  \bibinfo {author} {\bibfnamefont{M.~Z.}\ \bibnamefont{Iqbal},
  \bibfnamefont{M~Waqasand~Iqbal}}, \bibinfo {author} {\bibfnamefont{M.~F.}\
  \bibnamefont{Khan}}, \bibinfo {author} {\bibfnamefont{M.~A.}\
  \bibnamefont{Shehzad}}, \bibinfo {author}
  {\bibfnamefont{Y.}~\bibnamefont{Seo}}, \bibinfo {author}
  {\bibfnamefont{J.~H.}\ \bibnamefont{Park}}, \bibinfo {author}
  {\bibfnamefont{C.}~\bibnamefont{Hwang}},\ and\ \bibinfo {author}
  {\bibfnamefont{J.}~\bibnamefont{Eom}},\ }%
  \Doi{https://doi.org/10.1038/srep10699}{\emph{\bibinfo {title}
  {{High-mobility and air-stable single-layer WS2 field-effect transistors
  sandwiched between chemical vapor deposition-grown hexagonal BN films}}}},\
  \bibfield{journal}{%
  \bibinfo {journal} {Scientific Reports}\ }%
  \textbf{\bibinfo {volume} {5}},\ \bibinfo {pages} {10699} (\bibinfo {year}
  {2015}).%
  \bibAnnoteFile{Stop}{Iqbal2015}%
\bibitem{Huang2014}%
  \BibitemOpen
  \bibfield{author}{%
  \bibinfo {author} {\bibfnamefont{J.-K.}\ \bibnamefont{Huang}}, \bibinfo
  {author} {\bibfnamefont{J.}~\bibnamefont{Pu}}, \bibinfo {author}
  {\bibfnamefont{C.-L.}\ \bibnamefont{Hsu}}, \bibinfo {author}
  {\bibfnamefont{M.-H.}\ \bibnamefont{Chiu}}, \bibinfo {author}
  {\bibfnamefont{J.}~\bibnamefont{Zhen-Yu}}, \bibinfo {author}
  {\bibfnamefont{Y.-H.}\ \bibnamefont{Chang}}, \bibinfo {author}
  {\bibfnamefont{W.-H.}\ \bibnamefont{Chang}}, \bibinfo {author}
  {\bibfnamefont{Y.}~\bibnamefont{Iwasa}}, \bibinfo {author}
  {\bibfnamefont{T.}~\bibnamefont{Takenobu}},\ and\ \bibinfo {author}
  {\bibfnamefont{L.-J.}\ \bibnamefont{Li}},\ }%
  \Doi{10.1021/nn405719x}{\emph{\bibinfo {title} {{Large-Area Synthesis of
  Highly Crystalline WSe2 Monolayers and Device Applications}}}},\
  \bibfield{journal}{%
  \bibinfo {journal} {ACS Nano}\ }%
  \textbf{\bibinfo {volume} {8}},\ \bibinfo {pages} {923} (\bibinfo {year}
  {2014}).%
  \bibAnnoteFile{Stop}{Huang2014}%
\bibitem{Allein2014}%
  \BibitemOpen
  \bibfield{author}{%
  \bibinfo {author} {\bibfnamefont{A.}~\bibnamefont{Allain}}\ and\ \bibinfo
  {author} {\bibfnamefont{A.}~\bibnamefont{Kis}},\ }%
  \Doi{10.1021/nn5021538}{\emph{\bibinfo {title} {{Electron and Hole Mobilities
  in Single-Layer WSe2}}}},\ \bibfield{journal}{%
  \bibinfo {journal} {ACS Nano}\ }%
  \textbf{\bibinfo {volume} {8}},\ \bibinfo {pages} {7180} (\bibinfo {year}
  {2014}).%
  \bibAnnoteFile{Stop}{Allein2014}%
\bibitem{Note1}%
  \BibitemOpen
  \bibinfo {note} {This number is estimated by solving the BTE for each mode
  successively, setting the coupling to the other modes to zero. This is simply
  an educated estimation. Indeed, the process is not strictly valid
  quantitatively since the solution to the BTE including all phonon modes is
  not the sum of the contributions from each mode.}%
  \bibAnnoteFile{Stop}{Note1}%
\bibitem{Chakraborty2012}%
  \BibitemOpen
  \bibfield{author}{%
  \bibinfo {author} {\bibfnamefont{B.}~\bibnamefont{Chakraborty}}, \bibinfo
  {author} {\bibfnamefont{A.}~\bibnamefont{Bera}}, \bibinfo {author}
  {\bibfnamefont{D.~V.~S.}\ \bibnamefont{Muthu}}, \bibinfo {author}
  {\bibfnamefont{S.}~\bibnamefont{Bhowmick}}, \bibinfo {author}
  {\bibfnamefont{U.~V.}\ \bibnamefont{Waghmare}},\ and\ \bibinfo {author}
  {\bibfnamefont{A.~K.}\ \bibnamefont{Sood}},\ }%
  \Doi{10.1103/PhysRevB.85.161403}{\emph{\bibinfo {title} {{Symmetry-dependent
  phonon renormalization in monolayer MoS 2 transistor}}}},\
  \bibfield{journal}{%
  \bibinfo {journal} {Physical Review B}\ }%
  \textbf{\bibinfo {volume} {85}},\ \bibinfo {pages} {161403} (\bibinfo {year}
  {2012}).%
  \bibAnnoteFile{Stop}{Chakraborty2012}%
\bibitem{Takagi}%
  \BibitemOpen
  \bibfield{author}{%
  \bibinfo {author} {\bibfnamefont{S.}~\bibnamefont{Takagi}}, \bibinfo {author}
  {\bibfnamefont{A.}~\bibnamefont{Toriumi}}, \bibinfo {author}
  {\bibfnamefont{M.}~\bibnamefont{Iwase}},\ and\ \bibinfo {author}
  {\bibfnamefont{H.}~\bibnamefont{Tango}},\ }%
  \Doi{10.1109/16.337450}{\emph{\bibinfo {title} {On the universality of
  inversion layer mobility in Si MOSFET's: Part II-effects of surface
  orientation}}},\ \bibfield{journal}{%
  \bibinfo {journal} {IEEE Transactions on Electron Devices}\ }%
  \textbf{\bibinfo {volume} {41}},\ \bibinfo {pages} {2363} (\bibinfo {year}
  {1994}).%
  \bibAnnoteFile{Stop}{Takagi}%
\bibitem{Rode1971}%
  \BibitemOpen
  \bibfield{author}{%
  \bibinfo {author} {\bibfnamefont{D.~L.}\ \bibnamefont{Rode}}\ and\ \bibinfo
  {author} {\bibfnamefont{S.}~\bibnamefont{Knight}},\ }%
  \Doi{10.1103/PhysRevB.3.2534}{\emph{\bibinfo {title} {Electron Transport in
  GaAs}}},\ \bibfield{journal}{%
  \bibinfo {journal} {Phys. Rev. B}\ }%
  \textbf{\bibinfo {volume} {3}},\ \bibinfo {pages} {2534} (\bibinfo {year}
  {1971}).%
  \bibAnnoteFile{Stop}{Rode1971}%
\bibitem{Zhou2016}%
  \BibitemOpen
  \bibfield{author}{%
  \bibinfo {author} {\bibfnamefont{J.-J.}\ \bibnamefont{Zhou}}\ and\ \bibinfo
  {author} {\bibfnamefont{M.}~\bibnamefont{Bernardi}},\ }%
  \Doi{10.1103/PhysRevB.94.201201}{\emph{\bibinfo {title} {Ab initio electron
  mobility and polar phonon scattering in GaAs}}},\ \bibfield{journal}{%
  \bibinfo {journal} {Phys. Rev. B}\ }%
  \textbf{\bibinfo {volume} {94}},\ \bibinfo {pages} {201201} (\bibinfo {year}
  {2016}).%
  \bibAnnoteFile{Stop}{Zhou2016}%
\end{thebibliography}%

\end{document}